\documentclass[5p,sort,compress]{elsarticle}%,draft
\usepackage{fancyhdr}
\usepackage{amsfonts}
\usepackage{url}
\usepackage{amsmath}
\usepackage[labelfont=bf,justification=raggedright,singlelinecheck=false]{caption}
\usepackage{array}
\usepackage{tabu}
\usepackage{multirow}
\usepackage{graphicx}
\usepackage{amsmath}
\usepackage{algorithm}
\usepackage[noend]{algpseudocode}
\usepackage{pdflscape}
\usepackage{rotating}
\usepackage{wrapfig}
\usepackage[inline, shortlabels]{enumitem}
\usepackage{graphbox}
\usepackage{mathtools}
\usepackage{amsmath}
\newcommand*\rfrac[2]{{}^{#1}\!/_{#2}}
\newcolumntype{v}{>{\centering\arraybackslash}m{.18\linewidth} }
\usepackage{hyperref}
\hypersetup{
    colorlinks=true,
    linkcolor=blue,
    filecolor=magenta,      
    urlcolor=cyan,
}
\usepackage[labelfont=bf,   justification=justified,   format=plain]{caption} % 'format=plain' avoids hanging indentation, font=small,
\begin{document}
%\journal{Signal Processing}
\title{TRLG: Fragile blind quad watermarking for image tamper detection and recovery by providing compact digests with quality optimized using LWT and GA}
\author[fum]{Behrouz Bolourian Haghighi}
\ead{b.bolourian@stu.um.ac.ir}
\author[fum]{Amir Hossein Taherinia\corref{cor1}}
\ead{taherinia@um.ac.ir}
\author[fum]{Amir Hossein Mohajerzadeh}
\ead{mohajerzadeh@um.ac.ir}
\cortext[cor1]{Corresponding author}
\address[fum]{Computer Engineering Department, Ferdowsi University of Mashhad, Mashhad, Iran }
%-----------------------------------------------------------------------------------------------------------------------------------------------------------------------------------------------------------------------------------------------------
\begin{abstract}
In this paper, an efficient fragile blind quad watermarking scheme for image tamper detection and recovery based on lifting wavelet transform and genetic algorithm is proposed. TRLG generates four compact digests with super quality based on lifting wavelet transform and halftoning technique by distinguishing the types of image blocks. In other words, for each 2$\times$2 non-overlap blocks, four chances for recovering destroyed blocks are considered. A special parameter estimation technique based on genetic algorithm is performed to improve and optimize the quality of digests and watermarked image. Furthermore, CCS map is used to determine the mapping block for embedding information, encrypting and confusing the embedded information. In order to improve the recovery rate, Mirror-aside and Partner-block are proposed. The experiments that have been conducted to evaluate the performance of TRLG proved the superiority in terms of quality of the watermarked and recovered image, tamper localization and security compared with state-of-the-art methods. The results indicate that the PSNR and SSIM of the watermarked image are about 46 dB and approximately one, respectively. Also, the mean of PSNR and SSIM of several recovered images which has been destroyed about 90\% is reached to 24 dB and 0.86, respectively.
\end{abstract}
\begin{keyword}
Data hiding \sep Watermarking \sep Tamper detection and recovery \sep Texture analysis \sep Genetic algorithm \sep Lifting wavelet transform. % \sep Halftoning technique. 
\end{keyword}
\maketitle
%-----------------------------------------------------------------------------------------------------------------------------------------------------------------------------------------------------------------------------------------------------
\section{Introduction}
In today’s digital world, the communication networks like the Internet technology has rapidly developed and extended as a suitable channel for transferring all kinds of data, particularly multimedia data. Since the amount of multimedia data in Internet transmission increases extensively, the problem of copyright and integrity protection has become a very serious issue and arisen more attention nowadays \cite{ref11, ref8}. In other words, digital data can be easily copied or maliciously tampered by using various tools without loss of quality and perceived by the human visual system. Therefore, the secure strategies should be designed for solving these challenges. Among the solutions for these issues, digital watermarking techniques \cite{ref11, ref8, ref10} are the most popular until now.

Digital watermarking is a science and art that imperceptibly hides useful information into the digital media for various goals such as copyright protection \cite{ref26, ref27, ref29, ref30}, broadcast monitoring, authentication and etc \cite{ref11, ref8, ref10}. In recent years, watermarking has attracted much attention as an effective solution to guaranty the integrity and authenticity of digital images from being illegally modified \cite{ref9, ref1}. To do so, the watermarks information include authentication pattern and digest are embedded into host image without severely affecting the perceptual quality to detect and recover tampered regions in the receiver side. It should be noted, the host image is referred as the original image without the embedded watermark, while the image that is obtained by embedding the watermark into host image without serious destroying the quality is named as the watermarked image. Generally, these methods can be classified into two categories as fragile and semi-fragile techniques \cite{ref8, ref1}. The fragile scheme makes the hidden information invalid after any modifications in the content of the watermarked image. In another word, it can be introduced as a design of watermarks that become undetectable in the view of the slightest modification to the host signal. Therefore, fragile schemes are mainly used for authentication goals \cite{ref7, ref12, ref13, ref14, ref15, ref16, ref17, ref18, ref19, ref20, ref21, ref22, ref23}. On the other hand, a semi-fragile technique aims at making hidden information fragile to modify the content of the signal, and robust to all possible attacks such as compression, image processing operations and etc.

The fragile watermarking has its own specific requirements include increasing imperceptibility, capacity, and security \cite{ref8, ref9}. The imperceptibility denotes the idea that an embedded watermark must be invisible to the human visual system. In other words, the embedded information should keep the images’ visual quality. The size of information which embedded in the host is presented by capacity. Finally, the security of the watermarking system is referred as the safety of embedded watermark into the host, even though the hacker having full knowledge of embedding and detecting procedures. In this filed, security has become one of the most important and challenging problems for watermarking schemes.
%----------------------------------------------------------------------------------------
\subsection{Literature review}
In this subsection, a brief review of several fragile schemes which proposed in the last decade is presented. Also, the advantages and weaknesses of each method are described, and compared with each other. Nowadays, the fragile watermarking authentication schemes have been extended and developed extremely. These methods can be divided into two types. Some methods only focused on locating the suspicious regions in host image \cite{ref23, ref15, ref18, ref32}. On the contrary, more schemes can recovered tampered parts using the information which embedded in non-tampered information, clearly \cite{ref7, ref12, ref13, ref14, ref16, ref17, ref19, ref20, ref21, ref22}. 

In \cite{ref7}, an effective dual watermark method for image tamper detection and recovery was proposed. In this scheme, two chances are provided in the entire image to recover tampered regions for the first time. Consequently, the recovery rate and quality of the recovered image are efficiently optimized rather than previous methods. In addition, the hierarchical authentication is employed to detect the tampered regions. In \cite{ref15}, a probability-based tampering detection scheme for digital images was presented to reduce errors in the authentication phase. In another word,  a probability theory is used to enhance authentication accuracy. The experimental results show that the proposed scheme provides good accuracy in terms of detection precision. In \cite{ref32}, an image authentication scheme based on absolute moment block truncation coding was proposed. In this scheme, a hybrid mechanism was employed to hide the authentication watermark using AMBTC and improve the embedding efficiency. In embedding phase, the watermark is embedded into Bitmap or quantization levels based on the texture of blocks. The experimental results illustrate that the scheme can effectively thwart collage attack. Another self-embedding fragile watermarking scheme was presented in \cite{ref13}, as a novel image tamper localization and recovery algorithm based on watermarking technology. The security of this method has been increased by using non-linear chaotic sequence. In order to generate the digest, DCT is applied in coefficients of each 2$\times$2 block and embedded into another block according to the block mapping. A novel chaos-based fragile watermarking for image tampering detection and self-recovery was presented in \cite{ref12}. In this scheme, to determine blocks mapping, a new chaotic sequence generator as the cross chaotic map is employed. Hence, the security is increased due to the application of this map with many parameters, which can be used as keys. Similarly, two chances are considered to recover 2$\times$2 modified blocks. An effective Singular Value Decomposition based image tampering detection and self-recovery using active watermarking were proposed in \cite{ref14}. In this method, 12-bit tamper detection data were generated and embedded in a random block after being encrypted. One of the positive aspects of the proposed scheme to the previous schemes is the ability to detect tampered region under various security attacks include vector-quantization and collage attacks.

In \cite{ref16}, authors presented an efficient fragile watermarking scheme for image authentication and restoration based on Discrete Cosine Transform. In this scheme, the host is divided into 2$\times$2 non-overlapping blocks. Similar to most schemes, for each block 12 bits watermark is generated from the five Most Significant Bits of each pixel and are embedded into the three Least Significant Bits of the pixels corresponding to the mapped block. In addition, the proposed scheme uses two levels encoding for content correction bits generation. In \cite{ref19}, an image tamper detection and recovery scheme using adaptive embedding rules was presented. One of the major novelty of this method is used smoothness to distinguish the characteristics of image blocks. Accordingly, the different watermark embedding, tamper detection, and recovery strategies were designed and applied to different block types. Hence, information of authentication and recovery can be effectively embedded in a limited space to increase information hiding efficiency.  Experimental result showed that the proposed scheme causes less damage to the original image compared to the most fragile scheme. In the scheme \cite{ref20}, a DCT based effective self-embedding watermarking scheme for image tamper detection and localization with recovery capability was presented. In this scheme, as most schemes for each 2$\times$2 non-overlapping block, two authentication, and ten recovery bits are generated from the five Most Significant Bits of pixels.  The experimental results illustrate that the proposed scheme not only outperforms high-quality restoration, also removes the blocking artifacts. The authors of scheme \cite{ref18}, proposed image tamper detection scheme based on fragile watermarking and Faber-Schauder wavelet. The maximum coefficients of FSDWT are utilized with a logo to generate the watermark which is embedded in the Least Significant Bit of specified pixels in the host. In \cite{ref23}, a novel efficient reversible image authentication method using improved PVO and LSB substitution techniques was presented. In this scheme, instead of embedding the block-independent AC as the previous work, the proposed scheme embedded the hashed value of block features. In addition, a mechanism to deal with the overflow and underflow problems was considered. The proposed scheme \cite{ref18, ref23} achieved high image quality, low complexity computing, but the main drawback of this method is the inability to recover tempered regions.

Another scheme as improved image tamper localization using chaotic maps and self-recovery was proposed in \cite{ref17}. In this scheme, the authentication bits of a 2$\times$2 image block is generated using the chaotic maps. Thereinafter, for each non-overlapping block, two different sets of recovery bits of length 5 and 3 were computed and each one is embedded into randomly selected distinct blocks. In \cite{ref21}, a new fragile image watermarking with pixel-wise recovery based on overlapping embedding strategy was presented. In this work, the block-wise mechanism for tampering localization, and the pixel-wise mechanism for content recovery are considered. Compared to other methods, the proposed scheme can achieve superior performance of tampering recovery even for larger tampering rates. To achieve better performance of tampering recovery, authors in \cite{ref22} proposed hierarchical recovery for tampered images based on watermark self-embedding. In this scheme, the higher MSB layers of tampered parts have higher priority to be corrected than the lower MSB layers. Hence, the quality of the recovered image can be improved, especially for larger tampering rates. Experimental results demonstrate the effectiveness and superiority of the proposed scheme compared to previous methods.

In \cite{ref1}, a fragile and blind dual watermarking for image tamper detection and self-recovery based on Lifting Wavelet Transform and halftoning technique was proposed. In order to improve quality of the recovered image, two chances are provided by embedding a novel LWT-based digest and halftone version. In addition, to enhance the quality of the LWT-based digest, a new LSB$_{Rounding}$ technique was proposed. Experimental results prove the effectiveness, imperceptibility and real-time requirement of TRLH compared to another scheme which reviewed until now, especially in term of quality of watermarked and recovered image and security. In addition, TRLH not only outperforms high-quality restoration effectively but also removes the blocking artifacts and increase the accuracy of tamper localization due to use of very small size blocks. 

Totally, the fragile methods which proposed in recent years have low visual quality for watermarked and recovered image; Also, the low recovery rate under large tampering, weak localization, and poor security was observed. The most schemes have a severe security threat because of the independence between content and the watermark. In addition, the mostly schemes which proposed in recent years are vulnerable against vector quantization, collage and protocol attacks. 
%----------------------------------------------------------------------------------------
\subsection{Key contributions of TRLG}
In this paper, in order to perform better performance of visual quality for both watermarked and recovered image, and also improve security and overcome the mentioned challenges, an efficient fragile blind quad watermarking scheme for image tamper detection and recovery based on Lifting Wavelet Transform (LWT) and Genetic Algorithm (GA) is proposed. TRLG provides interesting extensions to the most important limitations of some of the previous state-of-the-art schemes. 

In TRLG, the digests classified into two categorize as primary and secondary digests. The two primary digests are generated based on LWT, and the rest two secondary digests are obtained by Floyd kernel of halftoning techniques. LWT (Haar, integer) \cite{ref31, ref33} is used because this transform uses the integer coefficients and has less computational time and memory requirement than traditional wavelet. In order to improve and optimize the quality of primary digests GA \cite{ref4, ref3} is employed. The utilizing GA avoids the exhausting searching and allows us to intelligently classify blocks of the image in terms of texture into flat or rough. Experimental results will show that the generated digest have better quality and decrease blocky artifact for recovered image compared to traditional digests which achieved based on averaging pixels, DCT-based or MSBs planes. 

Furthermore, In TRLG, to increase recovery rate and guaranty quality of recover image more and more, a novel mapping strategy for shuffling four digests is considered. Based on this technique, the coefficients of each digest is embedded in host image, so that the maximum distance between the coefficients of other digest and the initial position of original values is achieved. In TRLG, to enhance the security and raise detection accuracy a new chaotic map as CCS has been used. The irregular outputs are used to shuffle digest and improve the security of watermark. During watermark bits embedding process, first, authentication bits and digest are combined to form the watermark data in the LSBs by using LSB matching. Next, In order to avoid special tamperings such as vector-quantization, collage-attack, and protocol attack, the watermark is encrypted and permuted per block. In this way, a small non-overlapping block sized 2$\times$2 is used to improve the accuracy of localization. 

Moreover, In TRLG, to improve quality of the watermarked image, and also improve the security of watermarks, the embedded watermark in each block of the image is encrypted with a key that intelligently selected with GA. In other words, GA is applied to intelligently optimum and modified the watermark’s values of each block to decrease the difference between watermarked and original values, and also achieve the high level of security. Generally, applying optimization algorithms into watermarking techniques is practical and effective. Experimental results of other state-of-the-art methods are compared with TRLG, and it is revealed that the proposed scheme exhibits excellent quality for watermarked and recovered image, and as well as improve security. 

Generally, TRLG makes three main contributions. First, to the best knowledge of the authors, this is the first work that generating compact digests with quality optimization. Also, it is the first time that provides more than two chances for recover tampered regions. Second, in TRLG which is fragile scheme, generating digest and embedding watermark are modeled as a search and optimization problem. Third, combining chaotic maps, and utilized various keys to enhance the security of the watermarking system.
%---------------------------------------------------------------------------------------------------
\subsection{Road map}
The remainder of this paper is organized as follows: Section 2 briefly explains some background material for TRLG. In Section 3, the design and implementation of TRLG are described in detail. Next, the experimental evaluation scenario and details of comparison with the fragile state-of-the-art methods are described in Section 4. Finally, the conclusion and future scope of TRLG is found in Section 5.
%---------------------------------------------------------------------------------------------------
\section{Background}
In this section, some background material for the subsequent section is presented. First, the Chebyshev-Chebyshev (CCS) chaotic map is introduced. Next, a brief review of the Genetic algorithm (GA) is described. Finally, a new inverse halftoning method is presented.
%---------------------------------------------------------------------------------------------------
\subsection{Chebyshev-Chebyshev map (CCS)}
\label{sec:Chebyshev}    
The chaotic maps are the simple and efficient technique that is utilized in watermarking schemes for shuffling and encrypting the watermark.

The Logistic map is one of the popular and simplest 1D chaotic map which is used in this field, widely. The random sequence of this map is generated by Eq. (\ref{eq:Logistic}):
\begin{equation}
x_{n+1} = \mu\times x_n\times (1-x_n)
\label{eq:Logistic}
\end{equation}
where $\mu\in(0, 4]$ and $x_0$ is control parameter and initial value of map, respectively. This map has two main drawbacks: firstly, Its chaotic range is limited [3.57, 4], and secondly, $\mu$ beyond the range cannot generate chaotic behaviors \cite{ref2}. 

To overcome these issues, the fusion of two chaotic maps include Chebyshev and Logistic maps with well performance is proposed as Chebyshev-Chebyshev map in \cite{ref2}. The CCS is described by Eq. (\ref{eq:CCS}):
\begin{align}
x_{n+1} &= F(u, x_n, k) \nonumber \\
             &= G(u, x_n) \times H(k) - \lfloor G(u, x_n) \times H(k) \rfloor  \nonumber \\
G(u, x_n) &= cos((\mu + 1) \times arccos(x_{n})) \nonumber \\
H(k) &= 2^k, 8 \leq k \leq 20 
\label{eq:CCS}
\end{align}
where $\mu \in(0, 10]$ and $k$ are control parameters, and $x_0$ is the initial value of sequence. The chaotic performance of CCS is much better than single map.
%---------------------------------------------------------------------------------------------------
\subsection{Genetic algorithm}
\begin{figure}[t]
\center
\includegraphics[width=0.35\textwidth,trim=0cm 17.5cm 10cm 0.5cm,clip]{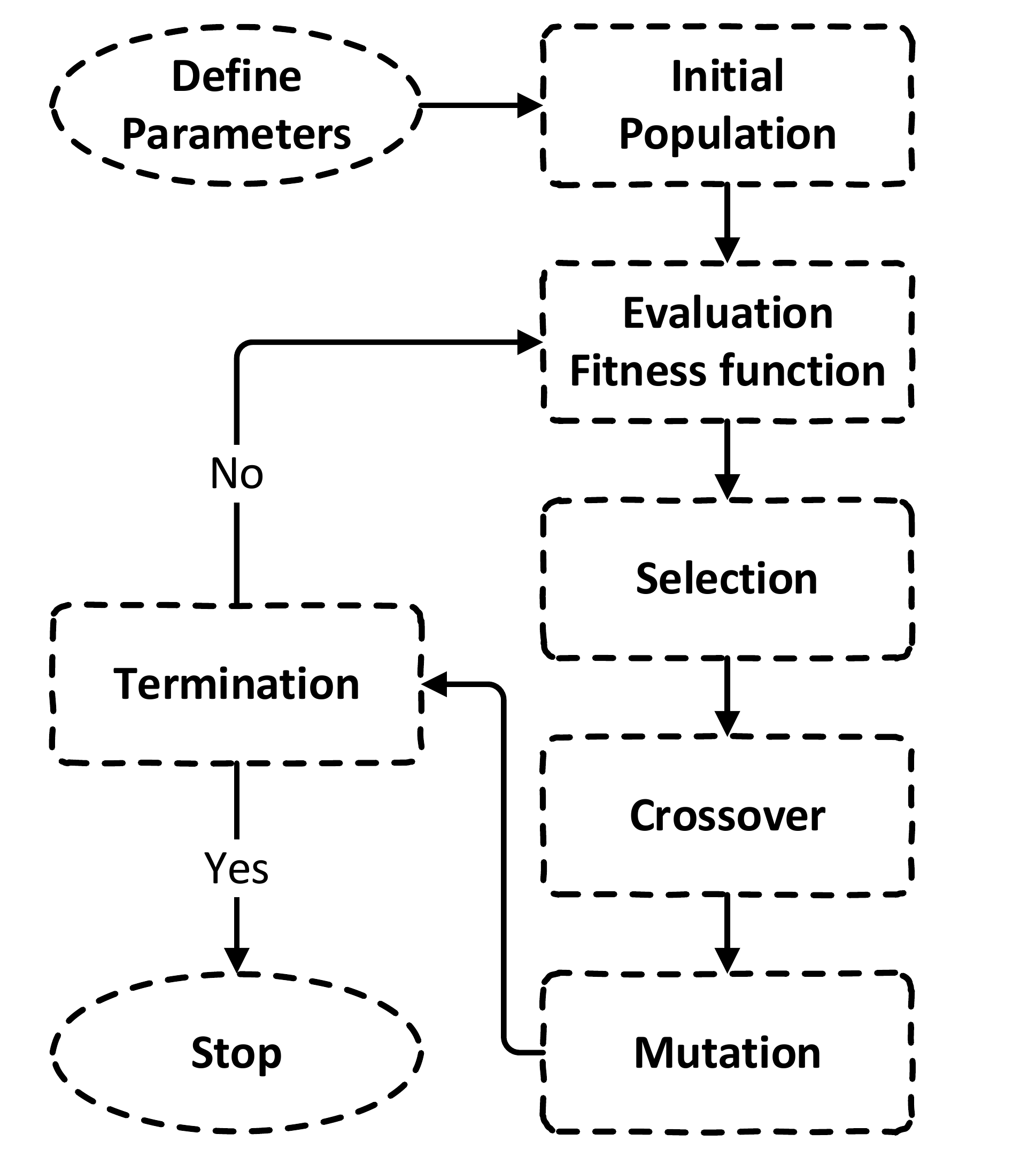}
\caption{The block diagram of Genetic Algorithm (GA).}
\label{fig:GA}
\end{figure}

Genetic algorithm (GA) is one of the famous optimization tools in artificial intelligent that introduced by Holland \cite{ref4, ref3}. It is a heuristic searching algorithm based on the mechanism of natural selection and genetics that find the best global minimum or maximum solutions in large space. The optimization problem based on GA is modeled by defining the chromosome, fitness function, and three main operators such as selection, crossover, and mutation. The whole steps of GA are shown in Fig. \ref{fig:GA}. 

The process is started with an initial population of chromosomes that represent the variables of the problem by an encoded binary string. The initial population is selected randomly from sets of possible solutions. The binary strings are adjusted to maximize or minimize the fitness values. To do so, a fitness function is utilized to measure the quality of each chromosome in the population. It should be noted, the fitness function should be carefully selected based on the requirement of the optimization problem. Next, GA tries to produce further possible solutions to achieve the desired optimization. In the other word, the next generation will be generated from a particular group of chromosomes to survive whose fitness values are high. Hence, three genetic operators are triggered to recombine the composition of the genes to create new chromosomes over successive generations. A brief summary for these basic operators can be summarized as follows:

\begin{description}[font=$\bullet$\scshape\bfseries,leftmargin=0cm]
\item \textbf{Selection:} In this step, the portion of fitter chromosomes are selected to generate new population, similar to the natural world. The chromosome that holds higher fitness value, subsequently, have the high chance to be survived. In another word, a part of the low fitness chromosomes is ignored through this natural selection step.
\item \textbf{Crossover:} In this step, pairs of optimal chromosomes among the survived chromosomes are chosen as parents to produce two new children. Evidently, the chromosomes with the higher fitness values generate more children. To do so, a crossover point is selected between the first and last chromosomes of the parent chromosomes. Next, two new children are generated by swapping the fraction of each chromosome after the crossover point. 
\item \textbf{Mutation:} Finally, to avoid GA get trapped on a local optimum and keeps GA from converging fast, the mutation operator is employed. To do so, some random positions of the chromosomes are flipped by changing 0 to 1 and vice versa.
\end{description}

At the end, the GA period is repeated until the desired termination criterion is satisfied or the number of iteration is reached.
%---------------------------------------------------------------------------------------------------
%---------------------------------------------------------------------------------------------------
\subsection{Halftone  technique}
Digital halftoning is a technique to generate halftone version of the image by homogeneously distributed of the black and white pixel from continues tone \cite{ref28, ref25, ref24}. In order to generate halftone version of the image, a Floyd kernel (Filter) is chosen \cite{ref33}. This kernel is illustrated in Eq. (\ref{eq:Filter}):
\begin{equation}
K = \frac{1}{16}\begin{bmatrix}
     &&\\
     &*&7\\
     3&5&1\\
     \end{bmatrix}
\label{eq:Filter}
\end{equation}
where $*$ represent current pixel.

One of the major applications of halftone technique is inverse halftoning. In this process, a halftone version of the image is used to reconstruct the continues tone version of the image. Noways, several methods have been proposed for this aim, but most of them have low quality for inverse version compared to the original. In TRLG, a novel and effective inverse halftoning technique base on Deep Convolution Neural Network that proposed in \cite{ref5} is utilized. In order to map a halftone version of the image to continues tone, a deep CNN as a nonlinear transform form is used. For this aim, a pre-trained deep CNN as a feature extractor is employed to construct the objective function for the training of the transformation CNN. The experimental results illustrate that it can create the inverse halftoned image with high image quality, compared to WInHD \cite{ref6} which is used in \cite{ref1}. For more information about the process of generating halftone version and the inverse method, refer to \cite{ref5}.
%---------------------------------------------------------------------------------------------------
\begin{figure*}[t]
\center
\includegraphics[width=0.95\textwidth,trim=1cm 13.2cm 1cm 1cm,clip]{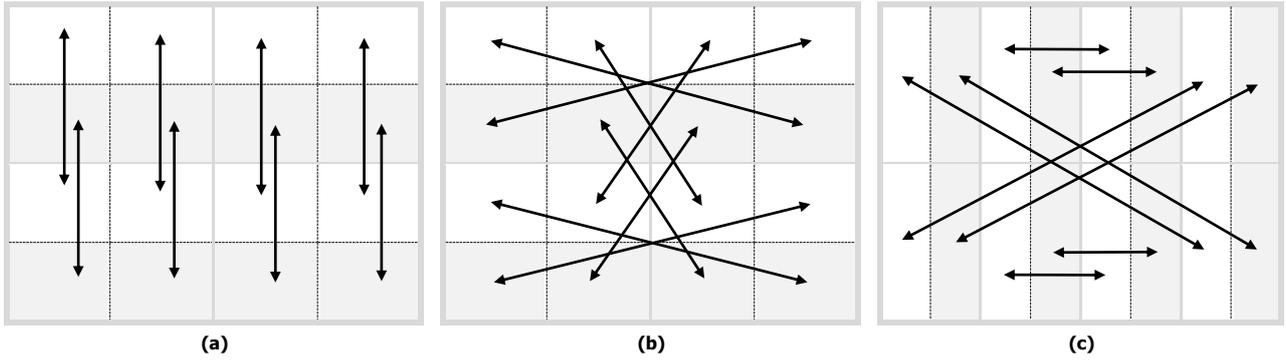}
\caption{The overall scheme of dividing digests. (a) Primary digest 2, (b) Secondary digest 1, (c) Secondary digest 2.}
\label{fig:map}
\end{figure*}
%---------------------------------------------------------------------------------------------------
\section{Proposed method}
\begin{figure*}[t]
\center
\includegraphics[width=0.95\textwidth,trim=9cm 12cm 10cm 9cm,clip]{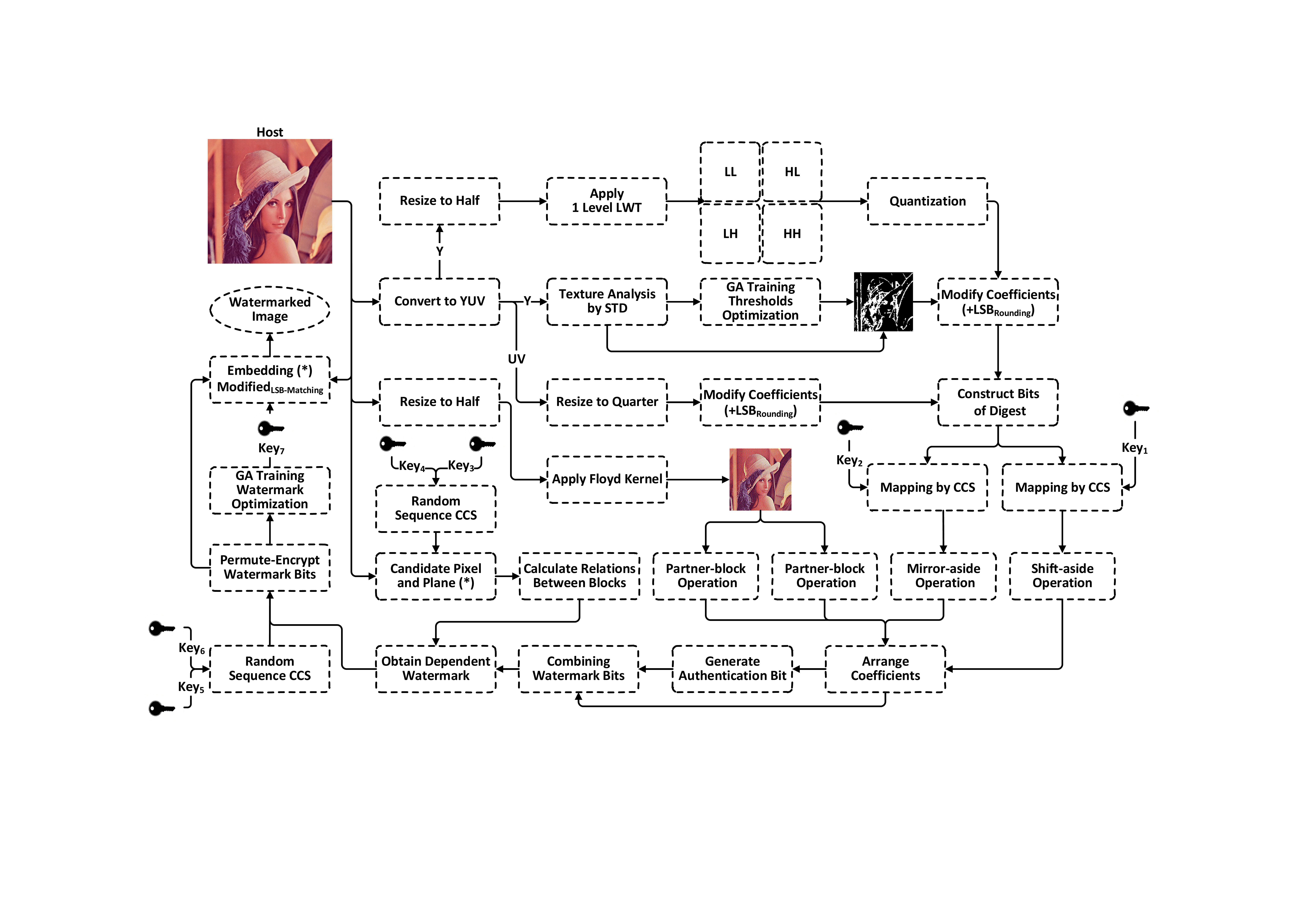}
\caption{Block diagram of generating and embedding watermarks.}
\label{fig:embedding}
\end{figure*}

\begin{figure*}[t]
\center
\includegraphics[width=0.95\textwidth,trim=13cm 9cm 12cm 9cm,clip]{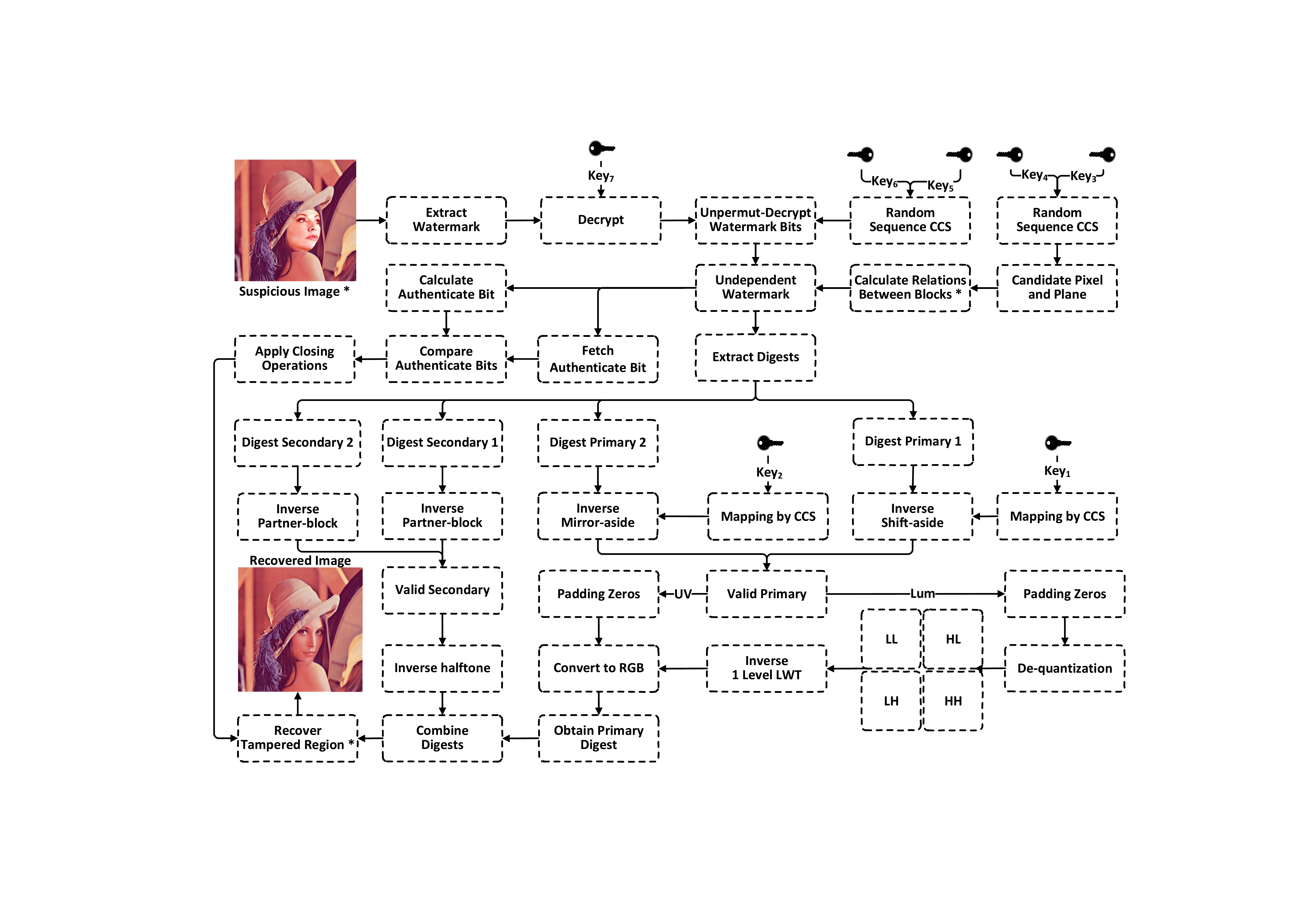}
\caption{Block diagram of extracting watermark and recovering tampered regions.}
\label{fig:extracting}
\end{figure*}

In this section, a fragile blind quad watermarking for image tamper detection and recovery by providing compact digests with quality optimized based on Lifting Wavelet Transform (LWT) and Genetic Algorithm (GA) is proposed. TRLG includes two main phases that are described below in details:

\begin{description}[font=$\bullet$\scshape\bfseries,leftmargin=0cm]
\item\textbf{Generating and embedding watermark:} In this phase, First, four digests are generated based on LWT and Floyd kernel. In the following, to improve recovery rate and increase security, each digest is shuffled and arranged separately by using a new chaotic map. Then, an authentication bit for each 2$\times$2 blocks is calculated based on the relation of pixels of the block and the digest that must be embedded in it. Finally, to form and embed watermark, first, digests and authentication bit should be combined, and then encrypting and embedding the watermark by using chaotic map, GA, and modified LSB-matching technique. The block diagram of this phase is shown in Fig. \ref{fig:embedding}.
\item\textbf{Tamper detection and recovery:} In this phase, to analyze the integrity of watermarked image received from the communication channels, first, the watermark is extracted and decrypted. Next, the tamper regions are marked based on the extracted and calculated authentication bit. Finally, four digests are reshuffled and reconstructed to recover tampered regions by valid parts of them. Fig. \ref{fig:extracting} illustrates the block diagram of tamper detection and recovery phase.
\end{description}

\subsection{Generating and embedding watermark}
Let's denoted the cover image as $host$ with the size of $M\times N$ (divisible by 4). TRLG is able to detect and recover 2$\times$2 modified blocks. Also, [$R, G, B$] and [$Y, U, V$] represent the color component of $host$ in RGB and YUV color spaces, respectively. If $host$ is in grayscale mode, chrominance components are meaningless and further processing is not needed for them. The procedure of generating and embedding watermark is described in details as below:

\subsubsection{Generating digests}
As mentioned before, four digests are considered in TRLG to recover tampered regions. These digests are classified as primary and secondary digests. The two primary digests are generated based on LWT and GA. Also, the two secondary digests are obtained by using the halftoning technique. 

The steps of generating primary digest are as follows:
\begin{enumerate}[1),itemsep=0mm]
\item The $Y$ component is resized to 50\% of original size, and a level of LWT is applied on the result to generate $LL$, $LH$, $HL$ and $HH$ bands.   
\item Quantizing coefficients of each band by Eq.  (\ref{eq:quantization}):
\begin{align}
Coef_{i, j} &= sign(Coef_{i, j})\times \lfloor\frac{|Coef_{i, j}|}{\mu}\rfloor \nonumber\\
\forall & i \in [1, \frac{M}{4}], j \in [1, \frac{N}{4}]
\label{eq:quantization}
\end{align} 
where $\mu (=2)$ and $Coef_{i, j}$ are quantization step and coefficients of wavelet bands, respectively. 
\item In this step, a texture analysis on each block is performed to intelligently generate the digest for any image with the various type. Hence, the type of each 4$\times$4 blocks of $Y$ is classified into two classes as texture and flat region based on Standard Deviation (STD) measure and GA. For this aim, first, STD is applied in $Y$ and denote result as $texture_{M\times N}$. Next, the optimal thresholds for separating blocks are obtained based on GA. The details of GA training will further explain in the Thresholds Optimization sub-section. At the end of GA training, a threshold matrix where denoted as $thresholds_{M \times N}$ is obtained. Then, the type of each block is marked as texture or flat region by Eq. (\ref{eq:texture}):
\begin{align}
\Gamma_{i, j} &=
  \begin{cases}
1 & \text{if } thresholds_{i, j} < texture_{i, j}\nonumber\\
0 & \text{otherwise}\\
  \end{cases}\\
& \forall  i \in [1, N], j \in [1,M]
\label{eq:texture}
\end{align}
where $\Gamma_{\rfrac{M}{4}\times \rfrac{N}{4}}$ ($\in$ [0, 1]) is illustrated the type of each block as two classes. 
\item In this step, the coefficients of each bands are modified according to Eq. (\ref{eq:modified}):
\begin{align}
LL_{i, j} &=
  \begin{cases}
    (LL_{i, j}+\vartheta)\&124_{(01111100)_{2}} &  \text{if }\Gamma_{i, j} = 1\\
    LL_{i, j}  & \text{otherwise}\\
  \end{cases} \nonumber \\
LH_{i, j} &=
  \begin{cases}
    (LH_{i, j}+\vartheta)\&60_{(00111100)_{2}} &  \text{if }\Gamma_{i, j} = 1\\
    (LH_{i, j}+\vartheta)\&28_{(00011100)_{2}}  & \text{otherwise}\\
  \end{cases} \nonumber \\
HL_{i, j} &=
  \begin{cases}
    (HL_{i, j}+\vartheta)\&60_{(00111100)_{2}} &  \text{if }\Gamma_{i, j} = 1\\
    (HL_{i, j}+\vartheta)\&28_{(00011100)_{2}} & \text{otherwise}\\
  \end{cases} \nonumber \\
HH_{i, j} &=
  \begin{cases}
    (HH_{i, j}+\vartheta)\&56_{(00111000)_{2}} &  \text{if }\Gamma_{i, j} = 1\\
    (HH_{i, j}+\vartheta)\&28_{(00011100)_{2}}  & \text{otherwise}\\
  \end{cases} \nonumber \\
& \forall  i \in [1, \rfrac{M}{4}], j \in [1, \rfrac{N}{4}]
\label{eq:modified}
\end{align}
where $\vartheta$ is the parameter of $LSB_{Rouning}$ technique which is proposed in TRLH \cite{ref1}. Based on this technique, the difference between two corresponding coefficient is reduced, and this leads to increase the quality of image digest. If two and three LSBs of coefficient must be ignored are zeros, $\vartheta$ is set as 2 and 4, respectively. %Lets denoted result as $LL_{\rfrac{M}{4}\times\rfrac{N}{4}}$, $LH_{\rfrac{M}{4}\times\rfrac{N}{4}}$, $HL_{\rfrac{M}{4}\times\rfrac{N}{4}}$, and $HH_{\rfrac{M}{4}\times\rfrac{N}{4}}$.
Totally, 20 bits (19 bits for describing coefficient of each band, and 1 bit for described the type of corresponded block) is obtained that represent $Y$ of digest. 
\item If $host$ is in grayscale mode, $U$ and $V$ components are resized to 25\% of original size. Then, the values of components are modified and updated by Eq. (\ref{eq:colormodified}):
\begin{align}
U_{i, j} &= (U_{i, j}\&254_{(11111110)_{2}}) >> 1 \nonumber \\
V_{i, j} &= (V_{i, j}\&254_{(11111110)_{2}}) >> 1 \nonumber \\
\forall i &\in [1, \rfrac{M}{4}],  j \in [1, \rfrac{N}{4}]
\label{eq:colormodified}
\end{align}
where $>>$ is bitwise right shift operation. 

Totally, 14 bits is obtained which described the 4$\times$4 blocks of chrominance components.
% and define result as $U_{\rfrac{M}{4}\times\rfrac{N}{4}}$ and $V_{\rfrac{M}{4}\times \rfrac{N}{4}}$ .
\end{enumerate}
Finally, 34 bits (20 bits for gray) are considered to represent 4$\times$4 blocks of gray or color images. It should be noted, although the size of the block is 4$\times$4, in the inverse procedure, TRLG can recover each block with 2$\times$2 precision. In another word, the digest which is proposed in TRLG has amazing quality rather than traditional method which are based on 2$\times$2 blocks or larger. This claim will be proved in Sec. \ref{sec:Experimental}.

At the end, lets define the result as $digest_{prim}$ which include [$\Gamma, LL, LH, HL, HH, U, V$]$_{\rfrac{M}{4}\times\rfrac{N}{4}}$.
A novel primary digest which proposed in TRLG is named as DLG.
\begin{figure*}[t]
\center
\includegraphics[width=0.95\textwidth,trim=1cm 11.5cm 1cm 2cm,clip]{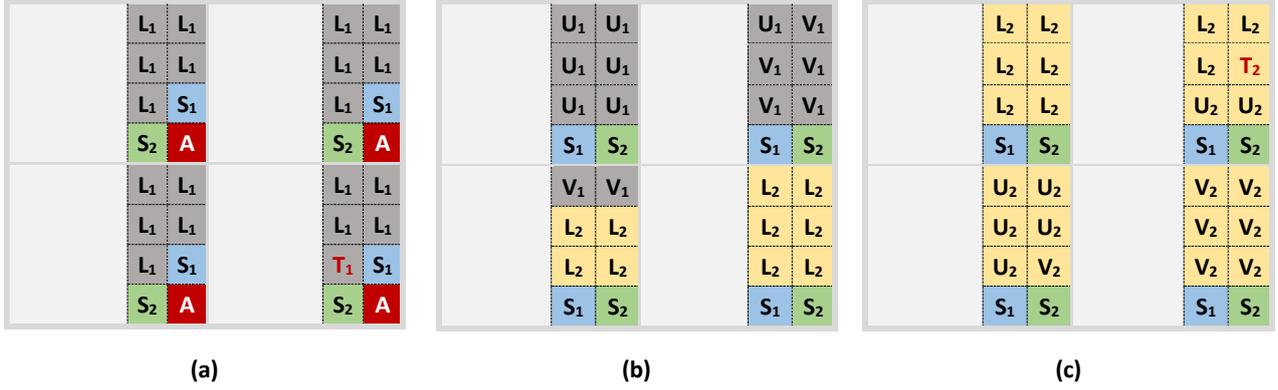}
\caption{The structure of embedding 8 bits in each 2$\times$2 blocks of the host. (a) red plane (grayscale), (b) green plane, (c) blue plane. L and $\text{\{U, V\}}$ mean Luminance and Chrominance, and T define texture.
S determine the secondary digest, and A belong to authentication bits.}
\label{fig:organize}
\end{figure*}
%----------------------------------------------------------
Thereinafter, to generate the secondary digest, the Floyd kernel is used. To do so, first, $R$, $G$ and $B$ are resized to 50\% of original size, and Floyd kernel is applied in each band, separately. Totally, for each 2$\times$2 blocks, 3 bits (a bit for gray) halftone is considered. 
Let define the result as $digest_{sec}$.

\noindent\textbf{Thresholds Optimization:} As can be seen, the threshold step is playing important role in DLG algorithm. In another word, the key challenge is how to classify block in terms of texture to achieve efficient digest with well quality. Therefore, to guarantee the quality of generated digest and select optimal thresholds, GA which is a well known modern optimization algorithm is employed. To do so, $Y$ is divided into non-overlap blocks of size $W\times H$ (128$\times$128). The overall GA-based generating digest is summarized in three steps:

\begin{enumerate}[1),itemsep=0mm]
\item First, the initial thresholds population is randomly created and converted into chromosomes. Next, the digest of the current block is generated by using the solutions in population.
\item The fitness function is evaluated between the current block of $Y$ and reconstructed primary digest that belongs to it for each corresponding solution by Eq. (\ref{eq:fitness1}):
\begin{equation}
Fitness function = SSIM
\label{eq:fitness1}
\end{equation}
where SSIM is the Structure Similarity Index.
\item Finally, GA operators include selection, crossover, and mutation are applied on each chromosome to generate next generation.
\end{enumerate}

These steps are continued for all blocks until a predefined condition is satisfied, or a constant number of generations is exceeded. Finally, the optimal thresholds are denoted as $threshold_{\rfrac{M}{W}\times \rfrac{N}{H}}$. At the end, $threshold$ is resized to the original size of $host$. 
%-------------------------------------------------------------------------------------------
\subsubsection{Scrambling digests}
As mentioned above, in TRLG, four digests are considered for tampering recovery. Therefore, the four schemes are designed to shuffle and place each part of four digests in the maximum possible distance from the original place in $host$, and other corresponded parts in rest digests. By using these strategies, the security and recovery rate will be increased and improved. Accordingly, if more than half of $host$ is manipulated, TRLG is able to recover tampered region efficiently. In Sec. \ref{sec:Experimental}, we will see, TRLG can efficiently recover tampering part with amazing quality when the watermarked image is manipulated under 80\% rate.

First of all, the coefficients of primary digest is shuffled to improve security by using novel chaotic map discussed in Sec. \ref{sec:Chebyshev}. The shuffling steps of primary digests are explained below in details:
\begin{enumerate}[1),itemsep=0mm]
\item Two copy of $digest_{prim}$ are taken, and named them as $\dot{d}_p$ and $\ddot{d}_p$.
\item Two sequences $\chi_1$ and $\chi_2$ are generated based on Eq. (\ref{eq:CCS}) with $key_1$ and $key_2$ by running  ${\frac{M}{4}\times \frac{N}{4}}$ times.
\item The permutation position matrix $\chi^{\prime}_1$ and $\chi^{\prime}_2$ is achieved by sorting the $\chi_1$ and $\chi_2$ in ascending order. 
\item Each plan of $\dot{d}_p$ and $\ddot{d}_p$ which is generated in previous section [$\Gamma, LL, LH, HL, HH, U, V$] are converted into 1D matrix as:
\begin{align*}
\dot{d}_p &= \{\dot{d}^1_p,  \dot{d}^2_p, ...,  \dot{d}_p^{{\frac{M}{4}\times \frac{N}{4}}} \}\\
\ddot{d}_p &= \{\ddot{d}^1_p,  \ddot{d}^2_p, ..., \ddot{d}_p^{{\frac{M}{4}\times \frac{N}{4}}} \}
\end{align*}
\item In this step, the shuffled digest pixel matrix $\dot{d}_p^{s}$ and $\ddot{d}_p^{s}$ are achieved by utilizing Eq. (\ref{eq:permute1}):
\begin{align}
\dot{d}_p^{s}(i) &=\dot{d}_p(\chi^{\prime}_1(i))    \nonumber \\
\ddot{d}_p^{s}(i) &=  \ddot{d}_p(\chi^{\prime}_2(i))      \nonumber \\
\forall i & \in [1,{\rfrac{M}{4}\times \rfrac{N}{4}}]
\label{eq:permute1}
\end{align}
\item Convert the $\dot{d}_p^{s}$ and $\ddot{d}_p^{s}$ to 2D matrix with size of ${\frac{M}{4}\times \frac{N}{4}}$.
\end{enumerate}

In the following, to improve recovery rate a Shift-aside technique \cite{ref1} is utilized for reordering the coefficients of $\dot{d}_p^{s}$, again. Accordingly, if the right or the left half side of $host$ totally tampered, the recovery phase is able to recover the tamper region which embedded in another side of $host$. In addition, each side is divided into two separate parts again, that makes the recovery phase more efficient when the tampered region is located at the center of $host$. It should be noted, these processes are applied on all plane in $\dot{d}_p^{s}$ include [$\Gamma, LL, LH, HL, HH, U, V$], and finally, $\dot{d}_p^{s}$ will be updated.

In order to reorder the coefficients of $\ddot{d}_p^{s}$, a new technique is proposed in TRLG as Mirror-aside operation. In Mirror-aside scheme, the coefficient of top and bottom half of the digests are swapped. To do so, first, $\ddot{d}_p^{s}$ is divided into four non-overlap blocks. Next, each block is divided into four non-overlap blocks, again. Fig. \ref{fig:map}(a) is illustrated the dividing process. As Shift-aside scheme, the determined location by CCS is reordered to placed into the corresponded quarter. Finally, the reordered $\ddot{d}_p^{s}$ is formed and updated.

Subsequently, to reorder the coefficients (Bits) of the secondary digest, first, two copy of $digest_{sec}$ are generated, as $\dot{d}_s$ and $\ddot{d}_s$. Next, the coefficients of these digest are reordered according to Fig. \ref{fig:map}(b) and Fig. \ref{fig:map}(c), respectively. Let named this strategy as Partner-block. Unlike primary digests, the secondary digests are not shuffled by any chaotic map. At the end, $\dot{d}_s^s$ and $\ddot{d}_s^s$ are achieved. As seen, the shuffling and reordering schemes for all digest in TRLG are designed to achieve maximum recovery rate in large tampering rate.
%----------------------------------------------------------------------------------------------
\subsubsection{Generating authentication bits}
\label{sec:Generatingauthenticationbits}
In TRLG, a bit is considered for authenticating each 2$\times$2 block. The process of generating authentication bit are explained below in details:
\begin{enumerate}[1),itemsep=0mm]
\item In first step, each band of $\dot{d}_p^{s}$ and $\ddot{d}_p^{s}$ which include [$\Gamma, LL, LH, HL, HH, U, V$] are converted into binary form. Next, the result are combined by Eq. (\ref{eq:digest}):
\begin{align}
\bar{d}_p{(i, j)} &=\overline{\dot{d}_p^{s}(i, j, k)}  \uplus \overline{\ddot{d}_p^{s}(i, j, k)} \nonumber \\
\forall i \in [1, &\rfrac{M}{4}], j \in [1, \rfrac{N}{4}], k \in  [1, 7]
\label{eq:digest}
\end{align}
where $k$ and $\uplus$ are represented index of each band in primary digest and  string joint operator, respectively. In this equation, $\bar{d}_p$ is a binary matrix with size of $\frac{M}{4}\times \frac{N}{4}$ that each cell contains 68 bits which belong to the information of two primary digests.

\item The bits of primary digests are formed to place into considered position in 4$\times$4 block according Fig. \ref{fig:organize} as:
\begin{align*}
\Delta_{(i,j,1)} &= {\bar{d}_p^k{(i, j)}}, \forall k \in [1, 2, ..., 20]\\
\Delta_{(i,j,2)} &= {\bar{d}_p^k{(i, j)}}, \forall k \in [21, 22, ..., 44]\\
\Delta_{(i,j,3)} &= {\bar{d}_p^k{(i, j)}}, \forall k \in [45, 46, ..., 68]\\
\forall i &\in [1, \rfrac{M}{4}], j \in [1, \rfrac{N}{4}]
\end{align*}
where $\Delta_{(i, j, p)} \{ p \in [1, 2, 3]\}$ is expressed the primary bits in each planes.
\item Next, the primary bits which must be embedded in each $2\times2$ blocks in all planes are denoted as:
\begin{align*}
\Omega_{(i, j)} &= \begin{bmatrix}
\omega_1&\omega_2\\
\omega_3&\omega_4
\end{bmatrix}\\
\forall i \in  & [1,\rfrac{M}{4}], j \in [1, \rfrac{N}{4}]
\end{align*}
where $\Omega$ and $\omega$ are represented $4\times4$ block, and its inner $2\times2$ sub-block, respectively, and $\omega$ is calculated by Eq. (\ref{eq:form}):
\begin{align}
\label{eq:form}
&\hspace{1cm}\Omega^n_{(i, j)}=\omega_n = \Delta^k_{(i,j,1)} \uplus \Delta^l_{(i,j,2)} \uplus \Delta^l_{(i,j,3)}\nonumber\\
&\hspace{1.5cm}\forall i \in  [1,\rfrac{M}{4}], j \in [1, \rfrac{N}{4}]\\
&\begin{cases}
\forall k \in [1, 2, ..., 5], l \in [1, 2, ..., 6] & \quad \text{if }  n = \text{1},\\
\forall k \in [6, 7, ..., 10], l \in [7, 8, ..., 12] & \quad \text{if }  n = \text{2},\\
\forall k \in [11, 12, ..., 15], l \in [13, 14, ..., 18] & \quad \text{if }  n  = \text{3},\\
\forall k \in [16, 17, ..., 20], l \in [19, 20, ..., 24] & \quad \text{if }  n  = \text{4}
\nonumber
\end{cases}
\end{align}

where $\Omega^n_{(i, j)}$ and $\uplus$ are represented, $n$th sub-block in $\Omega_{(i, j)}$ and string joint operator, respectively. 

At the end, $\Phi$ with size of $\frac{M}{2}\times\frac{N}{2}$ is generated as:
\begin{align*}
\Phi=
\begin{pmatrix}
\Omega^1_{1, 1}&\Omega^2_{1, 1} & \Omega^1_{1, 2}&\cdots&\Omega^1_{1, \frac{N}{4}}&\Omega^2_{1, \frac{N}{4}}\\
\Omega^3_{1, 1}&\Omega^4_{1, 1} & \Omega^3_{1, 2}&\cdots&\Omega^3_{1, \frac{N}{4}}&\Omega^4_{1,  \frac{N}{4}}\\
\Omega^1_{2, 1}&\Omega^2_{2, 1} & \Omega^1_{2, 2}&\cdots&\Omega^1_{2, \frac{N}{4}}&\Omega^2_{2,  \frac{N}{4}}\\
\vdots&\vdots&\vdots&\ddots&\vdots&\vdots \\
\Omega^1_{\frac{M}{4}, 1}&\Omega^2_{\frac{M}{4}, 1}&\Omega^1_{\frac{M}{4}, 2}&\cdots&\Omega^1_{\frac{M}{4}, \frac{N}{4}}&\Omega^2_{ \frac{M}{4}, \frac{N}{4}}\\
\Omega^3_{\frac{M}{4}, 1}&\Omega^4_{\frac{M}{4}, 1}&\Omega^3_{\frac{M}{4}, 2}&\cdots&\Omega^3_{\frac{M}{4}, \frac{N}{4}}&\Omega^4_{\frac{M}{4}, \frac{N}{4}}
\end{pmatrix}
\end{align*}
where each element in $\Phi$ contains 17 bits (or 5 bit for gray image) that belong to data of primary digests.
%-------------------------------------------------------------
\item In this step, the authentication bits for each $2\times2$ block is calculated by Eq. (\ref{eq:auth}):
\begin{align}
\label{eq:auth}
\tilde{A}_{i,j} &=\delta_{i,j}  \oplus \xi_{i,j}, \forall i \in [1, \rfrac{M}{2}], j \in [1, \rfrac{N}{2}] \nonumber\\
\delta_{i,j} &= [\displaystyle\sum_{n=1}^{k} \Phi^n_{i,j}\bmod2]  \oplus [\displaystyle\sum_{n=1}^{len}\overline{\gamma^n}\bmod 2] \nonumber\\ 
\xi_{i,j} &= \dot{d}_s^s{(i, j)} \oplus \ddot{d}_s^s{(i, j)}
\end{align}
where $\overline{\gamma}$ is a binary form of $\gamma$ obtains by Eq. (\ref{eq:index}):
\begin{align}
\gamma &= \zeta_i \oplus \zeta_{i-1}, \forall i \in [1, 2, ..., len] \nonumber\\ 
\zeta &= f(\Phi_{i,j})
\label{eq:index}
\end{align}
Here, $f$ is function as:
\begin{align*}
f(\bar{\chi}) &=
  \begin{cases}
    null & \quad \text{if } \bar{\chi}_i \text{ = 0}\\
    i  & \quad \text{if } \bar{\chi}_i \text{ = 1}
  \end{cases}
\end{align*}
where return index of each element in $\bar{\chi}$  equal by 1.
\end{enumerate}
\subsubsection{Combining watermarks bits}
After generating and shuffling primary and secondary digests, and also computing authentication bits, all bits are organized to be ready for embedding in $host$. In TRLG, 8 bits are embedded into  2 LSBs of each 2$\times$2 block. The bits arrangement of four digests and authentication bit is shown in Fig. \ref{fig:organize}. As shown, by assuming $host$ is colored, 20 bits are required for luminance (19+1 bits), and 14 bits are considered for chrominance. Totally, to provide a second chance for the primary digest, 68 bits space should be reserved for hiding information. In addition, 24 bits space are required for embedding two copies of the secondary digest. In other words, for each 2$\times$2 blocks, six bits are reserved for secondary digests. Totally, 92 bits for digests and 4 bits for authentication are combined to embed into 2 LSBs of each 4$\times$4 block in each plane. Subsequently, 20 bits (19+1 bits) are considered for the primary digest, and 8 bits are reserved for embedding two copies of secondary digests for gray images. Similarly, 28 bits digest and 4 bits authentication are combined for embedding into 2 LSBs of each 4$\times$4 block in the next phase.

Hence, Let the watermark must be embedded in each $2\times2$ block be $\Psi^k_{i, j}$ where achieved by using Eq. (\ref{eq:watermark}):
\begin{align}
\Psi^k_{i, j}=&
\begin{cases}
\Phi^n_{i, j, k} \uplus \dot{d}_s^s{(i, j, k)} \uplus \ddot{d}_s^s{(i, j, k)}  \uplus \tilde{A}_{i,j} &\quad \text{if } k= \text{1}\\
\Phi^n_{i, j, k} \uplus \dot{d}_s^s{(i, j, k)} \uplus \ddot{d}_s^s{(i, j, k)} &\quad \text{otherwise }\nonumber\\
\end{cases}\nonumber\\
&\begin{cases}
\forall n \in [1, 2, ..., 5] & \quad \text{if } k= \text{1},\nonumber\\
\forall n \in [6, 7, ..., 11] & \quad \text{if } k = \text{2},\nonumber\\
\forall n \in [12, 13, ..., 17] & \quad \text{if } k  = \text{3}
\end{cases}\\
\forall i &\in [1, \rfrac{M}{2}], j \in [1, \rfrac{N}{2}], k \in [1, 2, 3]
\label{eq:watermark}
\end{align}
where $k$ and $\uplus$ are represented index of each plane and  string joint operator, respectively. At the end of this phase, 8 bits which will be embedded into 2$\times$2 are encapsulated in each element of $\Phi^n$ to embed in next phase.
%----------------------------------------------------------------------------------------------
\subsubsection{Encrypting and embedding watermark}
In this phase, first, the watermark of each block must be depended to content of current block and its neighbors. Due to this strategy, TRLG is able to detect security tampering that applied based on collage, vector-quantization or protocol attacks.
The detail of this strategy is explained below:
\begin{enumerate}[1),itemsep=0mm]
\item Two sequences $\chi_1$ and $\chi_2$ are generated based on Eq. (\ref{eq:CCS}) with $key_3$ and $key_4$ by running  ${\frac{M}{2}\times \frac{N}{2}}$ times.
\item The candidate pixel $p'_c$ in each blocks can be calculated using Eq. (\ref{eq:candidate1}):
\begin{equation}
p'_c(i) = (\lfloor\chi_1(i)\times10^{14}\rfloor \bmod 4) + 1, \forall i
\label{eq:candidate1}
\end{equation}
Next, the candidate plane $p''_c$ in each blocks can be calculated using Eq. (\ref{eq:candidate2}):
\begin{equation}
p''_c(i) = (\lfloor\chi_2(i)\times10^{14}\rfloor \bmod 3) + 1, \forall i
\label{eq:candidate2}
\end{equation}
Now, $p'_c$ and $p''_c$ are converted into 2D matrix with size of ${\frac{M}{2}\times \frac{N}{2}}$.
\item The candidate version of $host$ is obtained by Eq. (\ref{eq:candidate3}):
\begin{align}
\label{eq:candidate3}
h_c(i, j) &= host^{p''_c{(i, j)}}_{p'_c{(i, j)}}\& 252_{(11111100)_{2}} \nonumber\\
\forall i &\in  [1,\rfrac{M}{2}], j \in [1, \rfrac{N}{2}]
\end{align}
This process should be repeated for all 2$\times$2 blocks.

\item In this step, the relations between 2$\times$2 sub-blocks of each 4$\times$4 block in $host$ is computed. To do so, $h_c$ of size $\frac{M}{2}\times\frac{N}{2}$ is partitioned into $\frac{M}{4}\times\frac{N}{4}$ non-overlapping blocks of 2$\times$2 pixels, and $h^i_c$ the $i$th block which is expressed as:
\begin{align*}
h^i_c = \begin{bmatrix}
h^i_c(1)&h^i_c(2)\\
h^i_c(3)&h^i_c(4)
\end{bmatrix}, \forall i \in [1, 2, 3, ..., \rfrac{M}{2}\times\rfrac{N}{2}]
\end{align*}
Now, the relations between pixels of $h^i_c$ is calculated by using Eq. (\ref{eq:relations}):
\begin{align}
\label{eq:relations}
R' &= \lfloor\arctan(\frac{h^i_c(1) - h^i_c(4)}{h^i_c(2) - h^i_c(3)}) \times 10^{14}\rfloor \bmod 256 \nonumber\\
R'' &= \lfloor\arctan(\frac{h^i_c(1) - h^i_c(3)}{h^i_c(2) - h^i_c(4)}) \times 10^{14}\rfloor \bmod 256 \nonumber\\
R''' &= \lfloor\arctan(\frac{h^i_c(1) - h^i_c(2)}{h^i_c(3) - h^i_c(4)}) \times 10^{14}\rfloor \bmod 256 \nonumber\\
R'''' &= \lfloor \text{DCT}(h^i_c)_\text{DC} \times 10^{14}\rfloor \bmod 256 
\end{align}
\item Finally, the dependent watermark for each 2$\times$2 block is achieved by Eq. (\ref{eq:relwatermark}):
\begin{align}
\label{eq:relwatermark}
\Psi'^k_{i, j} &= R'_{i, j} \oplus R''_{i, j} \oplus R'''_{i, j} \oplus R''''_{i, j} \oplus \Psi^k_{i, j}\nonumber\\
\forall i &\in [1, \rfrac{M}{2}], j \in [1, \rfrac{N}{2}], k \in [1, 2, 3]
\end{align}
\end{enumerate}
In the following, in order to improve security and guaranty the originality of watermark, and also to prevent the predictability of the arrange and value of bits, further process are done on $\Psi'^k_{i, j} $. To do so, the watermark bits are encrypted and permuted according to the following steps:
\begin{enumerate}[1),itemsep=0mm]
\item Two sequences $\chi_1$ and $\chi_2$ are generated based on Eq. (\ref{eq:CCS}) with $key_5$ and $key_6$ by running  ${\frac{M}{2}\times \frac{N}{2}}$ times.
\item The sequence of secret values to encrypt watermark bits is calculated by using Eq. (\ref{eq:encrypt}):
\begin{equation}
s'_v(i) = \lfloor\chi_1(i)\times10^{14}\rfloor \bmod 256, \forall i
\label{eq:encrypt}
\end{equation}
Next, the sequence of secret values to permute watermark bits is computed based on Eq. (\ref{eq:permute}):
\begin{equation}
s''_v(i) = (\lfloor\chi_2(i)\times10^{14}\rfloor \bmod 8) + 1, \forall i
\label{eq:permute}
\end{equation}
Finally, $s'_v$ and $s''_v$ are converted into 2D matrix with size of ${\frac{M}{2}
\times \frac{N}{2}}$.
\item In this step, the permuted and encrypted watermark is achieved according Eq. (\ref{eq:pe}):
\begin{equation}
\label{eq:pe}
\Psi''^k_{i, j}  = f(\Psi'^k_{i, j}  \oplus s'_v(i, j), s''_v(i, j)), \forall i, j, k
\end{equation}
where $\oplus$ is exclusive-or operator, and $f$ is permuted function which permute bits of watermark by Eq. (\ref{eq:pe1}):
\begin{equation}
f(v, p) =v[f'(p)]
\label{eq:pe1}
\end{equation}
where $f'$ is a 1D simple mapping sequence generator algorithm \cite{ref7}, which calculated by Eq. (\ref{eq:pe2}):
\begin{equation}
f'(\chi_i) = [(\chi_{i-1} \times k) \bmod N)]+1, \forall i
\label{eq:pe2}
\end{equation}
where $\chi_i \in [1, N]$, $k$ a secret key, and $N$ total number of bits ($k$ = 13, $N$ = 8).
\end{enumerate}

As said in the previous section, in TRLG to improve quality of the watermarked image, and also improve the security of watermarks, the embedded watermark in each block of $host$ are encrypted with a $key_7$ that intelligently selected with GA.  In other words, this strategy leads to decrease the difference between watermark and original values (LSBs), and also achieve high level of security. The details of GA training will further explain in the watermark optimization sub-section. At the end of the GA training, $key_7$ is achieved. Now, the watermark bits are encrypted again by Eq. (\ref{eq:encrypted2}):
\begin{equation}
\Psi'''^k_{i, j} = \Psi''^k_{i, j} \oplus key_7, \forall i, j, k
\label{eq:encrypted2}
\end{equation}
where $\oplus$ is exclusive-or operator.
Finally, the 24 bits (8 bits for gray) watermark are embedded into 2 LSBs of each 2$\times$2 non-overlap block of $host$ in each plane. In TRLG, to decrease the difference between watermarked and original pixel, a modified LSB-Mathching with considering statistical parameter of block is proposed. It should be noted, this strategy is applied on all pixels in each planes, expect the candidate pixels and planes [$p'_c, p''_c$] choosed in encryption phase. In Algorithm. \ref{ALG:LSBmathcing}, the pseudo code of this technique is showed in details.
\begin{algorithm}[t]
\caption{Modified LSB-Matching technique.}
\label{ALG:LSBmathcing}
\textbf{Input:} 1$\times$4  block, watermark (8 Bits)\\ 
\textbf{Output:} watermarked block
\begin{algorithmic}[1]
\Procedure{Matching}{$block$, $\Psi'''$}
\State $initial_{std}$ = STD$(block)$
\State $shift = 0$
\State $block_{w} = block$
\For{$i = 1$ to ${4}$}
\State $f = 1$
\If {$initial_{std} < \text{STD}(block_{w}$)}
\State $f = -1$
\EndIf
\State $\omega = (\Psi''' >> shift) \& 3$
\For{$j = 0$ to ${3}$}
\State $\xi = j \times f$
\State $p = (block(i)\&3) \pm \xi$
\If {$p =\omega$}
\State    $block_{w}(i) = block(i) \pm \xi$
\State     $break$
\EndIf
\EndFor
\State $shift = shift - 2$
\EndFor
\State \Return  $block_{w}$
\EndProcedure
\end{algorithmic}
\end{algorithm}
Finally, the watermark image as $host_w$ is achieved, and it can be transfer during communication channels. 

\noindent\textbf{Watermark Optimization:} In TRLG to maximize the similarity between watermark bits and LSBs of pixels in each 2$\times$2 block, and also to enhance security more and more GA is employed. This strategy can effectively balance the difference between watermark and original bits. To do so, the GA is applied to find optimal parameter as $key_7$ that will be used to decrease the difference between coefficients. The overall GA-based watermark optimization is summarized as below:
\begin{enumerate}[1),itemsep=0mm]
\item In the first step, the initial key population is randomly created, and convert them into chromosomes. Next, the watermarked image is generated based on the solutions in population.
\item The fitness function value is evaluated between the $host$ and $host_w$ by Eq. (\ref{eq:fitness}):
\begin{equation}
Fitness function = PSNR
\label{eq:fitness}
\end{equation}
where PSNR is Peak Signal to Noise Ratio.
\item In the last step, the operators of selection, crossover, and mutation are applied on each chromosome to generate next generation.
\end{enumerate}
These steps are continued until a predefined condition is satisfied, or a constant number of generations is exceeded. Finally, the optimal key as $key_7$ is achieved
%----------------------------------------------------------------------------------------------
%----------------------------------------------------------------------------------------------
\subsection{Tamper detection and recovery}
After receiving the suspicious watermarked image as $host_s$ through the public communication channels, In this phase, first, the tampered regions with 2$\times$2 accuracies are located and marked, and then are recovered by valid parts of four digests which embedded in $host$. The procedure of tamper detection and recovery is described in details as below:

\subsubsection{Extracting and decrypting watermark}
In this phase, the watermark bits are extracted from two LSBs of each 2$\times2$ blocks of $host_s$. Subsequently, the watermark bits are decrypted and depermuted to achieve the initial watermark bits. These process are explained below in details:
\begin{enumerate}[1),itemsep=0mm]
\item Firstly, the watermark bits are extacted from each 2$\times$2 block. Let denote the result as $\Psi'''^k_{i, j}$.
\item The watermark bits are decrypted based on $key_7$ by Eq. (\ref{eq:decrypted2}):
\begin{equation}
\Psi''^k_{i, j} = \Psi'''^k_{i, j} \oplus key_7, \forall i, j, k
\label{eq:decrypted2}
\end{equation}
where $\oplus$ is exclusive-or operator.
\item To locate and decrypt the watermark bits to initial position, the process are performed in inverse direction. Hence, the watermark is reconstructed by Eq. (\ref{eq:decrypted1}):
\begin{equation}
\Psi'^k_{i, j}  = f(\Psi''^k_{i, j}  \oplus s'_v(i, j), s''_v(i, j)), \forall i, j, k
\label{eq:decrypted1}
\end{equation}
where $\oplus$ is exclusive-or operator, and \{$s'_v$, $s''_v$\} are generated based $key_5$ and $key_6$ by utilizing Eqs. (\ref{eq:encrypt}, \ref{eq:permute}); and $f$ is depermuted function which depremute watermark bits by Eq. (\ref{eq:depermuted}):
\begin{equation}
f(v, p) =f'(p)[v]
\label{eq:depermuted}
\end{equation}
where $f'$ is achived based on Eq. (\ref{eq:pe2}).
\item Finally, the undependent watermark bits is generated by using Eq. (\ref{eq:relwatermark2}):
\begin{align}
\label{eq:relwatermark2}
\Psi^k_{i, j} &= R'_{i, j} \oplus R''_{i, j} \oplus R'''_{i, j} \oplus R''''_{i, j} \oplus \Psi'^k_{i, j}\nonumber\\
\forall i &\in [1, \rfrac{M}{2}], j \in [1, \rfrac{N}{2}], k \in [1, 2, 3]
\end{align}
where $\oplus$ is exclusive-or operator, and \{$R'$, $R''$, $R'''$, and $R''''$\} are computed according Eq. (\ref{eq:relations}). It should be noted, $p'_c$  and $p''_c$ are generated based on $key_3$ and $key_4$ by utilizing Eqs. (\ref{eq:candidate1}, \ref{eq:candidate2}).
\end{enumerate}
%sec:Generatingauthenticationbits
%-------------------------------------------------------------------------------------
\begin{table*}[t]
%\begin{sidewaystable}
\centering
\footnotesize
\caption{The PSNR and SSIM values of watermarked images for TRLG and related works. \\ Note: - means image is unavailable when using the previous scheme.}
\label{TABLE:compare_psnr_ssim}
\renewcommand{\arraystretch}{1.5}
\setlength{\tabcolsep}{4pt}
\scalebox{1} {
\begin{tabular*}{\textwidth}{ @{\extracolsep{\fill}}l@{}c@{}c@{}c@{}c@{}c@{}c@{}c@{}c@{}c@{}c@{}c@{}c@{}c@{}c@{}c@{}c@{}c@{}c@{}c@{}c@{}c@{}c@{}}
\cline{1-18}
\multicolumn{1}{c}{\multirow{4}{*}{Image}} & \multicolumn{4}{c}{TRLG}&\cite{ref7}&\cite{ref12}&\cite{ref13}&\cite{ref14} &\cite{ref15}&\multicolumn{2}{c}{\cite{ref16}}&\cite{ref19}& \multicolumn{2}{c}{\cite{ref20}}&\cite{ref21}&\cite{ref22}&\cite{ref32}\\
\cline{2-5}\cline{6-18}&\multicolumn{2}{c}{Color}&\multicolumn{2}{c}{Gray} &Gray &Color & Gray &Gray&Gray&Gray&Color&Gray&Gray& Color&Gray&Gray&Gray\\
\cline{2-3}\cline{4-5}\cline{6-18}
&PSNR&SSIM&PSNR&SSIM&\multicolumn{13}{c}{PSNR}\\
\cline{1-18}
Baboon&46.2362&0.9991&45.7945&0.9959&40.73&40.71&44.30&-&-&39.03&39.55&40.92&37.49&-&-&44.17&-\\ 
Barbara&46.2322&0.9980&45.8124&0.9903&40.72&-&44.26&-&-&-&-&40.94&-&-&-&-&-\\ 
Lena&46.4529&0.9995&45.8231&0.9883&40.68&40.73&44.16&44.22&44.18&39.31&39.80&40.95&38.06&37.59&44.27&-&40.69\\ 
Pepper&46.0257&0.9992&45.7965&0.9881&40.73&-&44.28&-&-&-&40.20&40.92&-&-&-&-&41.50\\ 
Gril&46.2269&0.9970&45.7908&0.9899&-&-&-&-&44.16&-&-&40.82&-&-&-&-&41.94\\ 
Lake&46.2211&0.9982&45.7947&0.9910&40.70&-&-&-&44.17&-&-&40.94&-&38.52&42.49&-&38.10\\ 
F16&46.2325&0.9904&45.8104&0.9861&-&40.86&-&43.39&44.16&-&-&-&-&-&43.85&44.11&40.77\\ 
House&46.2229&0.9972&45.7982&0.9896&-&-&-&-&-&-&39.16&-&-&38.25&-&-&-\\ 
Elaine&-&-&45.7827&0.9902&-&-&-&44.17&-&-&-&-&37.49&-&-&-&-\\ 
Goldhill&-&-&45.2293&0.9930&-&-&-&-&44.16&-&-&40.99&-&-&-&44.16&41.35\\ 
Boat&-&-&45.7610&0.9909&-&40.58&44.22&-&44.12&-&39.93&-&-&-&-&44.11&39.46\\ 
Camera&-&-&45.8084&0.9840&-&-&-&-&-&39.00&39.45&-&37.17&-&-&-&-\\ 
Toys&-&-&45.7968&0.9854&-&-&-&-&44.16&-&-&-&-&-&-&-&41.18\\ 
Zelda&-&-&45.7942&0.9866&40.71&-&44.21&-&-&-&-&40.83&-&-&-&-&-\\ 
Crowd&-&-&45.8469&0.9901&-&-&-&-&-&-&-&-&-&-&43.32&-&-\\ 
\cline{1-18}
\end{tabular*}}
%\end{sidewaystable}
\end{table*}
%--------------------------------------------------------------------------------------------------------------------
\begin{figure*}[t!]
\center
\setlength{\tabcolsep}{2pt}
\begin{tabular*}{1\textwidth}{lll}
\includegraphics[width=0.165\textwidth]{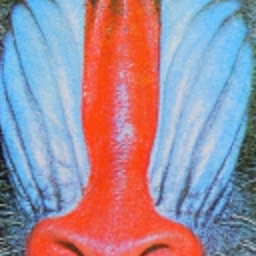}\includegraphics[width=0.165\textwidth]{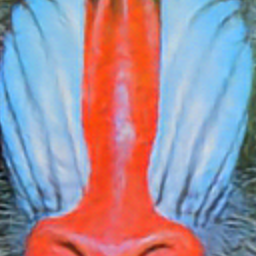} &
 \includegraphics[width=0.165\textwidth]{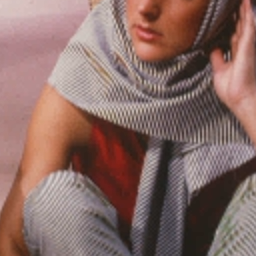}\includegraphics[width=0.165\textwidth]{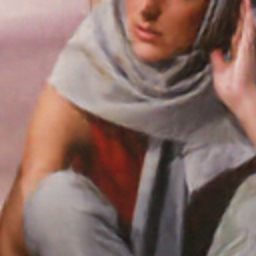}&
\includegraphics[width=0.165\textwidth]{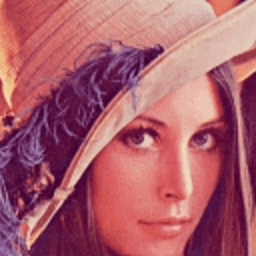}\includegraphics[width=0.165\textwidth]{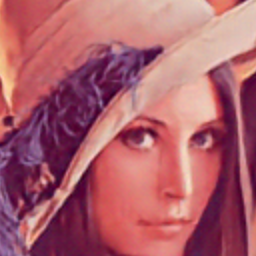}\\

\includegraphics[width=0.165\textwidth]{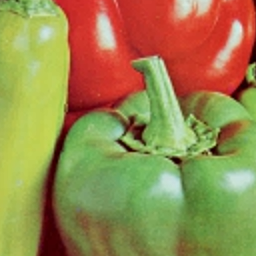}\includegraphics[width=0.165\textwidth]{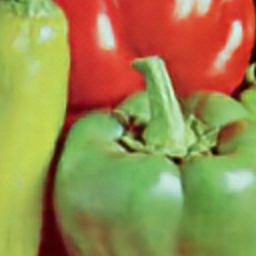} &
 \includegraphics[width=0.165\textwidth]{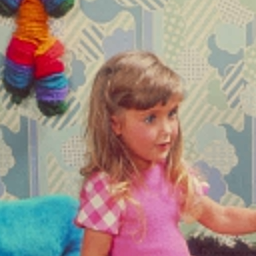}\includegraphics[width=0.165\textwidth]{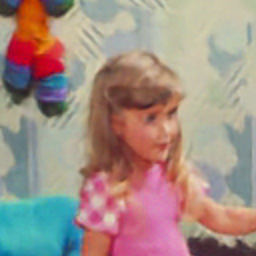} &
 \includegraphics[width=0.165\textwidth]{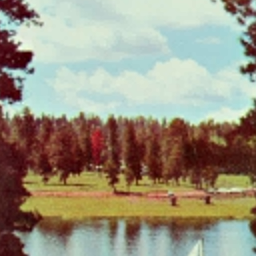}\includegraphics[width=0.165\textwidth]{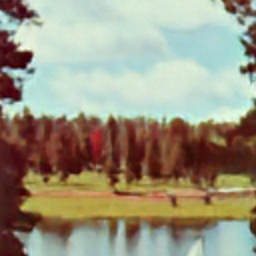} \\

\includegraphics[width=0.165\textwidth]{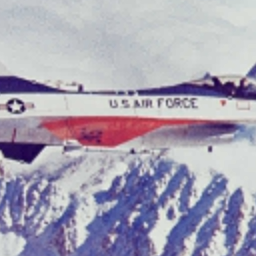}\includegraphics[width=0.165\textwidth]{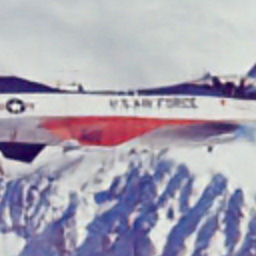} &
 \includegraphics[width=0.165\textwidth]{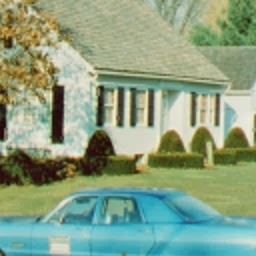}\includegraphics[width=0.165\textwidth]{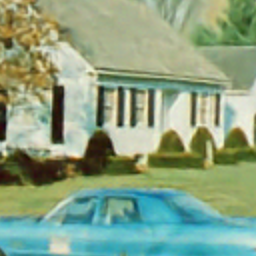} &
 \includegraphics[width=0.165\textwidth]{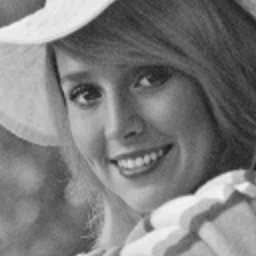}\includegraphics[width=0.165\textwidth]{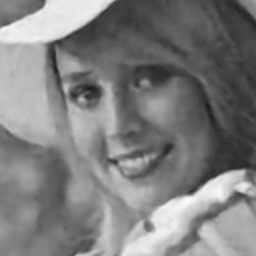} \\ 

\includegraphics[width=0.165\textwidth]{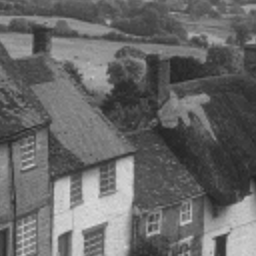}\includegraphics[width=0.165\textwidth]{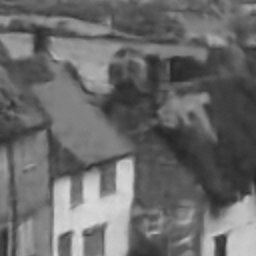}&
\includegraphics[width=0.165\textwidth]{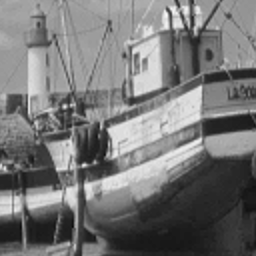}\includegraphics[width=0.165\textwidth]{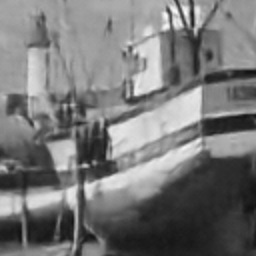}&
\includegraphics[width=0.165\textwidth]{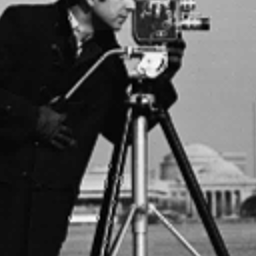}\includegraphics[width=0.165\textwidth]{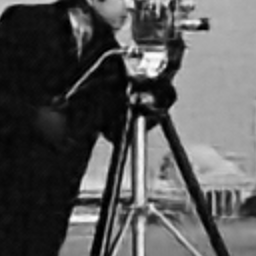}\\

\includegraphics[width=0.165\textwidth]{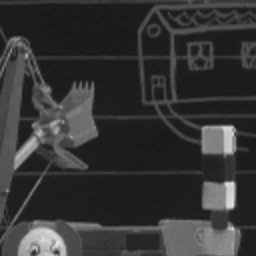}\includegraphics[width=0.165\textwidth]{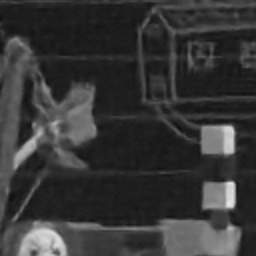}&
\includegraphics[width=0.165\textwidth]{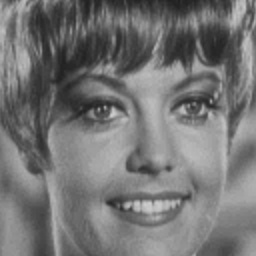}\includegraphics[width=0.165\textwidth]{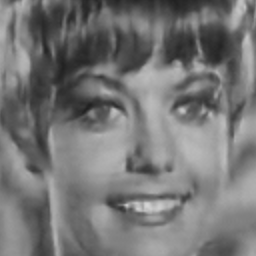}&
\includegraphics[width=0.165\textwidth]{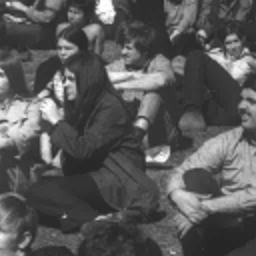}\includegraphics[width=0.165\textwidth]{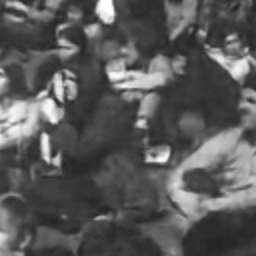}\\
\end{tabular*} 
\caption{Primary digest based on LWT, Inverse of secondary digest \cite{ref5}. (Primary digest\{PSNR, SSIM\}, Secondary digest\{PSNR, SSIM\}), Baboon: (\{22.6487, 0.7969\}, \{21.0739, 0.6934\}), Barbara(\{25.1693, 0.8373\}, \{23.8361, 0.7615\}), Lena(\{31.4157, 0.9814\}, \{28.5038, 0.9651\}), Pepper(\{28.8269, 0.9733\}, \{27.2044, 0.9610\}), Girl(\{30.8184, 0.9367\}, \{27.6036, 0.8705\}), Lake(\{27.0669, 0.9269\}, \{25.0003, 0.8844\}), F16(\{29.6441, 0.9276\}, \{27.2846, 0.8600\}), House(\{27.0235, 0.9277\}, \{25.3203, 0.8806 \}), Elaine(\{32.8147, 0.8029\}, \{29.8457, 0.7071\}), Goldhill(\{30.7620, 0.8245\}, \{27.1399, 0.6414\}), Boat(\{29.6655, 0.8499\}, \{26.0845, 0.6762\}), Camera(\{34.2506, 0.9590\}, \{29.0340, 0.8307\}), Toys(\{33.9918, 0.9365\}, \{29.1781, 0.8243\}), Zelda(\{36.6406, 0.9220\}, \{31.4033, 0.8349\}), Crowd(\{32.4380, 0.9384\}, \{26.8611, 0.7786\}).}
\label{fig:dbdigest}
\end{figure*}
%------------------------------------------------------------------------------------------------------------------------------------------------------------------
\begin{figure*}[t]
\center
\includegraphics[width=1\textwidth,trim=0cm 0cm 0cm 0cm,clip]{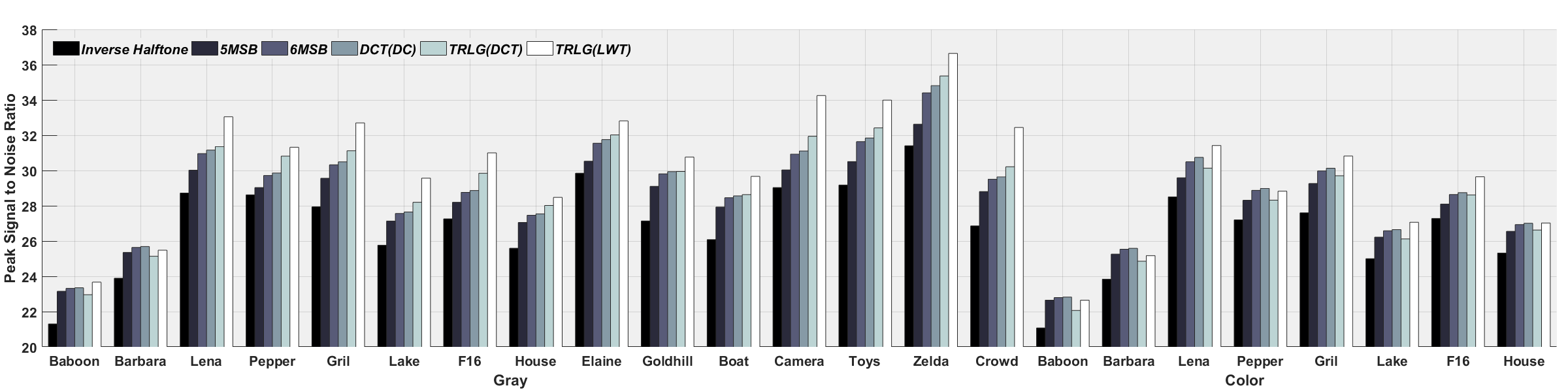}
\includegraphics[width=1\textwidth,trim=0cm 0cm 0cm 0cm,clip]{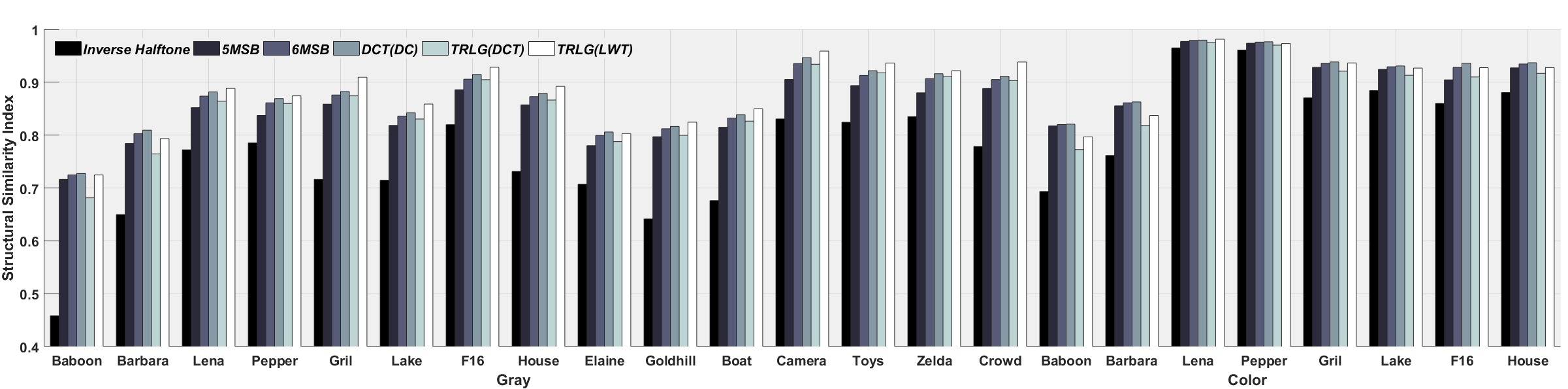}
\caption{The comparison between the proposed digests TRLG (DLG)  and other techniques in terms of quality (PSNR, SSIM).}
\label{fig:compared_digest}
\end{figure*}
%----------------------------------------------------------------------------------------------------------------------------------------------------------
\begin{figure*}[t!]
\center
\setlength{\tabcolsep}{2pt}
\begin{tabular}{ccccc}
\includegraphics[width=0.19\textwidth]{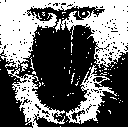} & 
\includegraphics[width=0.19\textwidth]{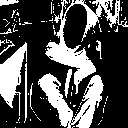} &
\includegraphics[width=0.19\textwidth]{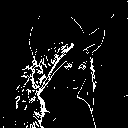} &
\includegraphics[width=0.19\textwidth]{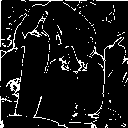} &
\includegraphics[width=0.19\textwidth]{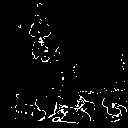} \\
\includegraphics[width=0.19\textwidth]{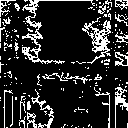} &
\includegraphics[width=0.19\textwidth]{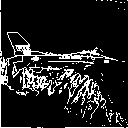} &
\includegraphics[width=0.19\textwidth]{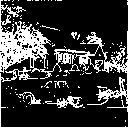} &
\includegraphics[width=0.19\textwidth]{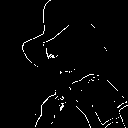} &
\includegraphics[width=0.19\textwidth]{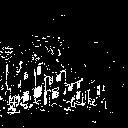} \\
\includegraphics[width=0.19\textwidth]{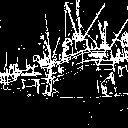} &
\includegraphics[width=0.19\textwidth]{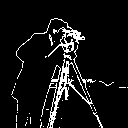} &
\includegraphics[width=0.19\textwidth]{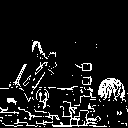} &
\includegraphics[width=0.19\textwidth]{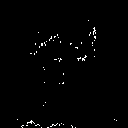} &
\includegraphics[width=0.19\textwidth]{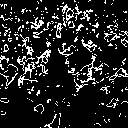} \\
\end{tabular}
\caption{The results of texture classification based on Standard Deviation and Genetic Algorithm for fifteen standard images.}
\label{fig:texture}
\end{figure*}

%------------------------------------------------------------------------------------------------------------------------------------------------------------------
\subsubsection{Authenticate received image}
In this phase, the authentication bits are fetched from each block, and then compared with generated bits using the previous procedure. Thereinafter, if the extracted and generated authentication bits of the block are matched, the block is marked as valid, otherwise it is invalid. The authentication steps are explained below in details:
\begin{enumerate}[1),itemsep=0mm]
\item Firstly, the authentication bit as $\tilde{A}_{i,j}$ is computed for all 2$\times$2 block according Sec. \ref{sec:Generatingauthenticationbits}. To do so,  $\dot{d}_s^s$, $\ddot{d}_s^s$, and $\Phi$ are fetched and formed from $\Psi^k_{i, j}$. Then, $\tilde{A}_{i,j}$ is calculated by using Eq. (\ref{eq:auth}). In addition, the authentication bit which embedded before is extracted and denote as $\tilde{A}^e_{i,j}$.
\item Next, the tampered 2$\times$2 blocks are recognized based on comparing the extracted and generated authentication bits by  Eq. (\ref{eq:tamperedregions}):
\begin{equation}
\label{eq:tamperedregions}
\varphi_{i, j} =
  \begin{cases}
    0 & \quad \text{if } \tilde{A}^e_{i,j} = \tilde {A}_{i,j}\\
    1  & \quad \text{otherwise}
  \end{cases}, \forall i, j
\end{equation}
\item Last, the closing morphology operator is applied on $\varphi$ as post-processing to fill gaps between tampered blocks that incorrectly mark as valid. A 5$\times$5 square is used as a structure element in this step.
\end{enumerate}
%------------------------------------------------------------------------------------------------------------------------------------------------------------------
\subsubsection{Reconstruct digest, Recover tampered regions}
After authentication phase, 2$\times$2 tampered block can be further recovered. To do so, first, the four primary and secondary digest are reconstructed and reshuffled to initial position. In the following, for each invalid block of $host_s$ according $\varphi$ the recovery steps are triggered to correct tampered regions. The reconstruct digest and recover tampered regions procedure includes the following steps:

\begin{enumerate}[1),itemsep=0mm]
\item First of all, the four primary and secondary digests are extracted from $\Psi^k_{i, j}$. Then, the digests are formed according Fig. \ref{fig:organize}, and define primary digests as $\{\dot{d}_p^{s}, \ddot{d}_p^{s}\}$, and the secondary digests as $\{\dot{d}_s^s, \ddot{d}_s^s\}$.
\item In the following, the valid part of $\{\dot{d}_p^{s}, \ddot{d}_p^{s}\}$ and $\{\dot{d}_s^s, \ddot{d}_s^s\}$ are marked and updated. In the other words, $\varphi_{i, j}$ is checked during extraction step to reconstruct each digests based on valid parts. 
\item To place the coefficients into initial positions, the inverse reshuffling are applied on the plane's coefficients of $\{\dot{d}_p^{s}, \ddot{d}_p^{s}\}$ include [$\Gamma, LL, LH, HL, HH, U, V$]. For this aim, each planes of $\{\dot{d}^s_p$, $\ddot{d}^s_p\}$ are converted into 1D matrix. Then, the coefficients of each digests are reordered and reshuffled based on Shift-aside and Mirror-aside operators, and two chaotic sequence generate based on Eq. (\ref{eq:CCS}) with $key_1$ and $key_2$ by Eq. (\ref{eq:reshuffling}):
\begin{align}
&\dot{d}_p(\chi^{\prime}_1(i)) =\dot{d}_p^{s}(i)    \nonumber \\
&\ddot{d}_p(\chi^{\prime}_2(i)) =  \ddot{d}_p^{s}(i)      \nonumber \\
&\forall i \in [1,{\rfrac{M}{4}\times \rfrac{N}{4}}]
\label{eq:reshuffling}
\end{align}
At the end, the $\dot{d}_p$ and $\ddot{d}_p$ is converted to 2D matrix with size of ${\frac{M}{4}\times \frac{N}{4}}$, and $\{\dot{d}^s_p$, $\ddot{d}^s_p\}$ is updated. 
\item Also, the inverse Partner-block reordering process are employed on $\{\dot{d}_s^s, \ddot{d}_s^s\}$ according to Fig. \ref{fig:map}(b) and Fig. \ref{fig:map}(c), respectively.  Let denote result as $\dot{d}_s$ and $\ddot{d}_s$.
\item Now, two unique digests are generated based on valid part of each digests by Eq. (\ref{eq:merge}):
\begin{align}
\label{eq:merge}
d_p(i, j, k) &= \dot{d}_p(i, j, k) \cup \ddot{d}_p(i, j, k) \nonumber \\
d_s(i, j, k) &= \dot{d}_s(i, j, k) \cup \ddot{d}_s(i, j, k) \nonumber \\
\forall i &\in [1, \rfrac{M}{2}], j \in [1, \rfrac{N}{2}], k \in [1, 2, 3]
\end{align}
where $\cup$ is union operator.
\item In this step, to reconstruct $d_p$ pad zeros bits to all coefficients include [$LL, LH, HL$, and $HH$] based on $\Gamma$ and LSBs ignored  according Eq. (\ref{eq:modified}). Also, pad a zeros to chrominance components include [$U, V$]. At the end, convert coefficients form binary to integer type.
\item Next, the invalid regions of chrominance components include [$U, V$] are reconstructed based on valid neighbors, and resized them to $\frac{M}{2}\times\frac{N}{2}$.
\item Now, inverse quantization are applied on all coefficients of each bands and update them by Eq. (\ref{eq:inversequantization}):
\begin{equation}
Coef_{i, j} = Coef_{i, j} \times \mu, \forall i, j
\label{eq:inversequantization}
\end{equation}
where $\mu$ and $Coef_{i, j}$ are quantization step and coefficients of wavelet bands, respectively. 
\item A level inverse LWT is applied on [$LL, LH, HL$, and $HH$], and reconstruct luminance as $Y$. Then, primary digest is converted to RGB, and denote result as $\Psi_p$.
\item The inverse halftone algorithm is employed on $d_s$ to generate the secondary digest from halftone version, and denote result as $\Psi_s$ \cite{ref5}.
\item Finally, the unique digest is achieved by combining valid parts of digests based on Eq.  (\ref{eq:seperate}):
\begin{equation}
\Psi i, j = \Psi_pi, j + \Psi_si, j , \forall i, j
\label{eq:seperate}
\end{equation}
where $+$ is combining operator. So, the unreconstructed pixels in $\Psi_pi, j$ can be reconstruct by $\Psi_si, j$.
\item $\Psi$ is resized to original size of $host_s$ by Eq. (\ref{eq:resize1}):
\begin{equation}
\Psi = R(\Psi + R(host_s, 0.5)\&\neg\varphi, 2)
\label{eq:resize1}
\end{equation}
where +, \&, and $\neg$ are combining, bitwise-and, and complement operators, respectively. Also, $R$ is represented the resize function using bi-cubic interpolation.
\item Now, the pixels of $\Psi$ which are not reconstructed due to the large size of the manipulation, can be recovered based on neighboring pixels.
\item Last, the recovered image is achieved by Eq. (\ref{eq:resize2}):
\begin{equation}
host_r = host_s\&\neg R(\varphi ,2) + \Psi \& R(\varphi ,2)
\label{eq:resize2}
\end{equation}
\end{enumerate}
%----------------------------------------------------------------------------------------------------------------------------------------------------------
\section{Experimental results}
\label{sec:Experimental}
In this section, a series of experiments are employed to prove superiority and efficiency of TRLG compared with other state-of-the-art schemes. All experiments were implemented on a computer with a 3.30 GHz Intel i5 processor, 4.00 GB memory, and Windows 10 operating system, and the programming environment was Matlab R2016b. Furthermore, each watermarked image is modified by Adobe Photoshop cc 2015 to make forgery image. A feasibility and effectiveness investigation for TRLG is conducted using a set of standard images with the size of 512$\times$512 include Baboon, Barbara, Lena, Pepper, Girl, Lake, F16, House, Elaine, Goldhill, Boat, Camera, Toys, Zelda, and Crowd. The number of objects and various types of texture such as edge, smooth, rough, etc. in these images leads to challenging watermarking methods. 

To demonstrate the superiority of TRLG, first, a set of experiments are reported to illustrate the extreme performance of TRLG in term of quality of the watermarked image and generated digests. The second set is conducted to demonstrate the performance of TRLG in terms of security and detect special tampering. The third set of experiments is performed to show the recovered rates and quality of the recovered image under various tampering. In the following, some types of tampering with security attacks are done on the watermarked image to visually present the performance of TRLG in the context of tamper detection and recovery. At the end, TRLG is compared with state-of-the-art schemes in various terms which play the main role in tamper detection and recovery schemes.

%-----------------------------------------------------------------------------------------------------------------------------------------------------------------------------------------------------------------------------
\subsection{Quality analysis of watermarked image and digests}
In this set of experiments, first, the quality of the watermarked image is analyzed by comparing with state-of-the-art schemes. To do so, the PSNR and SSIM values of the watermarked images which generated based on TRLG and other schemes that presented in recent years are reported and compared in Table \ref{TABLE:compare_psnr_ssim}. As seen, the PSNR values of the watermarked image of TRLG are extremely larger than other proposed schemes. Furthermore, to prove the performance of TRLG in term of quality, besides PSNR, the SSIM metric is employed. The measure of SSIM was presented based on the characteristics of the Human Visual System (HVS), that compared the information of structure, luminance, and contrast for watermarked image quality assessment. Subsequently, the SSIM metric of the watermarked image is close to one, that proves the efficiency of TRLG in term of quality of the watermarked image. It should be noted, unfortunately, the most of proposed schemes in recent years only focused on grayscale image, and prove the performance based on PSNR. However, in TRLG, the SSIM metric are reported for fifteen standard color and gray images. 
%--------------------------------------------------------------------------------------------------------------------------------------------------------------------------------------------
\begin{table}[b]
\footnotesize
\caption{Number of bits for each 4$\times$4 block (Accuracy 2$\times$2).\\
CR : Compression Ratio.}
\label{TABLE:volume}
\renewcommand{\arraystretch}{1.5}
\scalebox{1} {
\begin{tabular*}{\columnwidth}{@{\extracolsep{\fill}}@{}l@{}c@{}c@{}c@{}c@{}c@{}c@{}c@{}}
\cline{1-8}
&\multirow{2}{*}{Method}&Inverse&MSB&MSB&DCT&TRLG&TRLG\\
&&Halftone&5 Bits&6 Bits&DC&DCT&LWT\\
\cline{1-8}
\multirow{2}{*}{Bits}&Gray&4&20&24&40&20&20\\
&Color&12&60&72&120&34&34\\
\cline{2-8}
\multirow{2}{*}{CR}&Gray&32&6.4&5.3&3.2&6.4&6.4\\
&Color&32&6.4&5.3&3.2&11.3&11.3\\
\cline{1-8}
\end{tabular*}}
\end{table}

As said before, one of the novelties of TRLG is proposing compact digest with high quality as DLG. In addition, the halftone version of the image is used to provide the further chance for recovering tampered regions. To prove the superiority and efficiency of DLG, and also the inverse halftone version \cite{ref5}, first, PSNR and SSIM values and the zoom of digests that obtained based on each technique are illustrated in Fig. \ref{fig:dbdigest}. As can be seen, the visual distortions are very low, and no blocky artifact can be detected in digests. In addition, the quality of digests is compared by other traditional strategies such as an averaging and DCT based technique in Fig. \ref{fig:compared_digest}. In this way, the digest based on averaging technique is obtained by averaging 2$\times$2 blocks and fetch 5 or 6 Most Significant Bits of each block. Also, the DCT based digest is constructed by fetching DC coefficient from each 2$\times$2 block. Furthermore, the proposed digest is analyzed based on DCT instead of LWT. To do so, the DC and first two AC coefficients are considered for each block. Finally, all digests are resized to initial size (512$\times$512) and compared to the original image. It is observed that the essential metrics like PSNR and SSIM are effectively high for DLG compared to other techniques. However, there is a light degradation of the texture image like Baboon rather than other schemes. Since all digests can recover tampered region with 2$\times$2 accuracies, so the compassion is quite fair.
%--------------------------------------------------------------------------------------------------------------------------------------------------------------------------------------------
%------------------------------------------------------------------------------------------------------------------------------------------------------------
As described previously, GA is employed in TRLG to select best thresholds for classifying each block in term of texture. Accordingly, the main challenge of generating digests for various type of blocks are solved, clearly. In another word, an intelligent trade-off is achieved between the low and high frequency of each block and makes DLG more flexible for various type of images. The result of this strategy is shown in Fig. \ref{fig:texture}. Also, the volume of digests which generated based on each technique are listed in Table \ref{TABLE:volume}. The low volume and high quality of digests that generated based on DGL makes the TRLG more efficient for tamper detection based on watermarking techniques. In other words, the low volume of embedded watermark bits leads the low degradation on host image that can not be detected by the naked eye, and also useful for the watermarking scheme with low data payload or embedding capacity. Summarily, these are primary reasons for why TRLG has better quality for watermarked and recovered image rather than other schemes. Totally, the experimental results have proved the efficiency and superiority of DLG in terms of quality and compactness compared to other schemes.

%----------------------------------------------------------------------------------------------------------------------------------------------------------------------------------------
%----------------------------------------------------------------------------------------------------------------------------------------------------------
\begin{figure}[t]
\center
\begin{tabular}{cc}
\includegraphics[width=0.45\columnwidth]{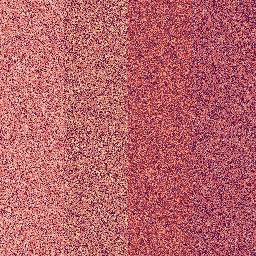} &
\includegraphics[width=0.45\columnwidth]{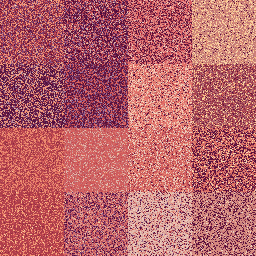} \\
(a) & (b) \\
\includegraphics[width=0.45\columnwidth]{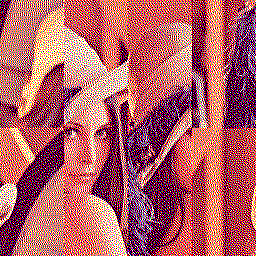} &
\includegraphics[width=0.45\columnwidth]{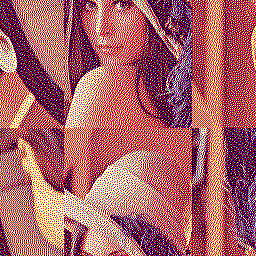} \\
(c) & (d)
\end{tabular}
\caption{Lena image scrambling. (a) Primary digest 1, (b) Primary digest 2, (c) Secondary digest 1, (d) Secondary digest 2.}
\label{fig:scramble}
\end{figure}
%---------------------------------------------------------------------------------------
\subsection{Analysis the security of watermark and detect tampering}
In this subsection, the sets of experiments are reported to prove the efficiency of TRLG in terms of security and detect special tampering. 

Evidently, the security of watermarking schemes plays important role in designing watermarking systems. The security of tamper detection and recovery methods are divided into two categories as scrambling (shuffling) digests and security of embedded watermark. In TRLG, a new chaotic method for determining blocks mapping, reduce the correlation between adjacent pixels and encrypting the watermark is applied. To shuffling the primary digests CCS with Shift-aside and Mirror-aside operations are employed. The CCS is one of the novels chaotic map which proposed by Pak et al \cite{ref2}. The CCS compared to other chaotic maps have three advantages. Firstly, it is one-dimensional. Secondly, have various control parameters which lead to an increase keyspace. Thirdly, implemented easily and have lower computation-cost. Generally, the chaotic algorithm is a good option for securing watermarking methods, because of sensitivity to initial conditions, uncertain and non-statistical prediction. In other words, these methods have a definite process, but appear to be random. So, by using chaotic maps the security of the watermarking system will be improved, and the location of embedding digests is seen random and unpredictable for the attacker, but it is clear and meaningful for the main recipient. As said before, two secondary digests are scrambled by Partner-block without using any chaotic map. The result of scrambling four digests are illustrated in Fig. \ref{fig:scramble}. 

%----------------------------------------------------------------------------------------------------------------------------------------------------------
\begin{table}[H]
\footnotesize
\caption{Analysis encryption measures for watermark - Test image Lena.}
\label{TABLE:security}
\renewcommand{\arraystretch}{1.5}
\scalebox{1} {
\begin{tabular*}{\columnwidth}{@{\extracolsep{\fill}}@{}l@{}c@{}c@{}c@{}c@{}c@{}c@{}c@{}}
\cline{1-7}
Measures&Entropy&STD&MAE&NPCR&UACI&EQ\\
\cline{1-7}
Initial&6.0475&66.2202&0&0.9961&0.3346&0\\
Dependent&7.8127&73.8491&19.7287&0.9961&0.3346&334.0078\\
Encrypt\&Permute&7.9972&73.7177&19.1009&0.9961&0.3346&312.6719\\
Encrypt (GA)&7.9972&73.8918&19.1541&0.9961&0.3346&313.0021\\
\cline{1-7}
\end{tabular*}}
\end{table}

%----------------------------------------------------------------------------------------------------------------------------------------------------------
The security of watermark is a vital problem in tamper detection schemes. In particular, the security of embedded watermark implies that the watermark should be difficult to remove or modify without damaging the watermarked image, even with a full knowledge of the embedding and extracting algorithm. As mentioned in the previous section, In TRLG to providing the security of watermark, first, the watermark of each block is depended to the content of block, second, the watermark is encrypted and permuted, and at last the watermark is encrypted by the key which is generated based on GA. 

To evaluate the performance of TRLG in term of security of watermark the statistical tests and security analysis are carried out in Table \ref{TABLE:security}. In this Table various measures include Entropy, Standard Deviation, Mean Absolute Error, Number of Pixels Change Rate, Unified Average Changing Intensity, Encryption Quality are reported. The results and security analysis are proved that TRLG has a high-security level and excellent performance in encryption.
%----------------------------------------------------------------------------------------------------------------------------------------------------------
\begin{figure}[t]
\center
\includegraphics[width=1\columnwidth,trim=0cm 0cm 0cm 0cm,clip]{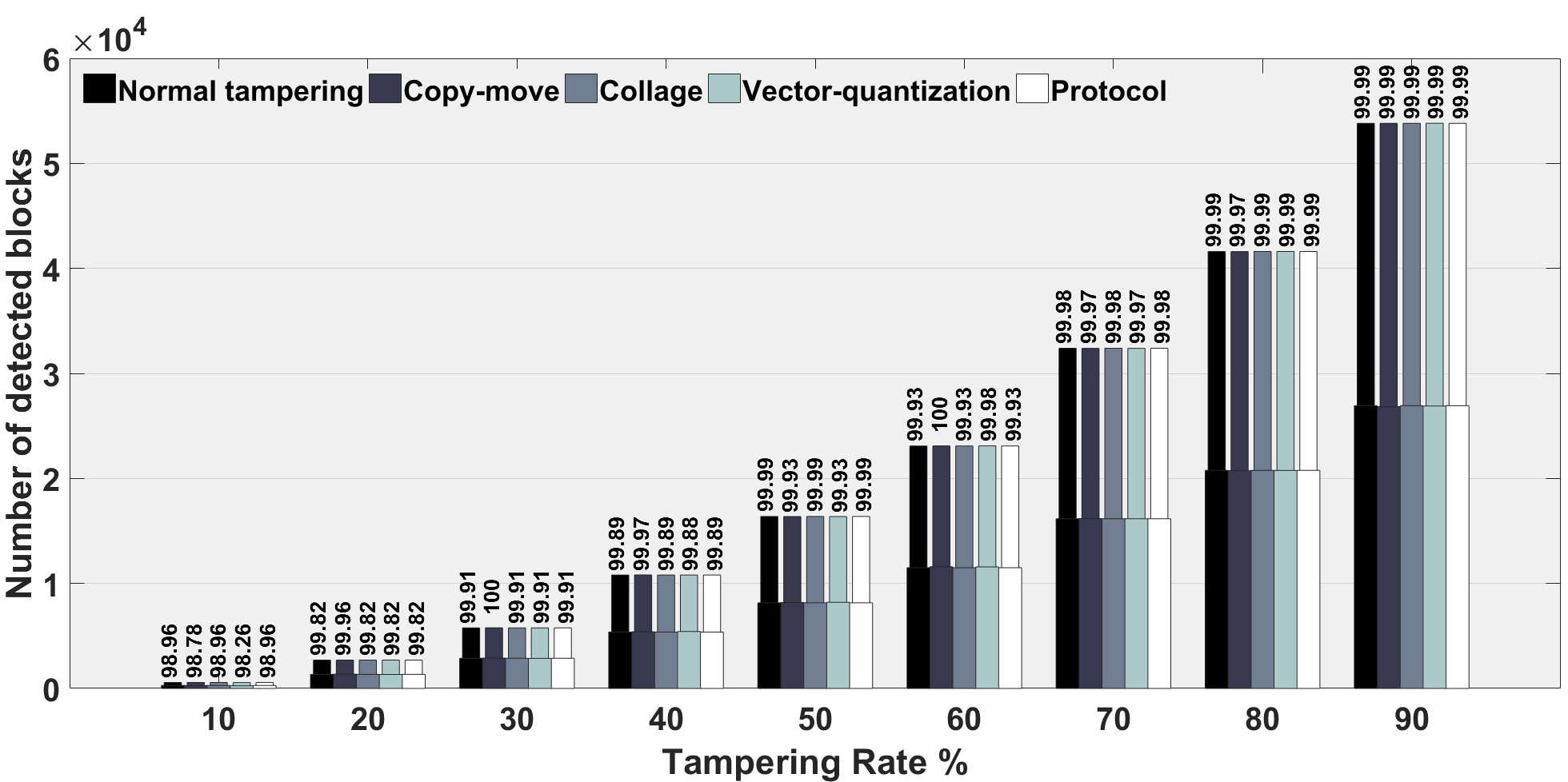}
\caption{The performance of TRLG in term of detect security tampering under various tampering rates. Test images Lena and Barbara. \\ Note: The thin bar shows post-processing (fill the gap by morphology).}
\label{fig:typeoftamper}
\end{figure}
%----------------------------------------------------------------------------------------------------------------------------------------------------------
\begin{figure*}[t!]
\center
\setlength{\tabcolsep}{1pt}
\begin{tabular}{ccc}
\includegraphics[width=0.33\textwidth,trim=2.2cm 0cm 3cm 0cm,clip]{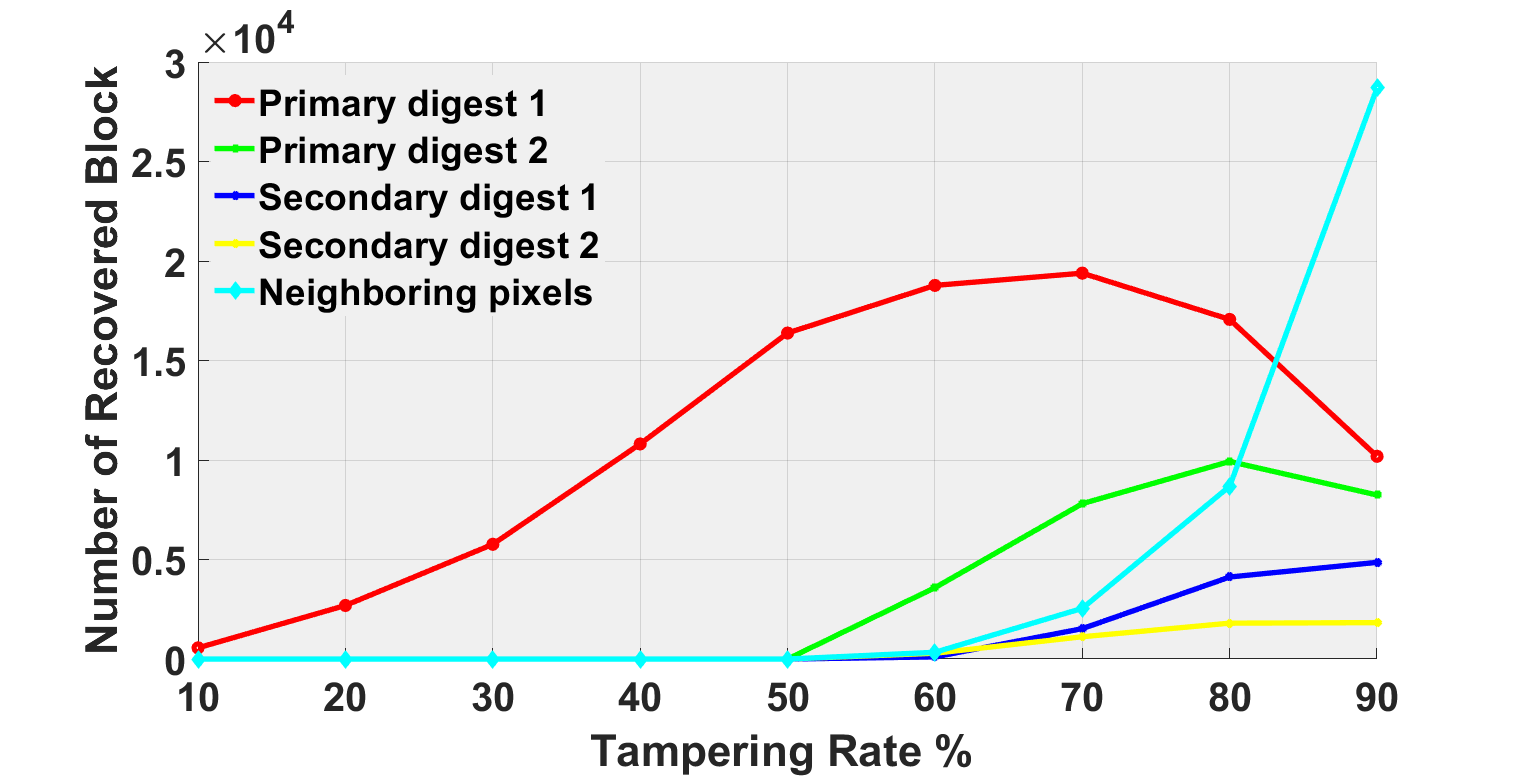} &
\includegraphics[width=0.33\textwidth,trim=2.2cm 0cm 3cm 0cm,clip]{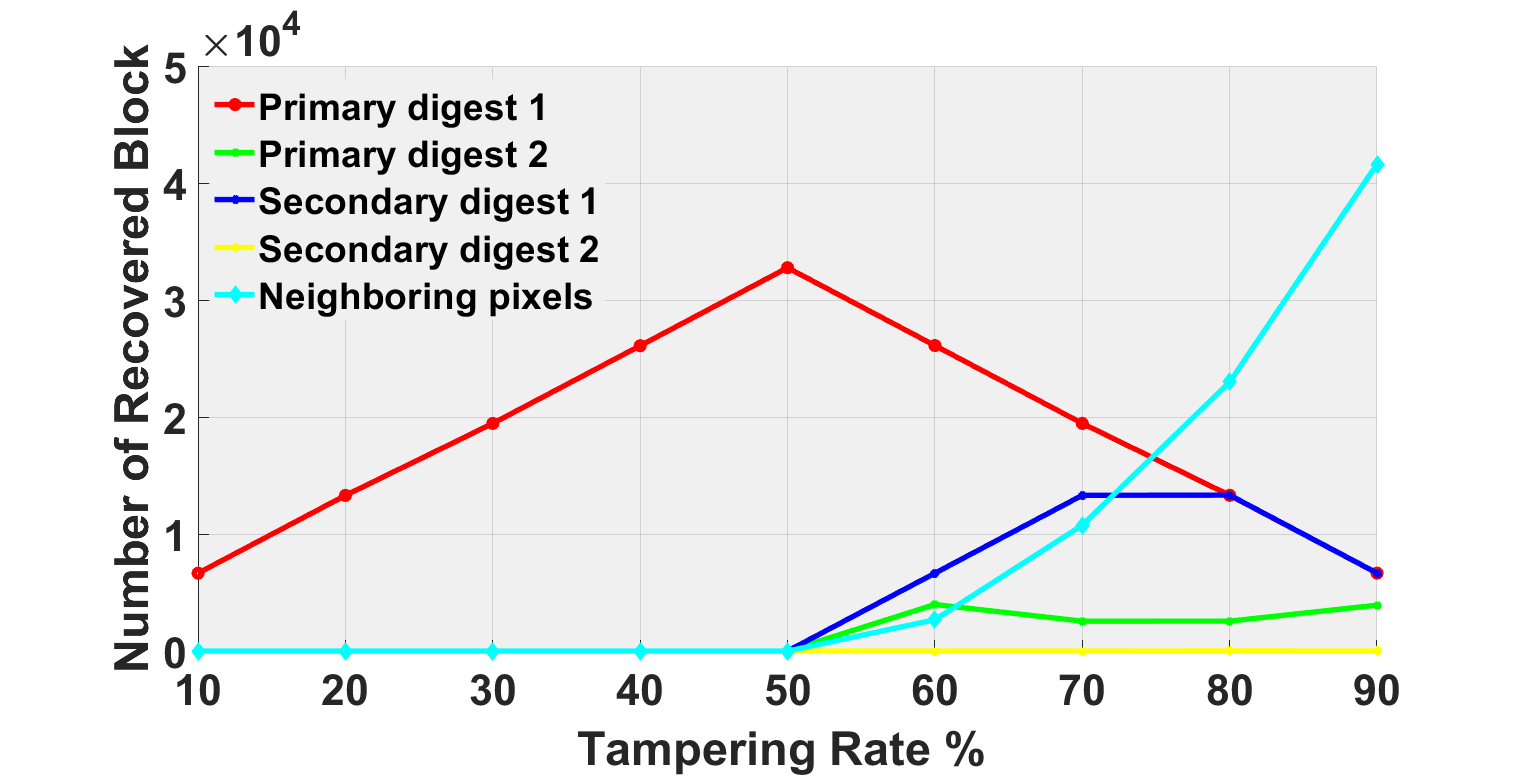} & 
\includegraphics[width=0.33\textwidth,trim=2.2cm 0cm 3cm 0cm,clip]{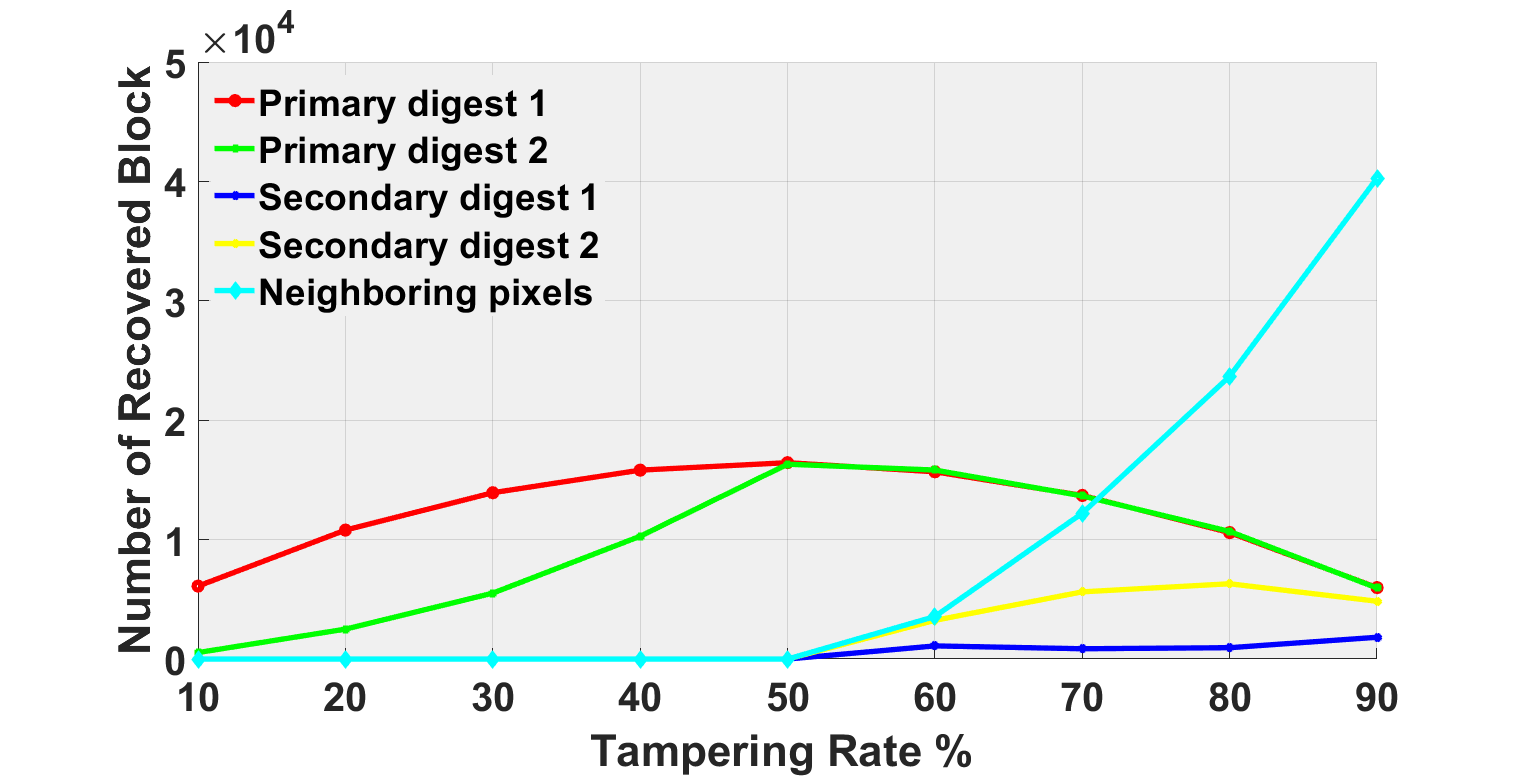}\\
(a) & (b) & (C)\\ 
\end{tabular}
\caption{The portion of digests in recovery (a) Center \% (height, width), (b) Left to right \%width, (c): Up to bottom \%height - Test image Lena.}
\label{fig:recoveryrate}
\end{figure*}
%----------------------------------------------------------------------------------------------------------------------------------------------------------
\begin{figure*}[t!]
\center
\setlength{\tabcolsep}{1pt}
\begin{tabular}{cc}
\includegraphics[width=0.49\textwidth,trim=2.5cm 0cm 3cm 0cm,clip]{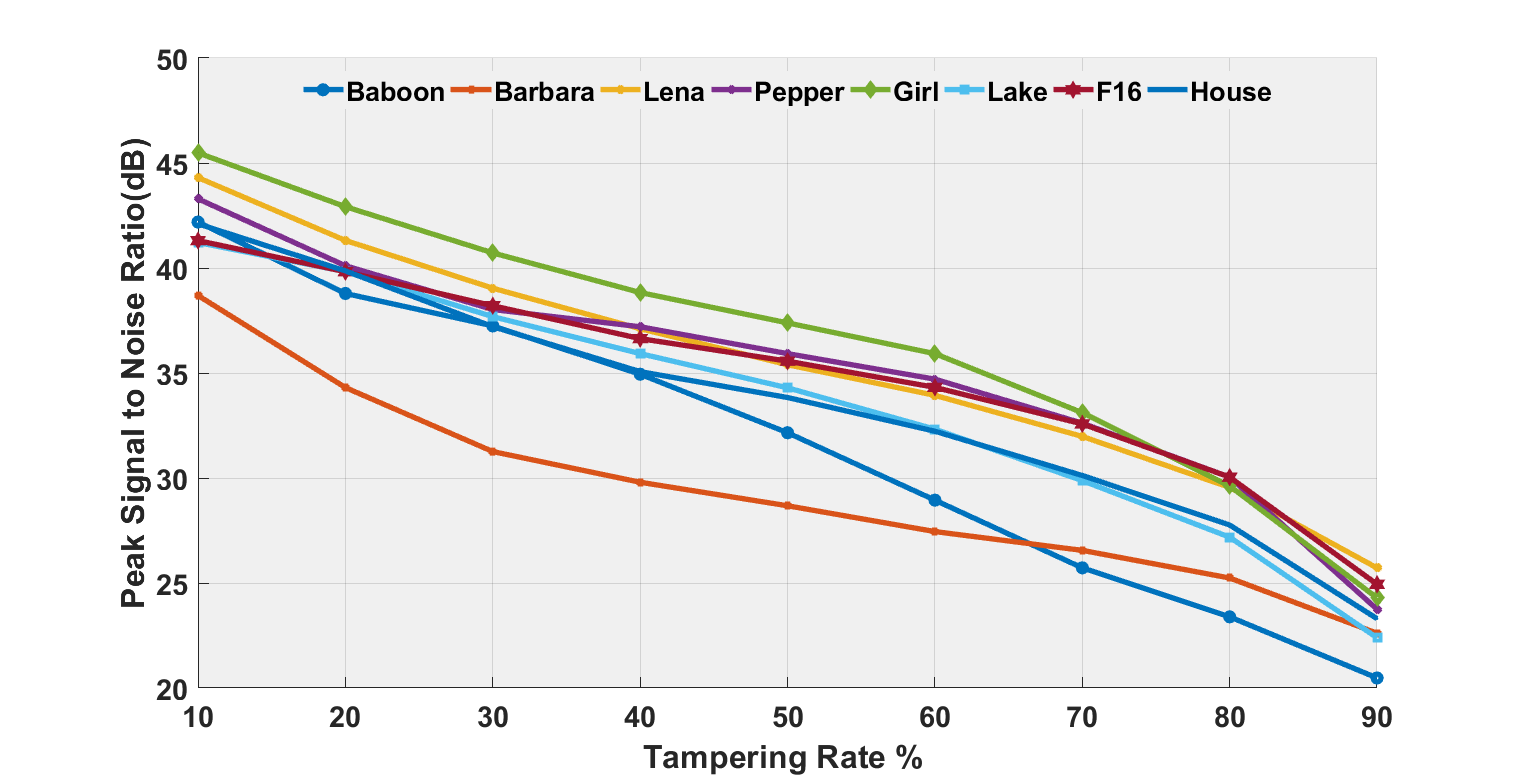} &
\includegraphics[width=0.49\textwidth,trim=2.5cm 0cm 3cm 0cm,clip]{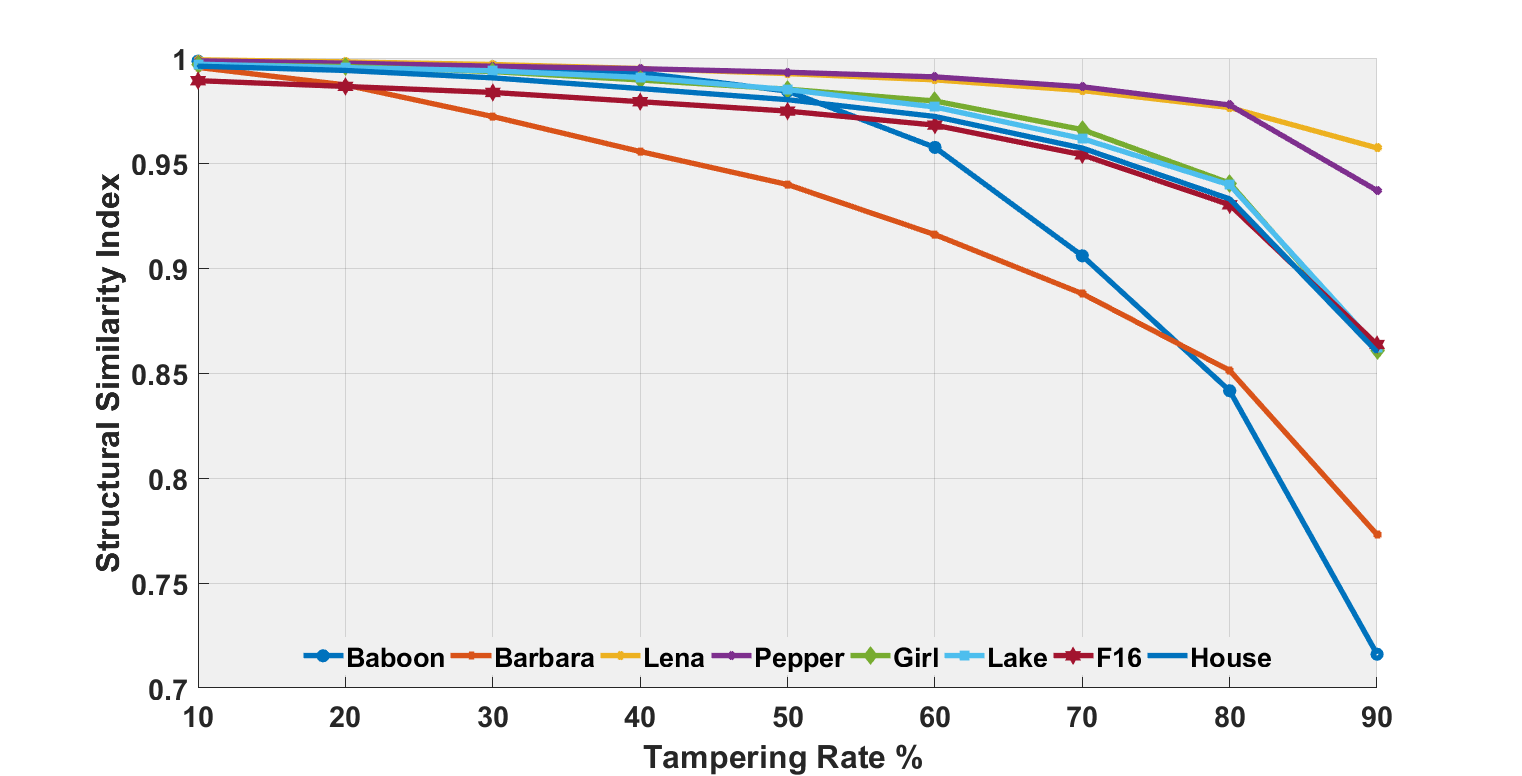}\\
(a) & (b)\\ 
\includegraphics[width=0.49\textwidth,trim=2.5cm 0cm 3cm 0cm,clip]{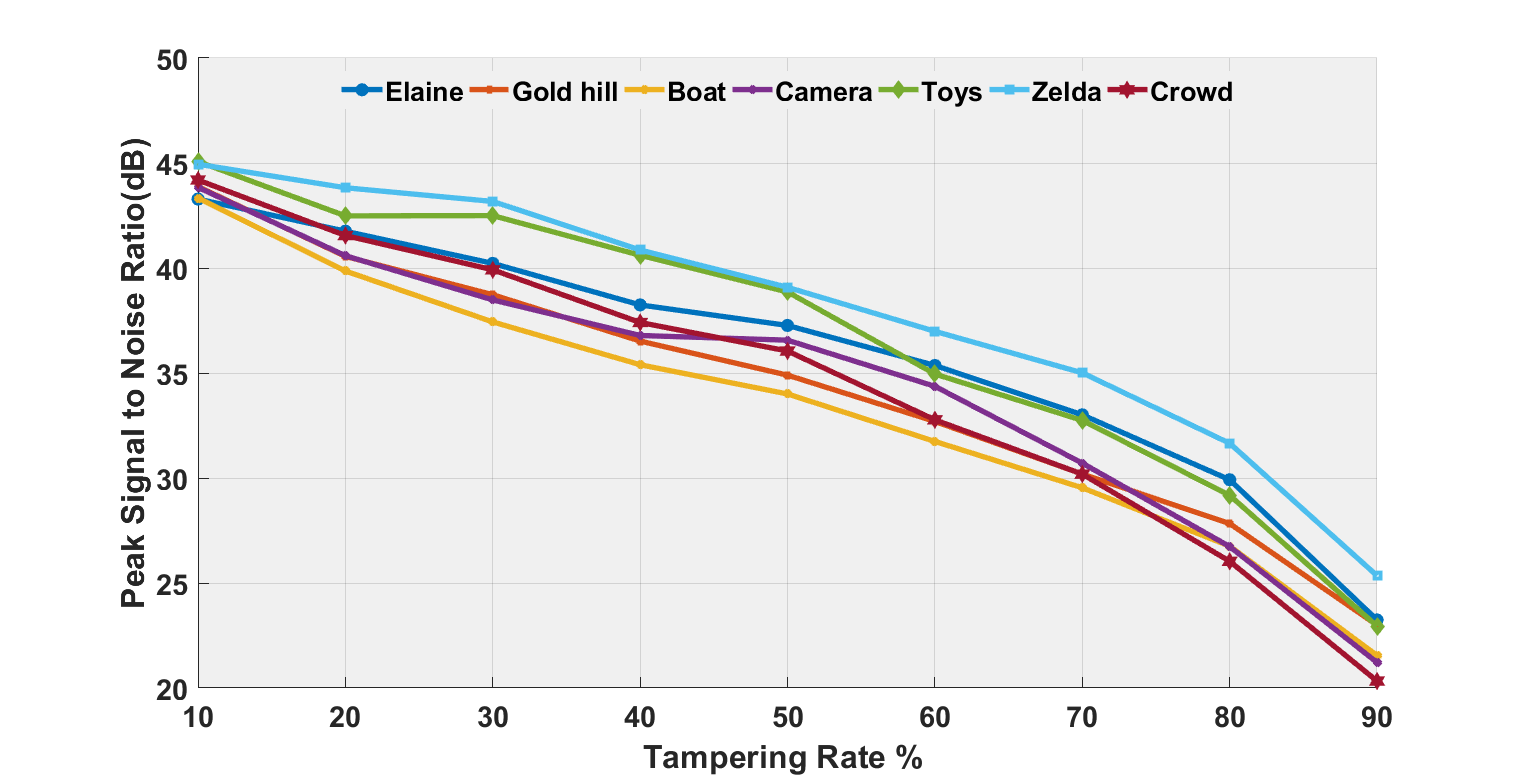} &
\includegraphics[width=0.49\textwidth,trim=2.5cm 0cm 3cm 0cm,clip]{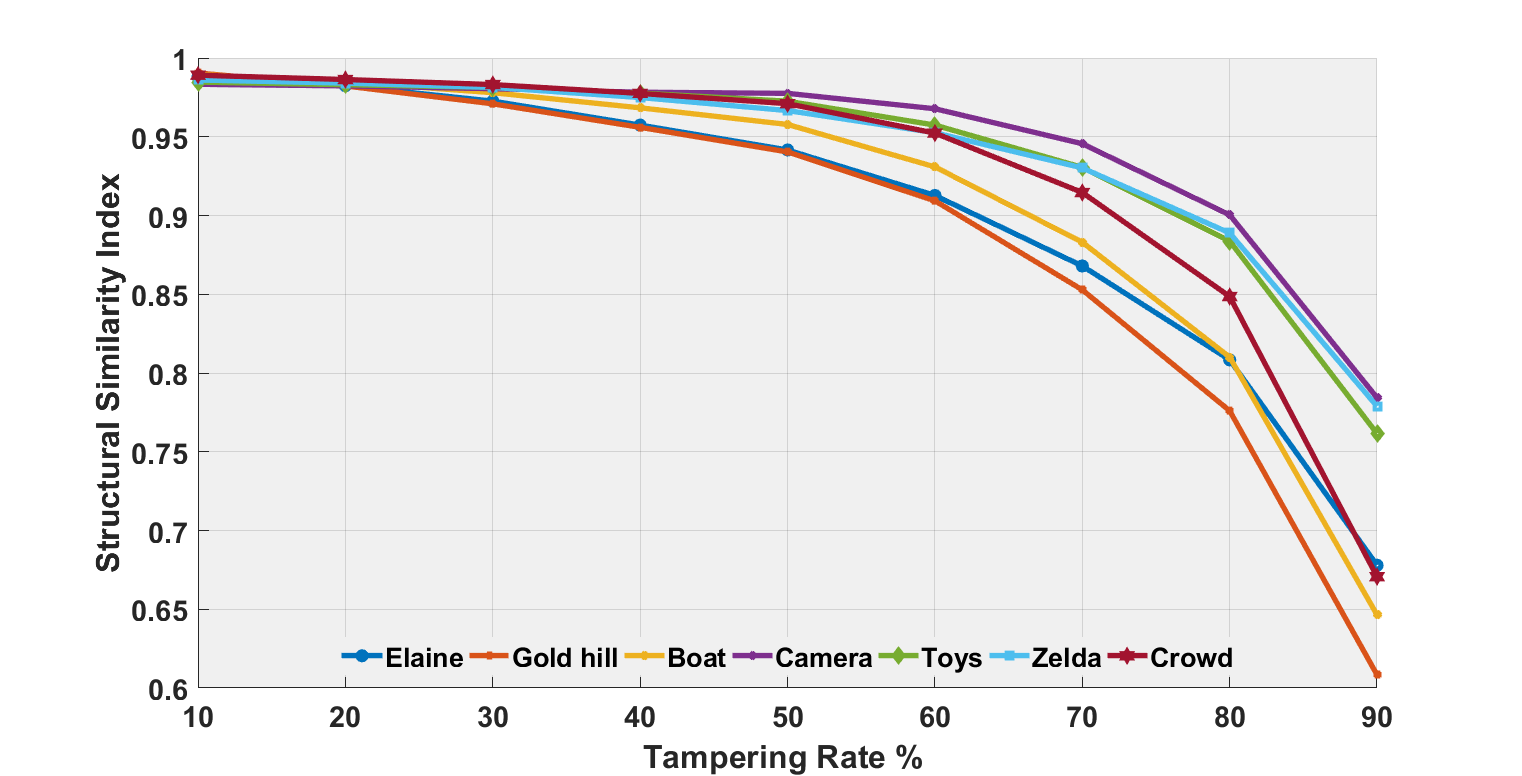}\\
(c) & (d)\\
\end{tabular}
\caption{The quality of the recovered image with different tampering rates. (a) 
Color-PSNR, (b) Color-SSIM, (c) Gray-PSNR, (d) Gray-SSIM.}
\label{fig:recoverypsnr}
\end{figure*}
%----------------------------------------------------------------------------------------------------------------------------------------------------------
\begin{figure*}[t!]
\center
\setlength{\tabcolsep}{2pt}
\begin{tabular}{ccccc}
\includegraphics[width=0.19\textwidth]{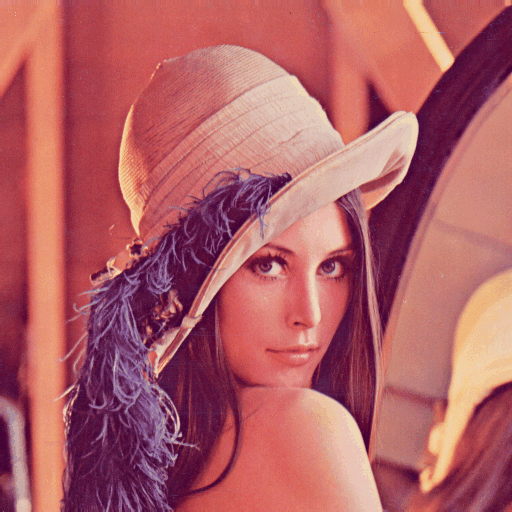} &
\includegraphics[width=0.19\textwidth]{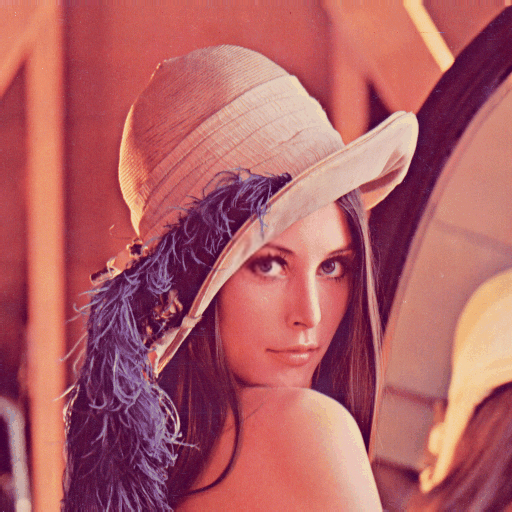} &
\includegraphics[width=0.19\textwidth]{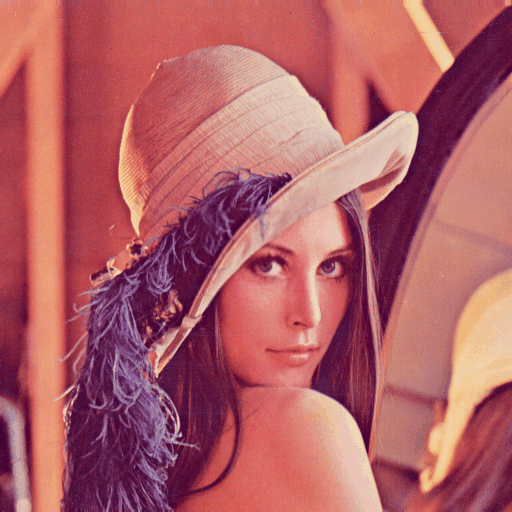} &
\includegraphics[width=0.19\textwidth]{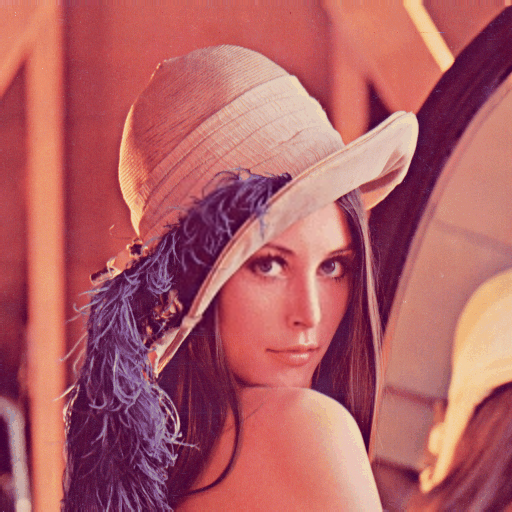} &
\includegraphics[width=0.19\textwidth]{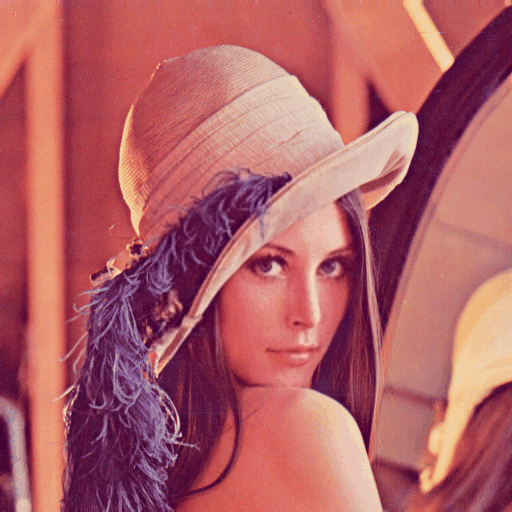}\\
(a) & (b) & (c) & (d) & (e)\\                             
\includegraphics[width=0.19\textwidth]{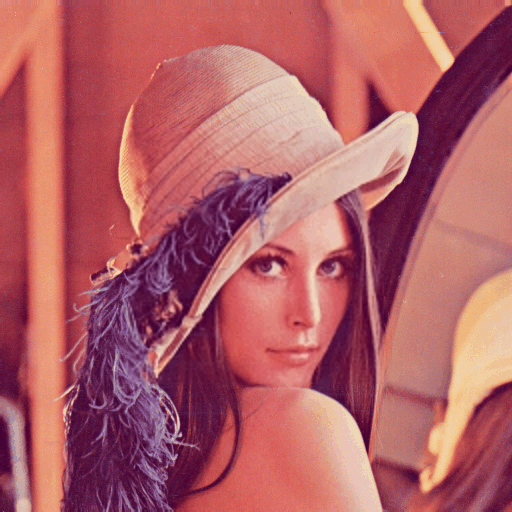} &
\includegraphics[width=0.19\textwidth]{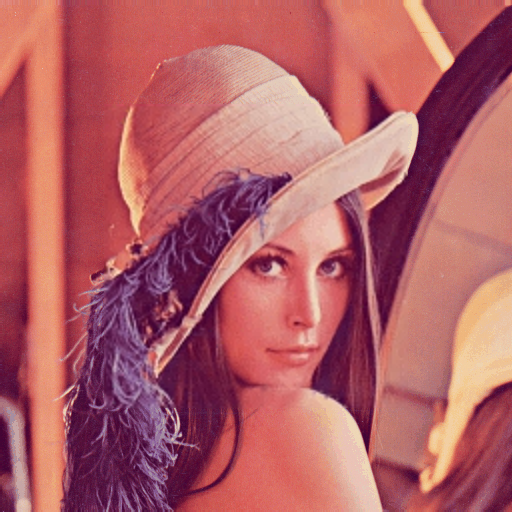} &
\includegraphics[width=0.19\textwidth]{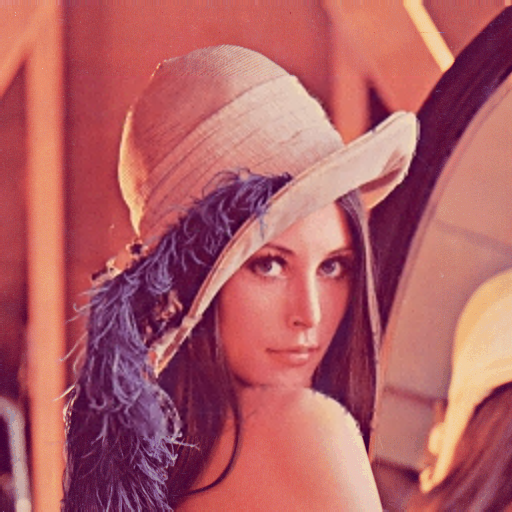} &
\includegraphics[width=0.19\textwidth]{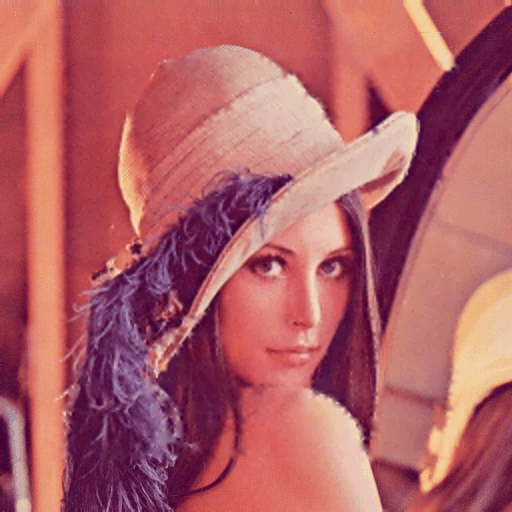} & 
\includegraphics[width=0.19\textwidth]{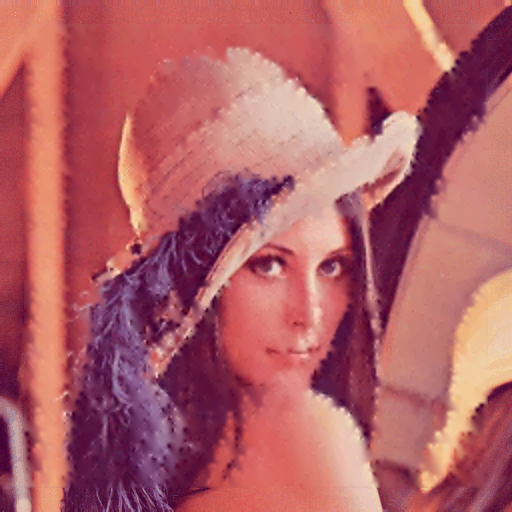} \\
(f) & (g) & (h) & (i) & (j)
\end{tabular}
\caption{The quality of recovered image under various tampering rates - Center (Tampering rate, PSNR, SSIM). (a) Watermarked image, (b) \{10\%, 44.8281, 0.9992\}, (c) \{20\%, 41.7046, 0.9983\}, (d) \{30\%, 39.4957, 0.9971\}, (e) \{40\%, 37.1910, 0.9951\}, (f) \{50\%, 35.4515, 0.9926\}, (g) \{60\%, 33.9525, 0.9896\}, (h) \{70\%, 32.0112, 0.9846\}, (i) \{80\%, 29.7049, 0.9768\}, (j) \{90\%, 25.7410, 0.9574\}.}
\label{fig:visualleanrecover1}
\end{figure*}
%----------------------------------------------------------------------------------------------------------------------------------------------------------
\begin{figure*}[t!]
\center
\setlength{\tabcolsep}{2pt}
\begin{tabular}{ccccc}
\includegraphics[width=0.19\textwidth]{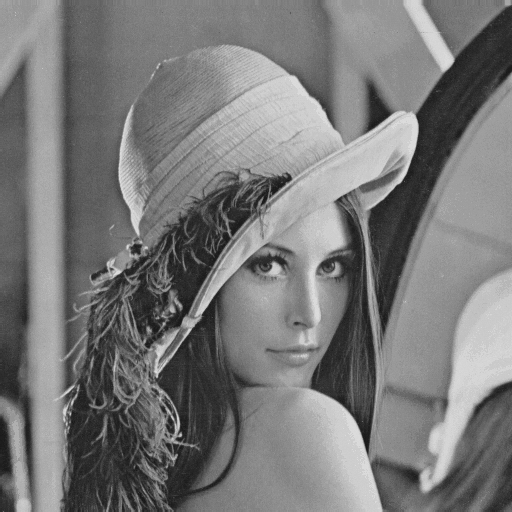} &
\includegraphics[width=0.19\textwidth]{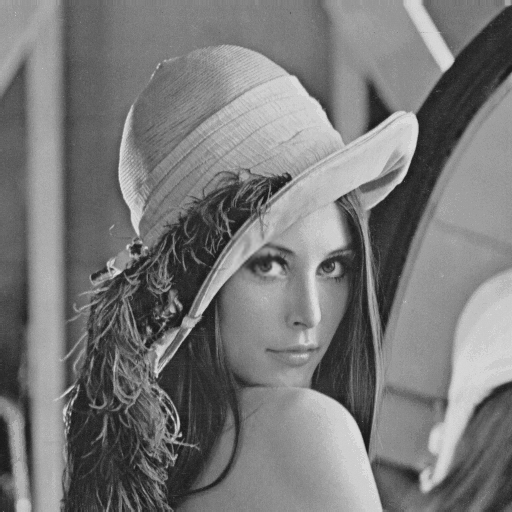} &
\includegraphics[width=0.19\textwidth]{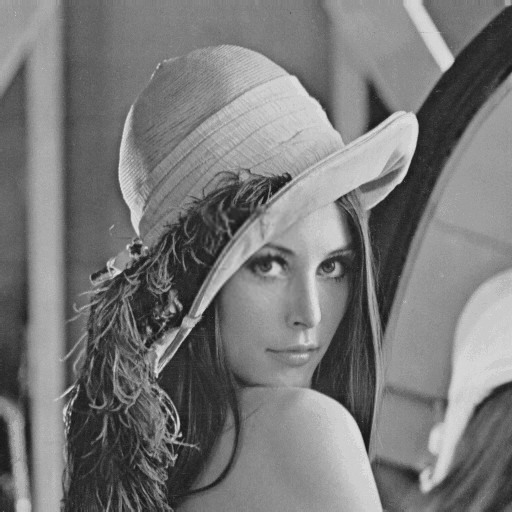} &
\includegraphics[width=0.19\textwidth]{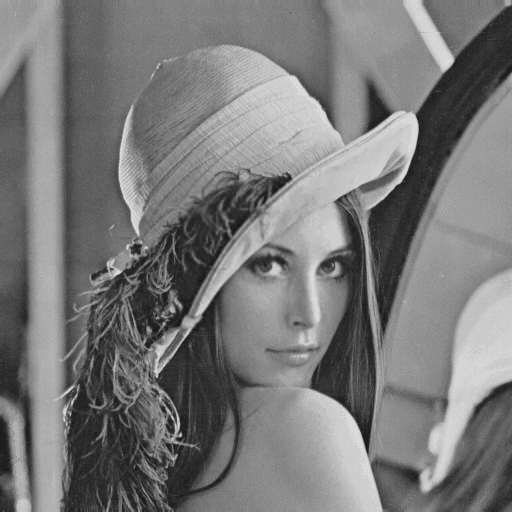} &
\includegraphics[width=0.19\textwidth]{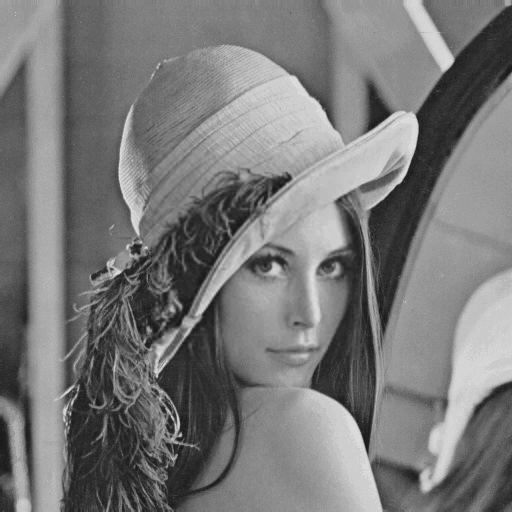}\\
(a) & (b) & (c) & (d) & (e)\\                             
\includegraphics[width=0.19\textwidth]{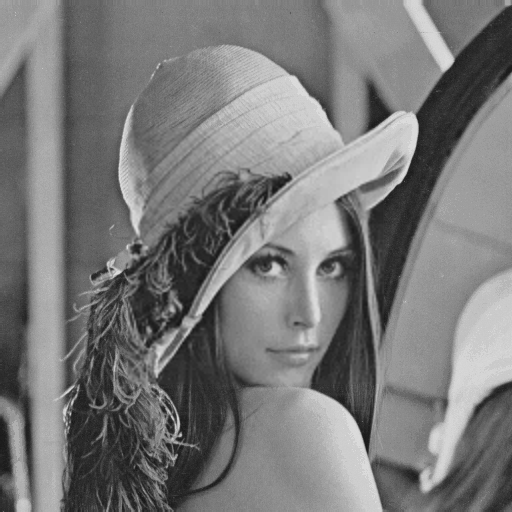} &
\includegraphics[width=0.19\textwidth]{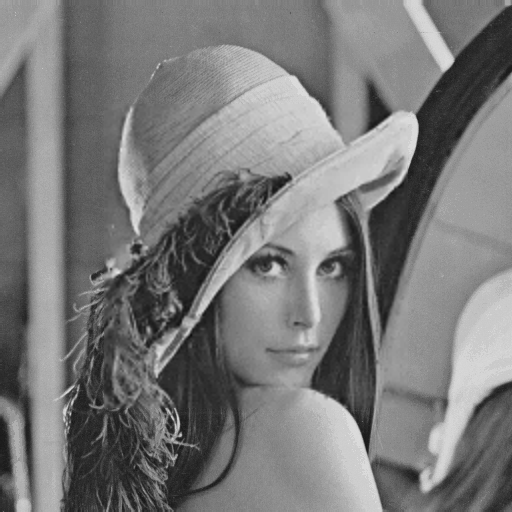} &
\includegraphics[width=0.19\textwidth]{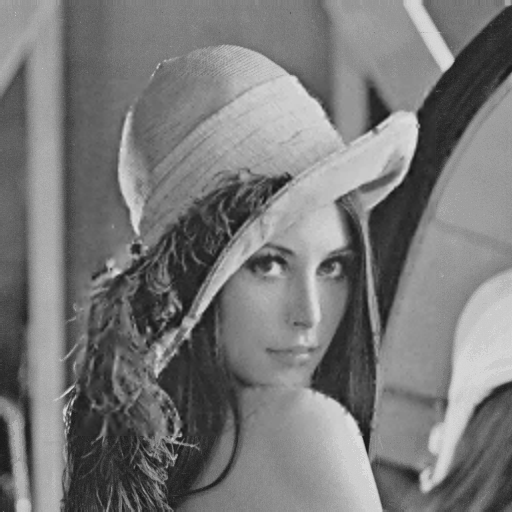} &
\includegraphics[width=0.19\textwidth]{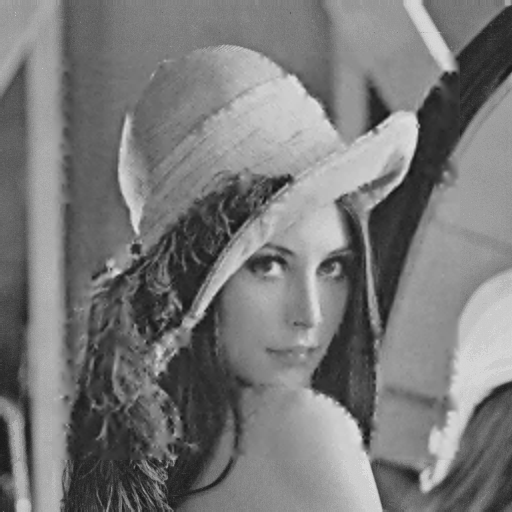} & 
\includegraphics[width=0.19\textwidth]{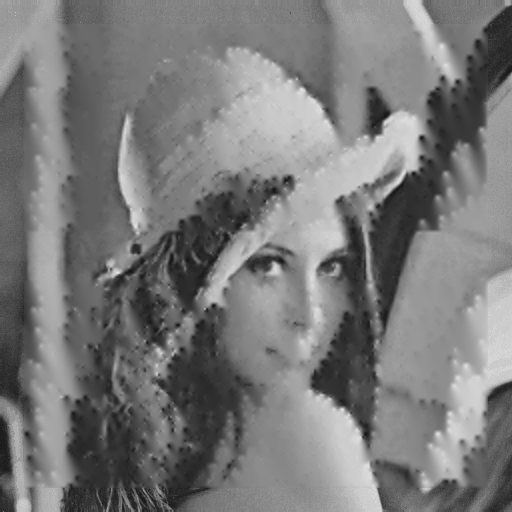} \\
(f) & (g) & (h) & (i) & (j)
\end{tabular}
\caption{The quality of recovered image under various tampering rates - Center (Tampering rate, PSNR, SSIM). (a) Watermarked image, (b) \{10\%, 44.7263, 0.9876\}, (c) \{20\%, 42.2214, 0.9841\}, (d) \{30\%, 40.3250, 0.9793\}, (e) \{40\%, 38.3121, 0.9709\}, (f) \{50\%, 36.6694,  0.9602\}, (g) \{60\%, 33.9932, 0.9371\}, (h) \{70\%, 31.5106, 0.8993\}, (i) \{80\%, 28.4436, 0.8470\}, (j) \{90\%, 22.1790, 0.7201\}.}
\label{fig:visualleanrecover2}
\end{figure*}
%----------------------------------------------------------------------------------------------------------------------------------------------------------

The various types of tampering which are applied on the watermarked image are divided into five categories as below:
\begin{enumerate}[1),itemsep=0mm]
\item \textbf{Normal tampering:} In this type of tampering an object added, removed, or modified on watermarked image from the desired image to produce the fake image.
\item \textbf{Copy move:} In copy move tampering, a part of the watermarked image is copied and pasted it somewhere else to generated forge image.
\item \textbf{Collage:} In collage attack, a forged image is generated by placing some part of the second watermarked image into the same spatial location in destination watermarked image. Accordingly, their relative spatial locations in the image were preserved. In this type, both watermarked image are generated based on same keys.
\item \textbf{Vector quantization:} In this mode, a spurious image is constructed by copying a section from the second watermarked image and past into the desired position in the watermarked image. It should be noted, similar to the collage both watermarked image are achieved based on same keys. These types of tampering are effective on the schemes that are block or pixel independent.
\item \textbf{Protocol:} One of the major tampering is protocol attack that the most of previous schemes are ignored it. The protocol attack involves replacing a part of an extra image into the watermarked image with keeping the Least Significant Bits of the block intact which carried the watermark data (deactivation of the watermark). In other words, it used the advantage of semantic deficits of the watermark‘s implementation. 
\end{enumerate}

The result of TRLG in term of tamper detection based on mentioned tampering under various rates are illustrated in Fig. \ref{fig:typeoftamper}. The experiments are shown the high accuracy of tamper detection under the security tampering. 
%----------------------------------------------------------------------------------------------------------------------------------------------------------
\subsection{Analysis recovery rates and quality of recovered image}
In this subsection, various experiments are conducted to prove the performance of TRLG in terms of recovery rate and quality of recovered image compared to state-of-the-art schemes. In TRLG, PSNR and SSIM values of the recovered image with respect to the original images are reported.

In the first set, to show the performance of TRLG in term of recovery rate, the number of recovered blocks by each digest is illustrated in Fig. \ref{fig:recoveryrate}. As seen, in TRLG the tampered regions under 50\% ratio are recovered without using neighboring pixels. In other words, the TRLG can recover tampered regions by using primary digest when the left, right, top, or bottom of the image is totally modified. Moreover, TRLG is able to recover most tamper regions by embedded digest when tampering rates is up to 80\%. Accordingly, the desired performance of Shift-aside, Mirror-aside, and Partner-block operations are proved.

In Fig. \ref{fig:recoverypsnr}, the curve of PSNR and SSIM values for the various recovered image with respect to the different tampering ratio are illustrated. As seen, in high tampering rates, the recovered images have satisfactory quality. Hence, TRLG has a good performance in term of recovery under various tampering rates for different type of image with extensive tampering. 

In the following, the recovered Lena image with quality measures under various tampering rates is visually presented in Figs. \ref{fig:visualleanrecover1} and \ref{fig:visualleanrecover2}. As it’s clear in this figures, TRLG has an extraordinary result. As mentioned before, four chances are provided in TRLG to recover the tampered regions. Meanwhile, attention to high and low frequencies, and also intelligently classify block in term of texture leads to remove blocky effect and increase the quality of recovered image under large tampering rates. So, it is easier for expert or any viewer to get more perceptual information from the recovered image. 

Next, the quality of recovered image relative to various tampered sizes and locations are listed in Table \ref{TABLE:recoveryratelcation}. From the results, it was found that TRLG has absolute superiority compared to other schemes. 

Finally, In Table \ref{TABLE:compare_recovery} a comparison between TRLG and some recent fragile schemes in terms of quality of recovered image relative to various tampered rates is provided. As seen, PSNR values of recovered images in the TRLG is significantly and effectively higher than the other methods. In other words, TRLG has the much better capability of tamper recovery particularly when the tampered regions are extremely large. Thus, TRLG is more flexible than previously existing schemes.

Totally, the satisfactory performance and superiority of TRLG in terms of recovery rate and quality of the recovered image are demonstrated in the above extensive experiments. As seen, both PSNR and SSIM, and also visual effects prove that TRLG has not only an extremely high accuracy of tampering localization but also a relatively very high recovery rate. Also, the blocking effects are removed because of using the small size of block and generate intelligent digests with high quality.
%----------------------------------------------------------------------------------------------------------------------------------------------------------
\begin{table*}[t]
\footnotesize
\caption{Quality of recovery image under various tampering rates - Test image Lena.}
\label{TABLE:recoveryratelcation}
\renewcommand{\arraystretch}{1.5}
\scalebox{1} {
\begin{tabular*}{\textwidth}{@{\extracolsep{\fill}}@{}l@{}c@{}c@{}c@{}c@{}c@{}c@{}c@{}c@{}c@{}c@{}c@{}}
\cline{1-12}
\multirow{2}{*}{Metric}&\multirow{2}{*}{Mode}&\multirow{2}{*}{Location}&\multicolumn{9}{c}{Tampering Rates \%}\\
\cline{4-12}
&&&10&20&30&40&50&60&70&80&90\\
\cline{1-12}
\cline{3-12}
\multirow{6}{*}{PSNR}&\multirow{3}{*}{Color}&Center&44.305&41.316&39.039&37.111&35.396&33.937&31.981&29.553&25.728\\
&&Left to Right&42.539&38.853&36.195&34.238&33.455&30.718&26.849&18.969&16.739\\
&&Up to Bottom&42.221&39.765&37.642&35.911&34.799&29.654&26.157&23.071&21.342\\
\cline{2-12}
&\multirow{3}{*}{Gray}&Center&44.156&41.836&40.221&38.173&36.554&33.834&31.489&28.419&22.169\\
&&Left to Right&43.666&40.511&37.418&35.597&34.832&27.019&21.836&18.313&15.603\\
&&Up to Bottom&43.679&41.153&38.634&36.182&34.381&28.446&26.042&23.005&19.264\\
\cline{1-12}
\multirow{6}{*}{SSIM}&\multirow{3}{*}{Color}&Center&0.9992&0.9982&0.9969&0.9950&0.9926&0.9896&0.9845&0.9764&0.9572\\
&&Left to Right&0.9986&0.9970&0.9939&0.9906&0.9884&0.9813&0.9698&0.9140&0.8705\\
&&Up to Bottom&0.9989&0.9976&0.9959&0.9939&0.9920&0.9824&0.9690&0.9476&0.9123\\
\cline{2-12}
&\multirow{3}{*}{Gray}&Center&0.9875&0.9839&0.9792&0.9705&0.9599&0.9363&0.8989&0.8467&0.7198\\
&&Left to Right&0.9805&0.9705&0.9582&0.9447&0.9321&0.8820&0.8252&0.7368&0.6129\\
&&Up to Bottom&0.9785&0.9640&0.9493&0.9296&0.9085&0.8695&0.8249&0.7559&0.6347\\
\cline{1-12}
\end{tabular*}}
\end{table*}
%--------------------------------------------------------------------------------------------------------------------------------------------------------------------
\begin{table}[H]
\footnotesize
\caption{PSNR comparison of recovered image between TRLG and related works.
\\ Note: - means recovered image is unavailable for current tampering.}
\label{TABLE:compare_recovery}
\renewcommand{\arraystretch}{1.5}
\scalebox{1} {
\begin{tabular*}{\columnwidth}{@{\extracolsep{\fill}}l@{}c@{}c@{}c@{}c@{}c@{}c@{}}
\cline{1-7}
\multicolumn{1}{l}{\multirow{2}{*}{Image}} & \multicolumn{1}{c}{\multirow{2}{*}{Method}} & 
\multicolumn{5}{c}{Tampering rate \%}\\
\cline{3-7}
&&10&20&30&40&50\\
\cline{1-7}
Baboon&TRLG&44.1772&42.5842&41.0369&38.4502&34.8195\\
&\cite{ref16}&39.92&37.00&35.22&34.12&33.16\\
&\cite{ref17}&32.69&29.93&28.29&27.19&26.28\\
&\cite{ref20}&38.05&35.26&33.61&32.51&31.67\\
Lena&TRLG&44.1560&41.8357&40.2214&38.1727&36.5542\\
&\cite{ref14}&43.02&37.92&33.01&32.23&31.14\\
&\cite{ref16}&45.09&40.58&38.25&36.84&35.79\\
&\cite{ref17}&40.20&36.57&33.38&32.14&29.32\\
&\cite{ref20}&39.57&36.15&34.35&33.00&31.94\\
&\cite{ref21}&34.53&31.95&30.75&29.90&-\\
Pepper&TRLG&44.0731&41.7370&40.3887&39.1925&38.0176\\
&\cite{ref22}&42.49&26.53&-&-&-\\
Lake&TRLG&42.0142&41.6306&39.7389&37.5469&36.2926\\
&\cite{ref17}&34.43&31.93&29.46&27.68&25.98\\
&\cite{ref21}&37.80&35.50&33.30&32.05&-\\
F16&TRLG&41.9860&40.2426&38.5662&36.9869&35.9469\\
&\cite{ref21}&36.52&34.60&33.40&32.32&-\\
Elaine&TRLG&43.2848&41.7539&40.2213&38.2445&37.2662\\
&\cite{ref20}&38.59&35.25&33.33&32.15&31.51\\
Goldhill&TRLG&43.8774&40.5553&38.7288&36.5216&34.8993\\
&\cite{ref22}&40.75&-&-&-&-\\
Boat&TRLG&43.3326&39.8584&37.4337&35.3961&34.0040\\
&\cite{ref22}&-&36.90&-&-&-\\
Camera&TRLG&43.8403&40.5883&38.4819&36.7889&36.5658\\
&\cite{ref16}&42.45&38.77&36.37&34.76&33.53\\
&\cite{ref17}&41.31&37.93&32.82&29.42&27.84\\
\cline{1-7}
\end{tabular*}}
\end{table}
%---------------------------------------------------------------------------------------------------------------------------------------------
\begin{figure*}[t!]
\center
\setlength{\tabcolsep}{1pt}
\begin{tabular}{ccccc}
\multirow{3}{*}[0.393in]{\includegraphics[width=0.248\textwidth]{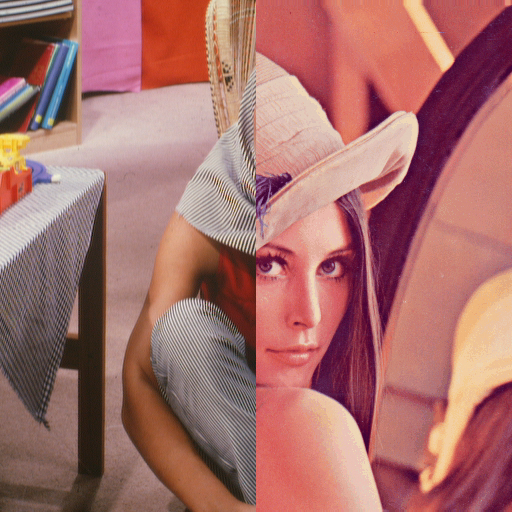}}&
\includegraphics[width=0.148\textwidth]{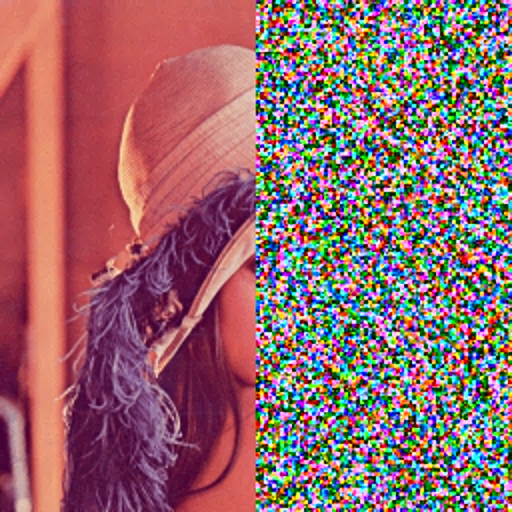} &
\includegraphics[width=0.148\textwidth]{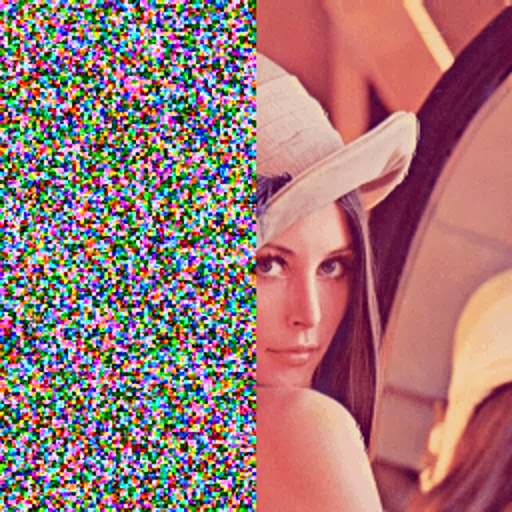} & 
\includegraphics[width=0.148\textwidth]{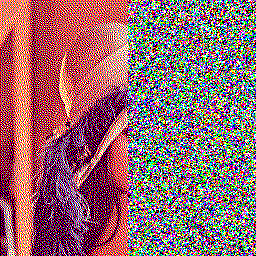}&\multirow{3}{*}[0.393in]{\includegraphics[width=0.248\textwidth]{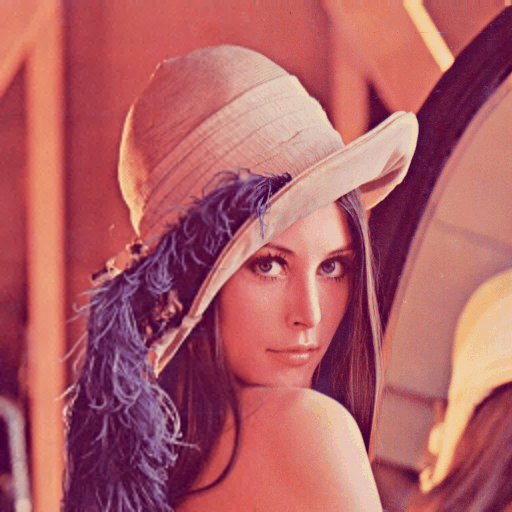}}\\
&(b) & (c) & (d)&\\ 
&\includegraphics[width=0.148\textwidth]{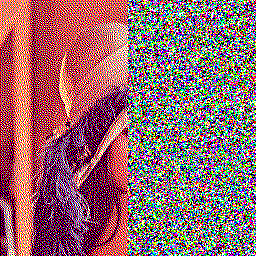} &
\includegraphics[width=0.148\textwidth]{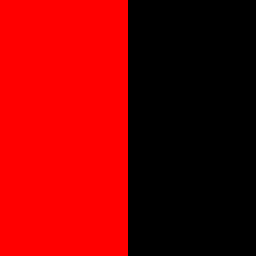} & 
\includegraphics[width=0.148\textwidth]{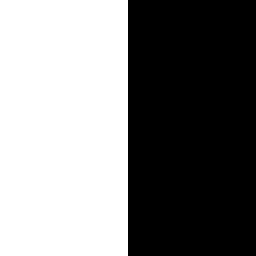}&\\
(a)&(e) & (f) & (g)&(h)\\ 
\end{tabular}
\caption{Tampering test of 50\% image splicing by collage attack. (a) Tampered image, (b) Primary digest 1, (c) Primary digest 2, (d) Secondary digest 1, (e) Secondary digest 2, (f) Recovery map, (g) Tamper detection, (h) Recovered image (PSNR=33.4826, SSIM=0.9885).}
\label{fig:tamperdetection1}
\end{figure*}
%----------------------------------------------------------------------------------------------------------------------------------------------------------
\begin{figure*}[t!]
\center
\setlength{\tabcolsep}{1pt}
\begin{tabular}{ccccc}
\multirow{3}{*}[0.393in]{\includegraphics[width=0.248\textwidth]{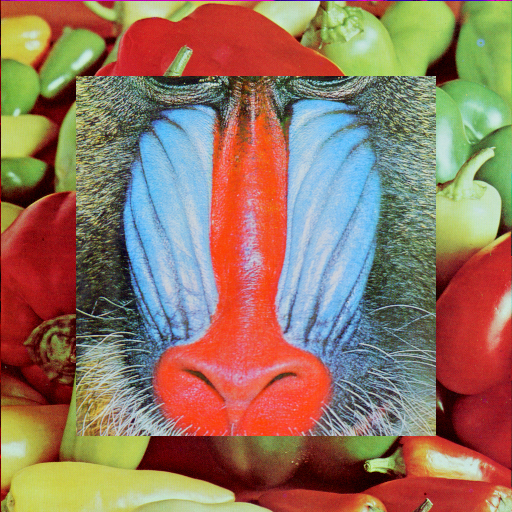}}&
\includegraphics[width=0.148\textwidth]{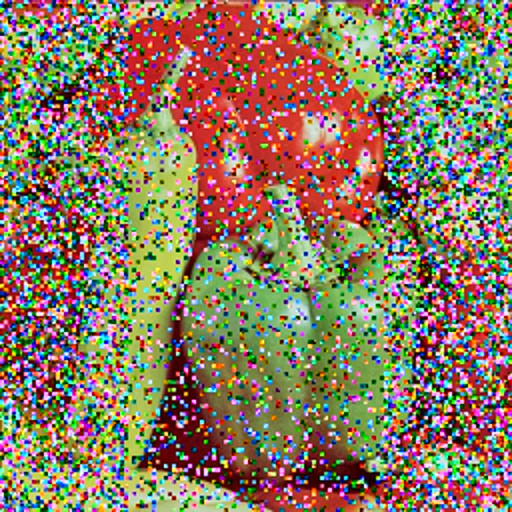} &
\includegraphics[width=0.148\textwidth]{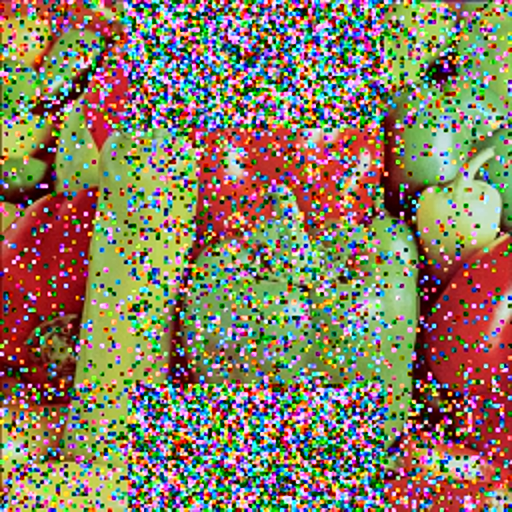} & 
\includegraphics[width=0.148\textwidth]{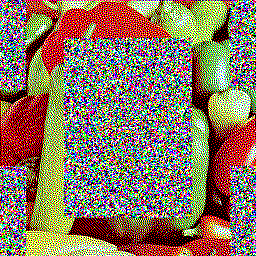}&\multirow{3}{*}[0.393in]{\includegraphics[width=0.248\textwidth]{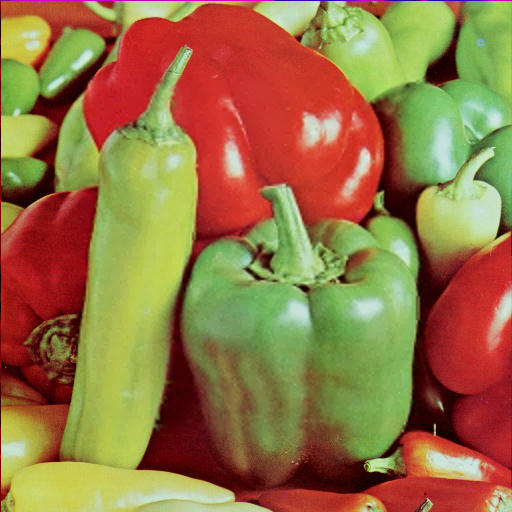}}\\
&(b) & (c) & (d)&\\ 
&\includegraphics[width=0.148\textwidth]{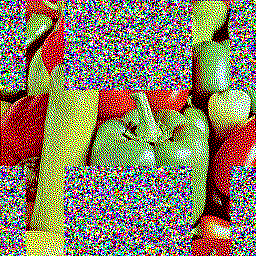} &
\includegraphics[width=0.148\textwidth]{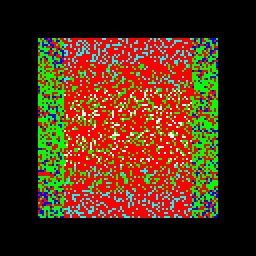} & 
\includegraphics[width=0.148\textwidth]{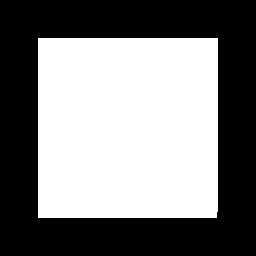}&\\
(a)&(e) & (f) & (g)&(h)\\ 
\end{tabular}
\caption{Tampering test of 70\% image splicing by protocol attack. (a) Tampered image, (b) Primary digest 1, (c) Primary digest 2, (d) Secondary digest 1, (e) Secondary digest 2, (f) Recovery map, (g) Tamper detection, (h) Recovered image (PSNR=32.5198, SSIM=0.9862).}
\label{fig:tamperdetection2}
\end{figure*}
%%----------------------------------------------------------------------------------------------------------------------------------------------------------
\begin{figure*}[t!]
\center
\setlength{\tabcolsep}{1pt}
\begin{tabular}{ccccc}
\multirow{3}{*}[0.393in]{\includegraphics[width=0.248\textwidth]{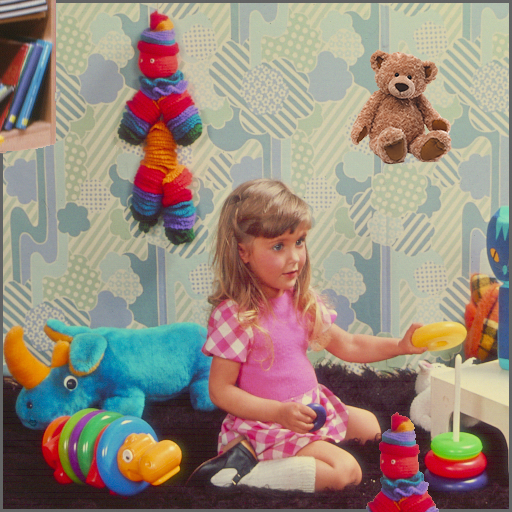}}&
\includegraphics[width=0.148\textwidth]{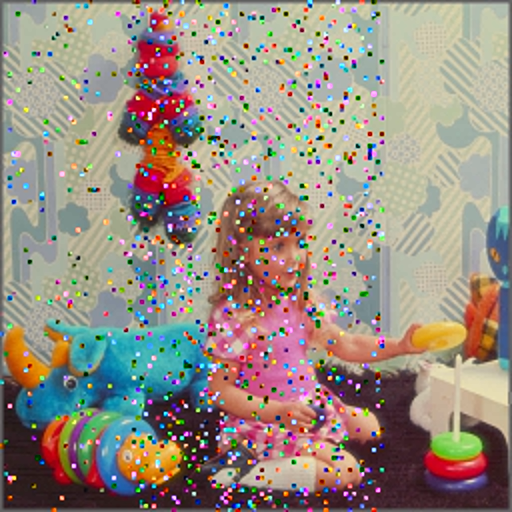} &
\includegraphics[width=0.148\textwidth]{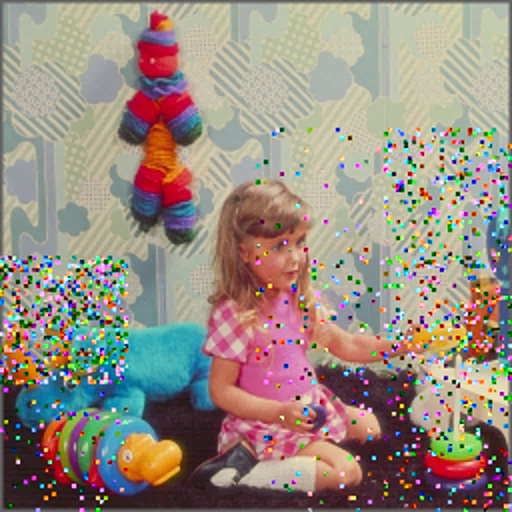} & 
\includegraphics[width=0.148\textwidth]{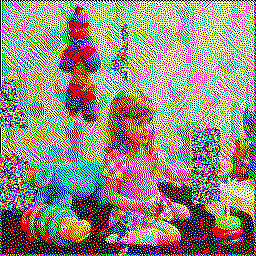}&\multirow{3}{*}[0.393in]{\includegraphics[width=0.248\textwidth]{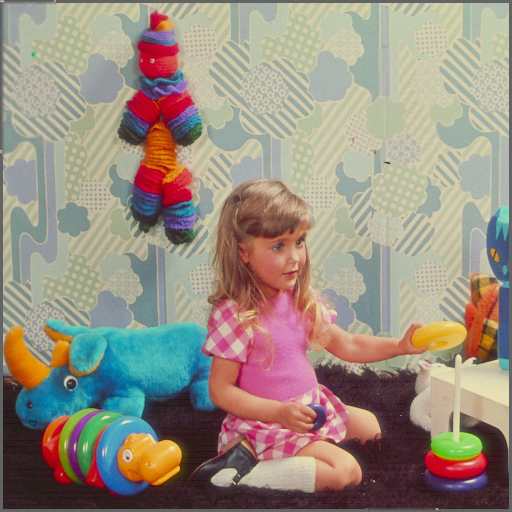}}\\
&(b) & (c) & (d)&\\ 
&\includegraphics[width=0.148\textwidth]{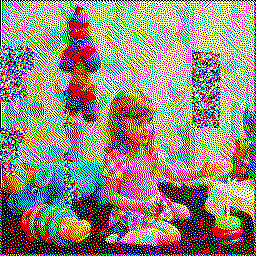} &
\includegraphics[width=0.148\textwidth]{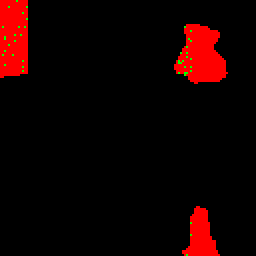} & 
\includegraphics[width=0.148\textwidth]{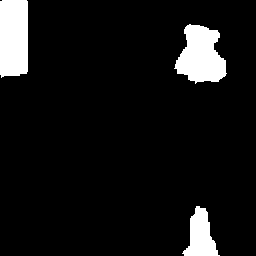}&\\
(a)&(e) & (f) & (g)&(h)\\ 
\end{tabular}
\caption{Tampering test of 10\% add object, copy move, and vector quantization. (a) Tampered image, (b) Primary digest 1, (c) Primary digest 2, (d) Secondary digest 1, (e) Secondary digest 2, (f) Recovery map, (g) Tamper detection, (h) Recovered image (PSNR=40.1968, SSIM=0.9893).}
\label{fig:tamperdetection3}
\end{figure*}
%----------------------------------------------------------------------------------------------------------------------------------------------------------
\begin{sidewaystable}
%\begin{table*}[t!]
\footnotesize
\caption{Compression between TRLG and recent schemes in various terms.\\
Note: Watermarked image (A:45$\leq$PSNR, B:42$\leq$PSNR$<$45, C:PSNR$<$42), RG and - means Random Generator and not supported, respectively.} %Recovered image (A:40$\leq$PSNR, B:35$\leq$PSNR$<$40, C:PSNR$<$35).
%RG means random generator, 
\label{TABLE:compressions_various_terms}
\renewcommand{\arraystretch}{1.5}
\scalebox{1} {
\begin{tabular*}{\textwidth}{@{\extracolsep{\fill} }@{}l@{}c@{}c@{}c@{}c@{}c@{}c@
{}c@{}c@{}c@{}c@{}c@{}c@{}c@{}c@{}c@{}}
\cline{1-16}
Features &TRLG &\cite{ref1}&\cite{ref7}&\cite{ref12}&\cite{ref13}&\cite{ref14} &\cite{ref15}&\cite{ref16}&\cite{ref17}&\cite{ref18}&\cite{ref19}& \cite{ref20}&\cite{ref21}&\cite{ref22}&\cite{ref32}\\
\cline{1-16}
Block size&2$\times$2&2$\times$2&2$\times$2&2$\times$2&2$\times$2&4$\times$4&2$\times$2&2$\times$2&2$\times$2&4$\times$4&4$\times$4&2$\times$2&3$\times$3&2$\times$2&2$\times$2\\
Embedding technique&LSBs&LSBs&LSBs&LSBs&LSBs&LSBs&LSBs&LSBs&LSBs&LSBs&LSBs&LSBs&LSBs&LSBs&Quantize\\
Support color&Yes&Yes&No&Yes&No&Yes&No&Yes&No&No&No&Yes&No&No&No\\
Watermarked image&A&A&C&C&B&B&B&C&C&A&C&C&B&B&C\\
%Recovered image&A&A&C&C&-&B&-&C&C&-&C&C&C&C&-\\
Generating digest&W+H&W+H&AVG&AVG&-&AVG&-&DCT&AVG&-&AVG&DCT&AVG&MSBs&-\\
Number of digest&4&2&2&2&-&2&-&1&2&-&1&1&1&1&-\\
Shuffling digest&CCS&ACM&1-DT&CCM&-&RG&-&1-DT&ACM&-&-&1-DT&-&-&-\\
Payload (bpp)&2&2&3&3&2&2&2&3&3&0.5&2.5&3&2&2&1\\
Watermark security&Yes&Yes&No&No&No&Yes&No&Yes&No&No&Yes&Yes&Yes&Yes&Yes\\
Copy-move&Yes&Yes&No&No&No&Yes&No&No&No&No&Yes&No&No&No&No\\
Vector-quantization&Yes&Yes&No&No&No&Yes&No&No&Yes&No&No&No&No&No&No\\
Collage attack&Yes&No&No&No&No&Yes&No&No&Yes&Yes&No&No&Yes&No&Yes\\
Protocol attack&Yes&No&No&No&No&No&No&No&No&No&No&No&No&No&No\\
Extraction process&Blind&Blind&Blind&Blind&Blind&Blind&Blind&Blind&Blind&S-blind&Blind&Blind&Blind&Blind&Blind\\
\cline{1-16}
\end{tabular*}}
%\end{table*}
\end{sidewaystable}
%----------------------------------------------------------------------------------------------------------------------------------------------------------

\subsection{Visual representation performance of TRLG }
In the last set of experiments, the context of tamper detection and recovery is visually presented to prove the excellent performance of TRLG. For this aim, some type of tampering such as image splicing and adding object based on copy move, collage, vector quantization and protocol attacks are applied on the watermarked image to illustrate the superiority and efficiency of TRLG. For this aim, Photoshop is used to apply these tampering on the watermarked images. These experiments are illustrated in Figs. \ref{fig:tamperdetection1}, \ref{fig:tamperdetection2}, and \ref{fig:tamperdetection3}. In these Figs, recovery map is demonstrated the recovered regions by primary digest 1 (red), primary digest 2 (green), secondary digest 1 (blue), secondary digest 2 (white), and neighboring pixels (cyan).

In Fig. \ref{fig:tamperdetection1}, the watermarked Lena is tampered by half of watermarked Barbara based on collage attack. As can be seen, the TRLG can detect and recover tampered parts, clearly. The PSNR and SSIM of the recovered image from 50\% spliced image is 33.4826 dB and 0.9885, respectively. Next, to simulate the performance of TRLG against protocol attack, 70\% of Baboon is pasted on watermarked Pepper in Fig. \ref{fig:tamperdetection2}. As said before, in protocol attack, the Least Significant Bits of the fake part is replaced by Least Significant Bits of destination image which carried watermarked bits. It can be found that, due to the novel and intelligent shuffling, the recovery rate of the recovered image in TRLG is extremely high. The PSNR and SSIM of the recovered image are 32.5198 dB and 0.9862, respectively. In Fig. \ref{fig:tamperdetection3}, the watermarked Girl is modified based on various tampering. First, the extra object is added. Thereinafter, a copy move and vector quantization are employed by copying parts of watermarked Barbara and watermarked Girl, respectively. It can be clearly seen that the tamper detection is highly accurate, and TRLG can recover the tampered region successfully. The PSNR and SSIM between the original image and the recovered image are 40.1968 dB and 0.9893, respectively.

Generally, the excellence and superiority of TRLG are proved based on various experiments. It is concluded that the essential metrics like PSNR and SSIM are effectively high for watermarked and recovered image. Also, TRLG can detect special tampering accurately. In another word, the overall performance of tampering recovery for TRLG is satisfactory.

%----------------------------------------------------------------------------------------------------------------------------------------------------------------------------------------
\subsection{Comparison with recent schemes in various terms}
There are a number of important characteristics that a tamper detection and recovery schemes should be considered. In this set of experiment, the various properties of TRLG are reported and compared with other fragile schemes in Table \ref{TABLE:compressions_various_terms}. These characteristics include the size of a block (localization), embedding technique, support color, quality of watermarked image, the technique of generating the digest, provide multiple chances, shuffling digest, payload, security of watermark, detect special tampering, and extraction process. As illustrated, It is clear that TRLG is superior, efficient and effective compared to other schemes.
%----------------------------------------------------------------------------------------------------------------------------------------------------------
\section{Conclusion and future works}
Digital watermarking is a science of hiding information in digital media. The embedded information should be later extracted for various goals such as authenticate ownership, tamper detection and recovery, and so on. The most methods have been proposed in recent years have single chance for recovering tampered regions, and poor quality for watermarked and recovered images. In addition, previous methods generate digest based on traditional averaging techniques without considering the characteristics of the block. How to design efficient digest in fragile tampering recovery scheme which works in spatial or frequency domain plays an important role.

In this paper, an efficient fragile blind quad watermarking scheme for image tamper detection and recovery based on lifting wavelet transform and genetic algorithm is proposed.  A novel compact digests generating with super quality has been proposed in TRLG. For this purpose, four digests are generated based on the LWT and halftoning technique. Generating compress digests with admissible quality can be referred as an optimization problem. Hence, the genetic algorithm is used to optimize the data which fetched from rough and smooth regions for generating the digest. To do so, the best thresholds are selected by GA for classifying the type of blocks. Totally, for each 2$\times$2 non-overlapping block two different digests as primary and secondary are used which each of them provides two chances for recover tampered regions. As been expected, the performance of TRLG in term of quality of the recovered image can be significantly improved, especially when the rate of tampering is extremely high. Experimental results have proved this claim. In order to guaranty, the security of TRLG, the combining modified Logistic map as CCS is used to shuffle the digests and encrypt the watermarks, respectively. As said before, in TRLG the watermarks embedding procedure is modeled as a search problem to minimize and optimize the difference between the original and watermark values. To do so, the genetic algorithm is applied, again. 

Experiment results have demonstrated the superiority of the TRLG in comparison to other state-of-the-art schemes, particularly in large tampering rates. PSNR and SSIM represent that TRLG has good imperceptibility after embedding process. Totally, TRLG achieves high quality for watermarked and recovered image without blocky artifact, low complexity, and well localization. Furthermore, TRLG has good results in terms of security, and detect security tamperings such as copy move, vector quantization, collage, and protocol attack. Finally, based on the advantages described above, TRLG is efficient, secure, safe, and applicable for blind and fragile applications.
Because of the watermark is embedded into LSBs planes, the watermark may be destroyed by image processing operations or other attacks. In the ongoing research, we will extend TRLG as semi-fragile scheme. Since the capacity in the frequency domain is lower than the spatial domain, a compact digest with super quality can be used to solve one of the main challenges that exist there.

%------------------------------------------------------------------------------------------------------------------------------------------------------------------------------------------------------------------------
%\section*{References}
%\bibliographystyle{elsarticle-num}
%\bibliography{bibliography}

\begin{thebibliography}{10}
\expandafter\ifx\csname url\endcsname\relax
  \def\url#1{\texttt{#1}}\fi
\expandafter\ifx\csname urlprefix\endcsname\relax\def\urlprefix{URL }\fi
\expandafter\ifx\csname href\endcsname\relax
  \def\href#1#2{#2} \def\path#1{#1}\fi

\bibitem{ref11}
F.~Y. Shih, Digital Watermarking and Steganography: Fundamentals and
  Techniques, Second Edition, CRC press, 2017.

\bibitem{ref8}
P.~Singh, R.~Chadha, A survey of digital watermarking techniques, applications
  and attacks, International Journal of Engineering and Innovative Technology
  (IJEIT) 2~(9) (2013) 165--175.

\bibitem{ref10}
P.~Korus, Digital image integrity – a survey of protection and verification
  techniques, Digital Signal Processing 71 (2017) 1 -- 26.
\newblock \href {http://dx.doi.org/https://doi.org/10.1016/j.dsp.2017.08.009}
  {\path{doi:https://doi.org/10.1016/j.dsp.2017.08.009}}.

\bibitem{ref26}
Z.~Shao, Y.~Shang, Y.~Zhang, X.~Liu, G.~Guo,
  \href{http://www.sciencedirect.com/science/article/pii/S0165168415003485}{Robust
  watermarking using orthogonal fourier–mellin moments and chaotic map for
  double images}, Signal Processing 120 (2016) 522 -- 531.
\newblock \href
  {http://dx.doi.org/https://doi.org/10.1016/j.sigpro.2015.10.005}
  {\path{doi:https://doi.org/10.1016/j.sigpro.2015.10.005}}.
\newline\urlprefix\url{http://www.sciencedirect.com/science/article/pii/S0165168415003485}

\bibitem{ref27}
C.~Wang, X.~Wang, C.~Zhang, Z.~Xia,
  \href{http://www.sciencedirect.com/science/article/pii/S0165168416303528}{Geometric
  correction based color image watermarking using fuzzy least squares support
  vector machine and bessel k form distribution}, Signal Processing 134 (2017)
  197 -- 208.
\newblock \href
  {http://dx.doi.org/https://doi.org/10.1016/j.sigpro.2016.12.010}
  {\path{doi:https://doi.org/10.1016/j.sigpro.2016.12.010}}.
\newline\urlprefix\url{http://www.sciencedirect.com/science/article/pii/S0165168416303528}

\bibitem{ref29}
S.~H. Soleymani, A.~H. Taherinia,
  \href{https://doi.org/10.1007/s11042-016-3734-2}{Double expanding robust
  image watermarking based on spread spectrum technique and bch coding},
  Multimedia Tools and Applications 76~(3) (2017) 3485--3503.
\newblock \href {http://dx.doi.org/10.1007/s11042-016-3734-2}
  {\path{doi:10.1007/s11042-016-3734-2}}.
\newline\urlprefix\url{https://doi.org/10.1007/s11042-016-3734-2}

\bibitem{ref30}
A.~H. Taherinia, M.~Jamzad, A robust image watermarking using two level dct and
  wavelet packets denoising, in: 2009 International Conference on Availability,
  Reliability and Security, 2009, pp. 150--157.
\newblock \href {http://dx.doi.org/10.1109/ARES.2009.132}
  {\path{doi:10.1109/ARES.2009.132}}.

\bibitem{ref9}
K.~Sreenivas, V.~Kamkshi~Prasad,
  \href{https://doi.org/10.1007/s13042-017-0641-4}{Fragile watermarking schemes
  for image authentication: a survey}, International Journal of Machine
  Learning and Cybernetics\href {http://dx.doi.org/10.1007/s13042-017-0641-4}
  {\path{doi:10.1007/s13042-017-0641-4}}.
\newline\urlprefix\url{https://doi.org/10.1007/s13042-017-0641-4}

\bibitem{ref1}
B.~Bolourian~Haghighi, A.~H. Taherinia, A.~Harati,
  \href{http://www.sciencedirect.com/science/article/pii/S1047320317301876}{Trlh:
  Fragile and blind dual watermarking for image tamper detection and
  self-recovery based on lifting wavelet transform and halftoning technique},
  Journal of Visual Communication and Image Representation 50 (2018) 49 -- 64.
\newblock \href {http://dx.doi.org/https://doi.org/10.1016/j.jvcir.2017.09.017}
  {\path{doi:https://doi.org/10.1016/j.jvcir.2017.09.017}}.
\newline\urlprefix\url{http://www.sciencedirect.com/science/article/pii/S1047320317301876}

\bibitem{ref7}
T.-Y. Lee, S.~D. Lin,
  \href{http://www.sciencedirect.com/science/article/pii/S003132030800174X}{Dual
  watermark for image tamper detection and recovery}, Pattern Recognition
  41~(11) (2008) 3497 -- 3506.
\newblock \href
  {http://dx.doi.org/https://doi.org/10.1016/j.patcog.2008.05.003}
  {\path{doi:https://doi.org/10.1016/j.patcog.2008.05.003}}.
\newline\urlprefix\url{http://www.sciencedirect.com/science/article/pii/S003132030800174X}

\bibitem{ref12}
X.~Tong, Y.~Liu, M.~Zhang, Y.~Chen,
  \href{http://www.sciencedirect.com/science/article/pii/S092359651200224X}{A
  novel chaos-based fragile watermarking for image tampering detection and
  self-recovery}, Signal Processing: Image Communication 28~(3) (2013) 301 --
  308.
\newblock \href {http://dx.doi.org/https://doi.org/10.1016/j.image.2012.12.003}
  {\path{doi:https://doi.org/10.1016/j.image.2012.12.003}}.
\newline\urlprefix\url{http://www.sciencedirect.com/science/article/pii/S092359651200224X}

\bibitem{ref13}
J.~Zhang, Q.~Zhang, H.~Lv,
  \href{http://www.sciencedirect.com/science/article/pii/S0030402613007031}{A
  novel image tamper localization and recovery algorithm based on watermarking
  technology}, Optik - International Journal for Light and Electron Optics
  124~(23) (2013) 6367 -- 6371.
\newblock \href {http://dx.doi.org/https://doi.org/10.1016/j.ijleo.2013.05.040}
  {\path{doi:https://doi.org/10.1016/j.ijleo.2013.05.040}}.
\newline\urlprefix\url{http://www.sciencedirect.com/science/article/pii/S0030402613007031}

\bibitem{ref14}
S.~Dadkhah, A.~A. Manaf, Y.~Hori, A.~E. Hassanien, S.~Sadeghi,
  \href{http://www.sciencedirect.com/science/article/pii/S0923596514001313}{An
  effective svd-based image tampering detection and self-recovery using active
  watermarking}, Signal Processing: Image Communication 29~(10) (2014) 1197 --
  1210.
\newblock \href {http://dx.doi.org/https://doi.org/10.1016/j.image.2014.09.001}
  {\path{doi:https://doi.org/10.1016/j.image.2014.09.001}}.
\newline\urlprefix\url{http://www.sciencedirect.com/science/article/pii/S0923596514001313}

\bibitem{ref15}
C.-S. Hsu, S.-F. Tu,
  \href{http://www.sciencedirect.com/science/article/pii/S0030401809013583}{Probability-based
  tampering detection scheme for digital images}, Optics Communications 283~(9)
  (2010) 1737 -- 1743.
\newblock \href
  {http://dx.doi.org/https://doi.org/10.1016/j.optcom.2009.12.073}
  {\path{doi:https://doi.org/10.1016/j.optcom.2009.12.073}}.
\newline\urlprefix\url{http://www.sciencedirect.com/science/article/pii/S0030401809013583}

\bibitem{ref16}
D.~Singh, S.~K. Singh, \href{https://doi.org/10.1007/s11042-015-3010-x}{Dct
  based efficient fragile watermarking scheme for image authentication and
  restoration}, Multimedia Tools and Applications 76~(1) (2017) 953--977.
\newblock \href {http://dx.doi.org/10.1007/s11042-015-3010-x}
  {\path{doi:10.1007/s11042-015-3010-x}}.
\newline\urlprefix\url{https://doi.org/10.1007/s11042-015-3010-x}

\bibitem{ref17}
K.~Sreenivas, V.~Kamakshiprasad,
  \href{http://www.sciencedirect.com/science/article/pii/S1047320317301761}{Improved
  image tamper localisation using chaotic maps and self-recovery}, Journal of
  Visual Communication and Image Representation 49 (2017) 164 -- 176.
\newblock \href {http://dx.doi.org/https://doi.org/10.1016/j.jvcir.2017.09.001}
  {\path{doi:https://doi.org/10.1016/j.jvcir.2017.09.001}}.
\newline\urlprefix\url{http://www.sciencedirect.com/science/article/pii/S1047320317301761}

\bibitem{ref18}
A.~Azeroual, K.~Afdel,
  \href{http://www.sciencedirect.com/science/article/pii/S1434841117300134}{Real-time
  image tamper localization based on fragile watermarking and faber-schauder
  wavelet}, AEU - International Journal of Electronics and Communications 79
  (2017) 207 -- 218.
\newblock \href {http://dx.doi.org/https://doi.org/10.1016/j.aeue.2017.06.001}
  {\path{doi:https://doi.org/10.1016/j.aeue.2017.06.001}}.
\newline\urlprefix\url{http://www.sciencedirect.com/science/article/pii/S1434841117300134}

\bibitem{ref19}
C.-S. Hsu, S.-F. Tu,
  \href{http://www.sciencedirect.com/science/article/pii/S0263224116300306}{Image
  tamper detection and recovery using adaptive embedding rules}, Measurement 88
  (2016) 287 -- 296.
\newblock \href
  {http://dx.doi.org/https://doi.org/10.1016/j.measurement.2016.03.053}
  {\path{doi:https://doi.org/10.1016/j.measurement.2016.03.053}}.
\newline\urlprefix\url{http://www.sciencedirect.com/science/article/pii/S0263224116300306}

\bibitem{ref20}
D.~Singh, S.~K. Singh,
  \href{http://www.sciencedirect.com/science/article/pii/S1047320316300566}{Effective
  self-embedding watermarking scheme for image tampered detection and
  localization with recovery capability}, Journal of Visual Communication and
  Image Representation 38 (2016) 775 -- 789.
\newblock \href {http://dx.doi.org/https://doi.org/10.1016/j.jvcir.2016.04.023}
  {\path{doi:https://doi.org/10.1016/j.jvcir.2016.04.023}}.
\newline\urlprefix\url{http://www.sciencedirect.com/science/article/pii/S1047320316300566}

\bibitem{ref21}
C.~Qin, P.~Ji, X.~Zhang, J.~Dong, J.~Wang,
  \href{http://www.sciencedirect.com/science/article/pii/S0165168417301251}{Fragile
  image watermarking with pixel-wise recovery based on overlapping embedding
  strategy}, Signal Processing 138 (2017) 280 -- 293.
\newblock \href
  {http://dx.doi.org/https://doi.org/10.1016/j.sigpro.2017.03.033}
  {\path{doi:https://doi.org/10.1016/j.sigpro.2017.03.033}}.
\newline\urlprefix\url{http://www.sciencedirect.com/science/article/pii/S0165168417301251}

\bibitem{ref22}
F.~Cao, B.~An, J.~Wang, D.~Ye, H.~Wang,
  \href{http://www.sciencedirect.com/science/article/pii/S0141938216301391}{Hierarchical
  recovery for tampered images based on watermark self-embedding}, Displays 46
  (2017) 52 -- 60.
\newblock \href
  {http://dx.doi.org/https://doi.org/10.1016/j.displa.2017.01.001}
  {\path{doi:https://doi.org/10.1016/j.displa.2017.01.001}}.
\newline\urlprefix\url{http://www.sciencedirect.com/science/article/pii/S0141938216301391}

\bibitem{ref23}
W.~Hong, M.~Chen, T.~S. Chen,
  \href{http://www.sciencedirect.com/science/article/pii/S0923596517301248}{An
  efficient reversible image authentication method using improved pvo and lsb
  substitution techniques}, Signal Processing: Image Communication 58 (2017)
  111 -- 122.
\newblock \href {http://dx.doi.org/https://doi.org/10.1016/j.image.2017.07.001}
  {\path{doi:https://doi.org/10.1016/j.image.2017.07.001}}.
\newline\urlprefix\url{http://www.sciencedirect.com/science/article/pii/S0923596517301248}

\bibitem{ref32}
C.-C. Lin, Y.~Huang, W.-L. Tai,
  \href{http://dx.doi.org/10.1007/s11042-015-3059-6}{A novel hybrid image
  authentication scheme based on absolute moment block truncation coding},
  Multimedia Tools and Applications 76~(1) (2017) 463--488.
\newblock \href {http://dx.doi.org/10.1007/s11042-015-3059-6}
  {\path{doi:10.1007/s11042-015-3059-6}}.
\newline\urlprefix\url{http://dx.doi.org/10.1007/s11042-015-3059-6}

\bibitem{ref31}
W.~Sweldens, The lifting scheme: A construction of second generation wavelets,
  SIAM Journal on Mathematical Analysis 29~(2) (1998) 511--546.

\bibitem{ref33}
B.~Bolourian~Haghighi, A.~H. Taherinia, R.~Monsefi, Trlf: An effective
  semi-fragile watermarking method for tamper detection and recovery based on
  lwt and fnn, arXiv preprint arXiv:1802.07119.

\bibitem{ref4}
J.~H. Holland, Genetic algorithms, Scientific american 267~(1) (1992) 66--73.

\bibitem{ref3}
D.~E. Goldberg, Genetic algorithms, Pearson Education India, 2006.

\bibitem{ref2}
C.~Pak, L.~Huang,
  \href{http://www.sciencedirect.com/science/article/pii/S0165168417300932}{A
  new color image encryption using combination of the 1d chaotic map}, Signal
  Processing 138 (2017) 129 -- 137.
\newblock \href
  {http://dx.doi.org/https://doi.org/10.1016/j.sigpro.2017.03.011}
  {\path{doi:https://doi.org/10.1016/j.sigpro.2017.03.011}}.
\newline\urlprefix\url{http://www.sciencedirect.com/science/article/pii/S0165168417300932}

\bibitem{ref28}
S.~H. Soleymani, A.~H. Taherinia,
  \href{https://doi.org/10.1007/s11042-016-4009-7}{High capacity image
  steganography on sparse message of scanned document image (smsdi)},
  Multimedia Tools and Applications 76~(20) (2017) 20847--20867.
\newblock \href {http://dx.doi.org/10.1007/s11042-016-4009-7}
  {\path{doi:10.1007/s11042-016-4009-7}}.
\newline\urlprefix\url{https://doi.org/10.1007/s11042-016-4009-7}

\bibitem{ref25}
J.~F. Jarvis, C.~N. Judice, W.~Ninke, A survey of techniques for the display of
  continuous tone pictures on bilevel displays, Computer Graphics and Image
  Processing 5~(1) (1976) 13--40.

\bibitem{ref24}
P.-E. Axelson, Quality measures of halftoned images (a review) (2003) 81.

\bibitem{ref5}
X.~Hou, G.~Qiu, Image companding and inverse halftoning using deep
  convolutional neural networks, arXiv preprint arXiv:1707.00116.

\bibitem{ref6}
R.~Neelamani, R.~D. Nowak, R.~G. Baraniuk, Winhd: Wavelet-based inverse
  halftoning via deconvolution, IEEE Transactions on Image Processing (2002)
  1--22.

\end{thebibliography}

\end{document}